\documentclass[letterpaper,11pt,fleqn]{article}
\usepackage{jheppub}
\usepackage{bm,amsmath,amssymb}

\long\def\comment#1{ }

\newcommand{\eqn}[1]{Eq.~\eqref{#1}}

\newcommand{\beq}{\begin{equation}}
\newcommand{\eeq}{\end{equation}}
\newcommand{\nn}{\nonumber\\}

\newcommand{\dif}{{\rm d}}
\newcommand{\rmd}{{\rm d}}
\newcommand{\rme}{{\rm e}}
\newcommand{\rmi}{{\rm i}}
\newcommand{\rmy}{{y}}
\newcommand{\rmtr}{{\rm tr}}
\newcommand{\rmTr}{{\rm Tr}}

\newcommand{\del}{\partial}

\newcommand{\lan}{\langle}
\newcommand{\ran}{\rangle}

\newcommand{\order}[1]{\mcal{O}{(#1)}}
\newcommand{\mcal}{\mathcal}
\newcommand{\wt}{\widetilde}

\newcommand{\bmk}{\bm{k}}
\newcommand{\bmx}{\bm{x}}
\newcommand{\bmy}{\bm{y}}
\newcommand{\bmu}{\bm{u}}
\newcommand{\bmv}{\bm{v}}
\newcommand{\bmz}{\bm{z}}

\newcommand{\bmw}{\bm{w}}

\newcommand{\barg}{\bar{\gamma}}
\newcommand{\atpi}{\frac{\bar{\alpha}}{2 \pi}}
\newcommand{\abar}{\bar{\alpha}}

\let\Oldcdot\cdot
\renewcommand{\cdot}{\mspace{-2mu}\Oldcdot\mspace{-2mu}}
\let\Oldtimes\times
\renewcommand{\times}{\mspace{-2mu}\Oldtimes\mspace{-2mu}}

\title{\Large JIMWLK evolution in the Gaussian approximation}

\author[a]{E.~Iancu}
\author[b]{and D.N.~Triantafyllopoulos}

\affiliation[a]{Institut de Physique Th\'{e}orique de Saclay,
F-91191 Gif-sur-Yvette, France}
\affiliation[b]{ECT*, European Center for Theoretical Studies in Nuclear
Physics and Related Areas,\\ Strada delle Tabarelle 286, I-38123 Villazzano
(TN), Italy}

\emailAdd{edmond.iancu@cea.fr}
\emailAdd{trianta@ectstar.eu}

\abstract{We demonstrate that the Balitsky--JIMWLK equations describing the
high--energy evolution of the $n$--point functions of the Wilson lines
(the QCD scattering amplitudes in the eikonal approximation) admit a
controlled mean field approximation of the Gaussian type, for any value
of the number of colors $N_c$. This approximation is strictly correct in
the weak scattering regime at relatively large transverse momenta, where
it reproduces the BFKL dynamics, and in the strong scattering regime
deeply at saturation, where it properly describes the evolution of the
scattering amplitudes towards the respective black disk limits. The
approximation scheme is fully specified by giving the 2--point function
(the $S$--matrix for a color dipole), which in turn can be related to the
solution to the Balitsky--Kovchegov equation, including at finite $N_c$.
Any higher $n$--point function with $n\ge 4$ can be computed in terms of
the dipole $S$--matrix by solving a closed system of evolution equations
(a simplified version of the respective Balitsky--JIMWLK equations) which
are local in the transverse coordinates.
For simple configurations of the projectile in the transverse plane, our
new results for the 4--point and the 6--point functions coincide with the
high--energy extrapolations of the respective results in the
McLerran--Venugopalan model.
One cornerstone of our construction is a symmetry property of the JIMWLK
evolution, that we notice here for the first time: the fact that, with
increasing energy, a hadron is expanding its longitudinal support
symmetrically around the light--cone. This corresponds to invariance
under time reversal for the scattering amplitudes.
}
\arxivnumber{1112.1104}
\vfill
\begin{document}
\maketitle

\section{Introduction}
\label{sect:intro}

The final state of an ultrarelativistic hadron--hadron collision, as
currently explored at RHIC and the LHC, is characterized by an extreme
complexity in terms of the number and the distribution of the produced
particles. The study of multiparticle correlations represents an
essential tool for organizing this complexity and extracting physical
information out of it. In particular, a recent measurement at RHIC of
di--hadron correlations in deuteron--gold collisions
\cite{Braidot:2011zj} revealed an interesting phenomenon --- the
azimuthal correlations are rapidly suppressed when increasing the
rapidity towards the fragmentation region of the deuteron  ---, which is
qualitatively
\cite{JalilianMarian:2004da,Nikolaev:2005dd,Baier:2005dv,Marquet:2007vb}
and even semi--quantitatively \cite{Albacete:2010pg,Stasto:2011ru}
consistent with the physical picture of gluon saturation in the nuclear
wavefunction. For this interpretation to be firmly established, one needs
a more precise understanding of the multi--particle correlations in the
high--energy scattering and, in particular, of their evolution with
increasing rapidity. This triggered new theoretical studies
\cite{Dumitru:2010ak,Dominguez:2011wm,Dominguez:2011gc,Dumitru:2011vk,Iancu:2011ns}
of many--body correlations in the color glass condensate (CGC), which is
the QCD effective theory for high--energy evolution and gluon saturation,
to leading logarithmic accuracy at least.

%(see
%Refs.~\cite{Iancu:2002xk,Iancu:2003xm,Weigert:2005us,JalilianMarian:2005jf,Gelis:2010nm}
%for reviews and references).

The central ingredient in the CGC theory is the JIMWLK (Jalilian-Marian,
Iancu, McLerran, Weigert, Leonidov, Kovner) equation
\cite{JalilianMarian:1997jx,JalilianMarian:1997gr,JalilianMarian:1997dw,Kovner:2000pt,Weigert:2000gi,Iancu:2000hn,Iancu:2001ad,Ferreiro:2001qy},
a functional renormalization group equation of the Fokker--Planck type
which describes the non--linear evolution of the gluon distribution in
the hadron wavefunction with increasing rapidity, or decreasing the gluon
longitudinal momentum fraction $x$. When applied to an asymmetric,
`dilute--dense', scattering (like a $pA$ collision), the JIMWLK evolution
can be equivalently reformulated as an infinite hierarchy of ordinary
evolution equations, originally derived by Balitsky
\cite{Balitsky:1995ub}, which refer to gauge--invariant correlations
built with products of Wilson lines. A `Wilson line' is a path--ordered
exponential of the color field in the target. It describes the scattering
between a parton from the projectile (the proton) and the dense gluonic
system in the target (the nucleus), in the eikonal approximation. Via the
optical theorem, the $n$--point functions of the Wilson lines can be
related to cross--sections for particle production in $pA$ collisions.
For instance, the single--inclusive quark (or gluon) production is
related to the $S$--matrix of a `color dipole' (the 2--point function of
the Wilson lines). Similarly, the production of a pair of partons with
similar rapidities is related to the `color quadrupole' (the 4--point
function). The suppression of azimuthal di--hadron correlations in d+Au
collisions at RHIC \cite{Braidot:2011zj} occurs in the right range of
transverse momenta, of the order of the nuclear saturation momentum
$Q_s\sim 1$~GeV, to be interpreted as a result of gluon saturation and
multiple interactions in the scattering of the quadrupole. Such
non--linear phenomena are mean field effects, which are likely to be
correctly described by the JIMWLK evolution, although the latter is known
to miss another important class of correlations --- those associated with
gluon number fluctuations in the dilute regime, or `Pomeron
loops'\footnote{The `Pomeron loops' are formally higher order effects
and, moreover, they are supressed by the running of the coupling
--- at least in the calculation of the single inclusive particle
production \cite{Dumitru:2007ew}. However, there is currently no reliable
estimate of their effects on correlations in multi--particle production.}
\cite{Iancu:2004es,Iancu:2004iy,Mueller:2005ut,Iancu:2005nj,Kovner:2005nq}.

Motivated by the above considerations, there were several recent studies
of the quadrupole evolution in the framework of the Balitsky--JIMWLK
equations \cite{Dominguez:2011gc,Dumitru:2011vk,Iancu:2011ns}. The
results in Ref.~\cite{Dumitru:2011vk} appeared as particularly
intriguing. In that paper, one has numerically solved the JIMWLK equation
by using its representation as a functional Langevin process
\cite{Blaizot:2002xy} and used the results to evaluate the quadrupole
$S$--matrix for different rapidities and for special configurations of
the 4 external points in the transverse plane. Remarkably, the results
thus obtained show a very good agreement with the heuristic extrapolation
to high energy of the corresponding results in the McLerran--Venugopalan
(MV) model \cite{McLerran:1993ni,McLerran:1993ka}. We recall that the MV
model refers to a large nucleus ($A\gg 1$) at not too high energy (where
the effects of the evolution are still negligible) and that in this model
the CGC weight function is taken to be a Gaussian: the only non--trivial
correlation of the color fields in the nucleus is their 2--point
function, the `unintegrated gluon distribution'. The `high--energy
extrapolation' alluded to above refers to using the MV expression for the
quadrupole $S$--matrix in terms of the dipole $S$--matrix
\cite{JalilianMarian:2004da,Dominguez:2011wm}, but with the latter taken
from the numerical solution to the JIMWLK equation at the rapidity of
interest.

Such extrapolations have often been used for phenomenological studies
\cite{Kovner:2001vi,Blaizot:2004wv,JalilianMarian:2004da,Marquet:2010cf,Kuokkanen:2011je,Dominguez:2011wm,Muller:2011bb},
but their justification from the viewpoint of the high--energy evolution
remained obscure. A Gaussian {\em Ansatz} has also been used for mean
field studies of the Balitsky--JIMWLK evolution
\cite{Iancu:2001md,Iancu:2002xk,Iancu:2002aq,Kovchegov:2008mk}. But these
previous studies have not convincingly addressed the issue of the
validity of the Gaussian approximation --- in particular, they did not
justify its suitability for describing higher $n$--point functions, such
as the quadrupole. (The only, qualitative, attempt in that sense is the
`random phase approximation' proposed in Ref.~\cite{Iancu:2001md}; see
the discussion in Sect.~\ref{sect:RPA} below.) In principle, there is no
contradiction between having a Gaussian weight function for the target
field and still generating non--trivial correlations in the scattering of
many--body projectiles: indeed, the scattering amplitudes are built with
Wilson lines, which are non--linear in the target field to all orders.
But within the context of the JIMWLK evolution, such Gaussian
approximations seem to be prohibited by the highly non--linear structure
of the evolution equation, which is the mathematical expression of gluon
saturation.

In spite of this theoretical prejudice, the numerical results in
Ref.~\cite{Dumitru:2011vk} suggest that a Gaussian approximation to the
JIMWLK evolution may nevertheless work. Another piece of evidence in that
sense emerges from the recent analytic study in Ref.~\cite{Iancu:2011ns}.
There, we have constructed an approximate version of the Balitsky--JIMWLK
hierarchy which is simple enough to allow for explicit solutions. Then we
have showed that, for the special configurations of the quadrupole
considered in Ref.~\cite{Dumitru:2011vk}, these approximate solutions
coincide with the respective predictions of the MV model extrapolated to
high energy. But in that context too, the similarity with the MV model
appears as merely an `accident', with no deep motivation: the simplified
hierarchy proposed in Ref.~\cite{Iancu:2011ns} is generated by the
`virtual' piece of the JIMWLK Hamiltonian, which is non--linear in the
target field and therefore seems incompatible with a Gaussian
approximation. Moreover, the approximations in Ref.~\cite{Iancu:2011ns}
have been justified only in the limit where the number of colors $N_c$ is
large (formally, $N_c\to\infty$). This does not explain the observation
in Ref.~\cite{Dumitru:2011vk} that the numerical solutions to the JIMWLK
evolution for $N_c=3$ are better reproduced by the finite--$N_c$ version
of the MV model (with $N_c=3$, of course) than by its large--$N_c$ limit.

Our purpose in the present analysis is to clarify such `coincidences' and
`apparent contradictions' by resolving the aforementioned tensions
between the simplified hierarchy proposed in Ref.~\cite{Iancu:2011ns},
the Gaussian approximation, and the large--$N_c$ limit. The results that
we shall obtain can be summarized as follows. We shall demonstrate that
the JIMWLK equation admits indeed an approximate Gaussian solution for
the CGC weight function, that this solution is unique within the limits
of its accuracy, and that it is tantamount to a simplified system of
evolution equations, which are linear (while being consistent with
unitarity) and local in the transverse coordinates. In the limit where
$N_c\to\infty$, these new equations reduce to those previously proposed
in Ref.~\cite{Iancu:2011ns}. The ultimate outcome of our analysis is a
global approximation to the Balitsky--JIMWLK hierarchy, which is valid
for any $N_c$ and allows one to construct explicit, analytic, solutions
for all the $n$--point functions of the Wilson lines. These approximate
solutions are strictly correct in the limiting regimes at very large
($k_\perp\gg Q_s(Y)$) and, respectively, very small ($k_\perp\ll Q_s(Y)$)
transverse momenta, and provide a smooth (infinitely differentiable)
interpolation between these limits. Here, $Q_s(Y)$ denotes the saturation
momentum in the target at a rapidity $Y$ equal with the rapidity
separation between the target and the projectile.

To describe our results in more detail, let us first explain the
distinction between `real' and `virtual' terms in the Balitsky--JIMWLK
equations. The `real' terms describe the evolution of the projectile via
the emission of small--$x$ gluons, whereas the `virtual' terms express
the probability for the projectile {\em not} to evolve, {\em i.e.} not to
radiate such (small--$x$) gluons. The `virtual' terms dominate the
evolution in the approach towards the unitarity (or `black disk') limit,
since in that regime the scattering is strong and the projectile has more
chances to survive unscattered if it remains `simple' --- {\em i.e.}, if
it does not evolve by emitting more gluons. By the same token, the
`virtual' terms control the evolution of the many--body correlations
which, within the context of JIMWLK, are built exclusively via
non--linear effects (multiple scattering and gluon recombination) in the
regime of strong scattering. More precisely, the `real' terms are
important for that process too --- they include the non--linear effects
responsible for unitarity and saturation ---, but deeply at saturation
their role becomes very simple: they merely prohibit the emission of new
gluons with low transverse momenta $k_\perp\lesssim Q_s(Y)$. Thus, one
can follow the evolution of correlations at saturation by keeping only
the `virtual' terms in the Balitsky--JIMWLK equations, but supplementing
them with a phase--space cutoff which expresses the effect of the `real'
terms. (This is strictly correct in a `leading--logarithmic
approximation' to be detailed in Sect.~\ref{sect:RPA}.) Moreover, since
the simplified equations thus obtained are {\em linear}, they can be
extended to also cover the BFKL evolution in the weak scattering regime
at $k_\perp\gg Q_s(Y)$. Indeed, in that regime and to the accuracy of
interest, the $n$--point functions of the Wilson lines reduce to linear
combinations of the dipole scattering amplitude, with the latter obeying
the BFKL equation. The BFKL dynamics involves both `real' and `virtual'
terms, but it can be effectively taken into account by tuning the kernel
in the `virtual' terms --- namely, by requiring this kernel to approach
the solution to the BFKL equation at large $k_\perp$.

The above considerations, to be substantiated by the subsequent analysis,
explain why it is possible to approximate the Balitsky--JIMWLK equations
by simpler equations which are linear and whose overall structure is
inherited from the `virtual' terms in the original equations. Similar
considerations have underlined our previous construction in
Ref.~\cite{Iancu:2011ns}, but their generalization to finite $N_c$ (that
we shall provide in this paper) turns out to be highly non--trivial.

Another subtle aspect of our present analysis is the recognition of the
fact that the simplified equations that we shall propose (for either
finite or infinite $N_c$) correspond to a Gaussian approximation for the
CGC weight function. {\em A priori}, the association of a linear system
of equations with a Gaussian approximation may look natural, but in the
present case this is complicated by the fact that, as alluded to before,
the `virtual' piece of the JIMWLK Hamiltonian is non--linear in the
target field to all orders. Such a non--linear structure seems to
preclude any Gaussian solution. The resolution of this mathematical
puzzle turns out to be interesting on physical grounds, as it sheds new
light on the physical picture of the JIMWLK evolution. Namely, we shall
show that the Wilson lines within the `virtual' Hamiltonian do not
represent genuine non--linear effects associated with saturation, rather
they express the physical fact that, with increasing energy, {\em the
longitudinal support of the target expands symmetrically around the
light--cone}. That is, in contrast to a widespread opinion in the
literature (see e.g.
\cite{Iancu:2002xk,Iancu:2002aq,Blaizot:2002xy,Iancu:2003xm,Kovchegov:2008mk}),
which was based on a misinterpretation of the mathematical structure of
the JIMWLK equation, the gluon distribution in the target expands {\em
simultaneously} towards larger and respectively smaller values of $x^-$,
in such a way to remain symmetric around $x^-=0$. (We assume the target
to propagate along the positive $x^3$ axis at nearly the speed of light
and we define $x^-=(x^0-x^3)/\sqrt{2}$.) In turn, this symmetry has
physical consequences for the multi--partonic scattering amplitudes: it
implies that the $n$--point functions of the Wilson lines with $n\ge 4$
obey a special permutation symmetry --- the {\em mirror symmetry} ---
which expresses their invariance under time reversal.

To summarize, a Gaussian weight function which is symmetric in $x^-$ and
whose kernel is energy--dependent and interpolates between the solution
to the BFKL equation at high transverse momenta $k_\perp\gg Q_s$ and the
JIMWLK (or `dipole') kernel at low momenta $k_\perp\ll Q_s$, provides a
reasonable approximation to the JIMWLK equation, which is strictly
correct in the limiting regimes alluded to above (for any value of
$N_c$). Within its limits of validity, this approximation is essentially
unique: different constructions for the kernel can differ from each other
only in the transition region around saturation, which is anyway not
under control within the present approximation.

In practice, it is convenient to trade this kernel for the dipole
$S$--matrix, which in turn can be obtained either as the solution to the
Balitsky--Kovchegov (BK) equation
\cite{Balitsky:1995ub,Kovchegov:1999yj}, or by solving a
self--consistency condition similar to that in Ref.~\cite{Iancu:2002xk}.
(The differences between these two expressions for the kernel should be
viewed as an indicator of the stability of the approximation scheme.)
Then the $n$--point functions with $n\ge 4$ (quadrupole, sextupole etc)
can be determined in terms of the 2--point function (the dipole
$S$--matrix) by solving the evolution equations associated with the
Gaussian weight function. These equations become particularly simple at
large $N_c$, where they reduce to the equations proposed in
Ref.~\cite{Iancu:2011ns} and can be explicitly solved for arbitrary
configurations of the $n$ external points in the transverse plane.

For finite $N_c$ and for generic configurations, the equations are more
complicated, as they couple the evolution of the various $n$--point
functions with the same value of $n$. (For instance, the quadrupole mixes
under the evolution with a system of two dipoles.) Yet, explicit
solutions can be obtained under the simplifying assumption that the
kernel of the Gaussian is a {\em separable} function of the rapidity and
the transverse coordinates (plus an arbitrary function of $Y$; see
Sect.~\ref{sect:eqsMF} for details). This is certainly not the case for
the actual kernel (say, as given by the solution to the BK equation), but
it is a good piecewise approximation to it and it is furthermore true for
the MV model\footnote{By `rapidity--dependence' within the MV model, we
more precisely mean the dependence upon the longitudinal coordinate
$x^-$. Within the JIMWLK evolution, there is a one--to--one
correspondence between $Y$ and $x^-$ (see the discussion in
Sect.~\ref{sect:mirror} below).}, that we shall take as our initial
condition at low energy. So, not surprisingly, the expressions for the
$n$--point functions that we shall obtain within this scenario are
formally similar to the respective predictions of the MV model. One can
reverse this last argument as follows: given that the Gaussian weight
function is a good, piecewise approximation to the JIMWLK evolution, as
we shall demonstrate, and that the kernel of this Gaussian can be taken
to be separable within the relevant kinematical regimes, we expect the
predictions of this approximation to be very close to those of the MV
model extrapolated to high energy.

For special configurations which are highly symmetric, exact solutions
can be obtained at finite $N_c$ even without assuming separability. We
shall study various examples of this type for the 4--point function and
the 6--point function, and thus find some surprising factorization
properties, that would be interesting to test against numerical solutions
to the JIMWLK equation. For one particular configuration of the 6--point
function, the exact numerical result is already known
\cite{Dumitru:2011vk} and our respective analytic solution appears to
agree with it quite well.

\section{The Balitsky--JIMWLK evolution equations}
\setcounter{equation}{0} \label{sect:evolution}

In this section, we shall briefly review the general formulation of the
JIMWLK evolution and then use the evolution equations satisfied by the
dipole and the quadrupole $S$--matrices in order to illustrate various
properties of the evolution, which are important for what follows: the
role and origin of the `real' and `virtual' terms, the factorization of
multi--trace observables at large $N_c$, and, especially, the symmetric
expansion of the longitudinal support of the target and the ensuing,
`mirror', symmetry of the $n$--point functions with $n\ge 4$.

\subsection{JIMWLK evolution: a brief reminder}

The color glass condensate is an effective theory for the small--$x$ part
of the wavefunction of an energetic hadron: the gluons carrying a small
fraction $x\ll 1$ of the hadron's longitudinal momentum are described as
a random distribution of classical color fields generated by sources with
larger momentum fractions $x'\gg x$. Given the high--energy kinematics,
in particular the fact that the distribution of the color sources is
`frozen' by Lorentz time dilation, this color field can be chosen (in a
suitable gauge) to have a single non--zero component, namely $A^\mu_a(x)
=\delta^{\mu+}\alpha_a(x^-,\bmx)$ for a hadron moving along the positive
$z$ axis\footnote{We use standard definitions for the light--cone
coordinates: $x^\mu=(x^+,x^-,\bmx)$, with $x^\pm=(t\pm z)/\sqrt{2}$ and
$\bmx=(x,y)$. The field $\alpha_a$ is independent of the light--cone time
$x^+$, because of Lorentz time dilation.}  (a `right mover'). All the
correlations of this field are encoded into a functional probability
distribution, the `CGC weight function' $W_Y[\alpha]$, which contains
information about the evolution of the color sources with increasing
`rapidity' $Y\equiv \ln(1/x)$, from some initial value $Y_0$ up to the
value $Y$ of interest. In the high energy regime where $\alpha_s
(Y-Y_0)\gtrsim 1$ and to leading logarithmic accuracy with respect to the
large logarithm $Y-Y_0=\ln(x_0/x)$, this evolution is described by a
functional renormalization group equation for $W_Y[\alpha]$, known as the
JIMWLK equation. The latter can be given a Hamiltonian form,
 \beq\label{jimwlk}
 \frac{\del}{\del Y}W_Y[\alpha]
 =  H W_Y[\alpha]\,,
 \eeq
where $H$ is the JIMWLK Hamiltonian --- a second--order, functional
differential operator, whose most convenient form for the present
purposes is that given in \cite{Hatta:2005as} and reads
 \beq\label{H}
 H = -\frac{1}{16 \pi^3} \int_{\bmu\bmv\bmz}
 \mcal{M}_{\bmu\bmv\bmz}
 \left(1 + \wt{V}^{\dagger}_{\bmu} \wt{V}_{\bmv}
 -\wt{V}^{\dagger}_{\bmu} \wt{V}_{\bmz}
 -\wt{V}^{\dagger}_{\bmz} \wt{V}_{\bmv}\right)^{ab}
 \frac{\delta}{\delta \alpha_{\bmu}^a}
 \frac{\delta}{\delta \alpha_{\bmv}^b},
 \eeq
where we use the notation $\int_{\bmu \dots} \equiv \int \dif^2 \bmu
\dots$ to simplify writing, $\mcal{M}$ is the `dipole kernel',
 \beq
 \mcal{M}_{\bmu\bmv\bmz}\,\equiv\, \frac{(\bm{u}-\bm{v})^2}
 {(\bm{u}-\bm{z})^2(\bm{z}-\bm{v})^2}\,,\eeq
and $\wt{V}^{\dagger}$ and $\wt{V}$ are Wilson lines in the adjoint
representation:
 \beq\wt{V}^{\dagger}_{\bmx}\,\equiv\,{\mbox P}\exp \left[
 \rmi g\int \rmd x^-\alpha_a(x^-,\bmx)T^a \right], \label{Vadj}
 \eeq
with P denoting path--ordering in $x^-$. The above form of the
Hamiltonian is valid only when acting on gauge--invariant functionals of
$\alpha_a$, which will always be the case throughout our analysis. In
fact, the observables of interest are gauge--invariant products of Wilson
lines (see below).

The functional derivatives in \eqn{H} are understood to act at the
largest value of $x^-$, that is, at the upper end point of path--ordered
exponentials like that in \eqn{Vadj} (see e.g. \eqn{donV}). These
derivatives do not commute with each other, but their commutator is
proportional to $\delta_{\bmu\bmv}$ (cf. \eqn{algebra}) and thus vanishes
when multiplied by $ \mcal{M}_{\bmu\bmv\bmz}$; hence, there is no
ambiguity concerning the ordering of the functional derivatives in
\eqn{H}. One can also notice that the last two terms in the JIMWLK
Hamiltonian, {\em i.e.} those proportional to $\wt{V}^{\dagger}_{\bmu}
\wt{V}_{\bmz}$ and to $\wt{V}^{\dagger}_{\bmz} \wt{V}_{\bmv}$
respectively, are in fact identical to each other, as it can be checked
by exchanging $\bmu\leftrightarrow\bmv$ and $a\leftrightarrow b$ and by
using the property $\wt{V}^{\dagger\,ac}_{\bmz}= \wt{V}^{ca}_{\bmz}$ for
color matrices in the adjoint representation. To fully specify the
problem, one also needs an initial condition for \eqn{jimwlk} at $Y=Y_0$;
at least for a sufficiently large nucleus, this initial condition is
provided by the McLerran--Venugopalan (MV) model
\cite{McLerran:1993ni,McLerran:1993ka} (see Sect.~\ref{sect:MV} below).

Physical observables, like scattering amplitudes for external
projectiles, are represented by gauge invariant operators
$\hat{\mcal{O}}[\alpha]$ built with the field $\alpha_a$, whose target
expectation values are computed via functional averaging with the CGC
weight function:
  \beq \langle \hat{\mathcal O}\rangle_Y \equiv \int
{\mathcal D}\alpha \, {\mathcal O}[\alpha]\, W_Y[\alpha].
 \label{average}
 \eeq
By taking a derivative in this equation with respect to $Y$, using
\eqn{jimwlk}, and integrating by parts within the functional integral
over $\alpha$, one obtains an evolution equation for the observable, in
which the JIMWLK Hamiltonian acts on the operator
$\hat{\mcal{O}}[\alpha]$ :
 \beq\label{general}
 \frac{\del \lan \hat{\mcal{O}} \ran_Y}{\del Y}
 = \lan H \hat{\mcal{O}} \ran_Y\,.
 \eeq
Unlike the JIMWLK equation \eqref{jimwlk}, this is not a functional
equation anymore, but an integro-differential equation. However, due to
the non--linear structure of the Hamiltonian \eqref{H} with respect to
the field $\alpha_a$, \eqn{general} is generally not a closed equation
--- the action of $H$ on $\hat{\mcal{O}}$ generates additional
operators in the right hand side ---, but just a member of an infinite
hierarchy of coupled equations --- the Balitsky--JIMWLK equations.
Although mathematically equivalent, the functional equation
\eqref{jimwlk} and the Balitsky--JIMWLK hierarchy offer complementary
perspectives over the high--energy evolution. \eqn{jimwlk} depicts the
evolution of the {\em target} via the emission of an additional gluon
with rapidity between $Y$ and $Y+\rmd Y$, in the background of the color
field $\alpha$ generated via previous emissions, at rapidities $Y' \le
Y$. The Wilson lines within the structure of the Hamiltonian \eqref{H}
describe the scattering between this new gluon and the background field,
in the eikonal approximation. The Balitsky--JIMWLK hierarchy rather
refers to the evolution of the {\em projectile}, more precisely, of the
operator which describes its scattering off the target. This scattering
is again computed in the eikonal approximation, so the operator
$\hat{\mcal{O}}$ is naturally built with Wilson lines --- one such a line
for each parton within the projectile.

\subsection{Evolution equations for the dipole and the quadrupole}
\label{sect:eqs}

To be more explicit, we shall consider two specific projectiles: a `color
dipole' made with a quark--antiquark ($q\bar q$) pair and a `color
quadrupole' made with two $q\bar q$ pairs. In both cases, the overall
color state of the partonic system is a color singlet. The $S$--matrix
operators describing the forward scattering of these projectiles off the
CGC target read
 \beq\label{Sdipole}
 \hat{S}_{\bmx_1\bmx_2} \equiv \hat{S}_{\bmx_1\bmx_2} ^{(2)}=
 \frac{1}{N_c}\,\rmtr({V}^{\dagger}_{\bmx_1} {V}_{\bmx_2})\,,
 \eeq
for the color dipole and, respectively,
 \beq\label{Squadrupole}
 \hat{Q}_{\bmx_1\bmx_2\bmx_3\bmx_4} \equiv
 \hat{S}_{\bmx_1\bmx_2\bmx_3\bmx_4}^{(4)}=
 \frac{1}{N_c}\,
 \rmtr({V}^{\dagger}_{\bmx_1} {V}_{\bmx_2}{V}^{\dagger}_{\bmx_3}
 {V}_{\bmx_4})\,,
 \eeq
for the color quadrupole. In these equations, ${V}^{\dagger}$ and ${V}$
are Wilson lines similar to those in \eqn{Vadj}, but in the fundamental
representation. The results that we shall obtain for these two partonic
systems will be easy to extend to projectiles made with $n$ $q\bar q$
pairs, for which
\beq\label{S2n}
 \hat{S}_{\bmx_1\bmx_2 ...\bmx_{2n\!-\!1}\bmx_{2n}}^{(2n)} =\,
 \frac{1}{N_c}\,
 \rmtr({V}^{\dagger}_{\bmx_1} {V}_{\bmx_2}...
 {V}^{\dagger}_{\bmx_{2n\!-\!1}}{V}_{\bmx_{2n}}).
 \eeq
As we shall see, within the high--energy evolution, such single--trace
operators mix with the multi--trace operators, of the form
 \beq\label{multitrace}
 \hat{\mcal{O}} = \,\frac{1}{N_c}\,\rmtr({V}^{\dagger}_{\bmx_1} {V}_{\bmx_2}...)
 \frac{1}{N_c}\,\rmtr({V}^{\dagger}_{\bmy_1} {V}_{\bmy_2}...)
 \frac{1}{N_c}\,\rmtr({V}^{\dagger}_{\bmz_1} {V}_{\bmz_2}...).
 \eeq
In order to construct evolution equations according to \eqn{general}, we
need the action of the functional derivatives w.r.t. $\alpha_a$ on the
Wilson lines. This reads (with $
\delta_{\bmx\bmu}=\delta^{(2)}(\bmx-\bmu)$)
 \beq\label{donV}
 \frac{\delta}{\delta \alpha^a_{\bmu}}\,
 V_{\bmx}^{\dagger} =
 \rmi g \delta_{\bmx\bmu}\, t^a V_{\bmx}^{\dagger},
 \qquad
 \frac{\delta}{\delta \alpha^a_{\bmu}}\,
 V_{\bm{x}} =
 - \rmi g  \delta_{\bmx\bmu} V_{\bmx}\, t^a.
 \eeq
By using these rules within Eqs.~\eqref{general} and \eqref{H}, it is
straightforward to derive the evolution equations satisfied by
$S$--matrices for the dipole and the quadrupole. The respective
derivations can be found in the literature (see e.g. the Appendix
Ref.~\cite{Iancu:2011ns}), but here we shall nevertheless indicate a few
intermediate steps (on the example of the dipole evolution), to emphasize
the origin of the various terms in the equations. To that aim, it is
useful to view the JIMWLK Hamiltonian \eqref{H} as the sum of two pieces,
$H=H_{\rm virt}+H_{\rm real}$, where $H_{\rm virt}$ corresponds to the
first two terms in \eqn{H} and the $H_{\rm real}$ corresponds to the last
two terms there. This division between `virtual' and `real' terms refers
to the evolution of the {\em projectile} (see the physical discussion
after \eqn{BK} below) and should not be confounded with the corresponding
division for the evolution of the {\em target}
\cite{Iancu:2002xk,Iancu:2003xm}. By acting with these Hamiltonian pieces
on the dipole $S$--matrix, one finds (with $\abar\equiv\alpha_s
N_c/\pi$).
 \beq\label{virt}
H_{\rm virt}\, \hat{S}_{\bmx_1\bmx_2} = -\frac{\abar}{2\pi}\, \left(1 -
\frac{1}{N_c^2} \right) \int_{\bmz} \mcal{M}_{\bmx_1\bmx_2\bmz}
\hat{S}_{\bmx_1\bmx_2}.
\eeq
and respectively (recall that the last two terms in \eqn{H}, which define
$H_{\rm real}$, are actually identical with each other)
\begin{align}\label{real}
 H_{\rm real}\, \hat{S}_{\bmx_1\bmx_2}
 &= \,\frac{\abar}{\pi}\,
 \int\limits_{\bm{z}}
 \mcal{M}_{\bmx_1\bmx_2\bmz}
 \Big[\big(\wt{V}_{\bmx_1}^{\dagger}\big)^{ac}\,
 \wt{V}_{\bmz}^{cb}\,
 \rmtr \big(t^a\, V_{\bmx_1}^{\dagger} V_{\bmx_2} t^b  \big)\Big]
 \nn
 &=
 \atpi\,
 \int\limits_{\bm{z}}
 \mcal{M}_{\bmx_1\bmx_2\bmz}
 \Big(\hat{S}_{\bmx_1\bmz} \hat{S}_{\bmz\bmx_2}
 -\frac{1}{N_c^2}\, \hat{S}_{\bmx_1\bmx_2}\Big),
 \end{align}
where the second line follows after reexpressing the adjoint Wilson line
in terms of fundamental ones, according to
 \beq\label{adjtofun}
 \big(\wt{V}^{\dagger}\big)^{ac} t^a =
 \wt{V}^{cb} t^b = V^{\dagger} t^c\, V.
 \eeq
and then using the Fierz identity
 \beq\label{Fierz1}
 \rmtr \big(t^a A \,t^a B \big)
 =\frac{1}{2}\,\rmtr A \, \rmtr B
 -\frac{1}{2 N_c}\, \rmtr(AB).
 %\nn \rmtr \big(t^a A \big)
 %\rmtr \big(t^a B \big)
 %&=&\frac{1}{2}\,\rmtr(A B)
 %-\frac{1}{2 N_c}\, \rmtr(A) \rmtr(B).
 \eeq
By adding together the above results, one sees that the terms
proportional to $1/N_c^2$, that would be suppressed at large $N_c$,
exactly cancel between `real' and `virtual' contributions, and we are
left with
 \beq\label{BK}
 \frac{\del \lan \hat{S}_{\bmx_1\bmx_2} \ran_Y}{\del Y}=
 \frac{\abar}{2\pi}\, \int_{\bmz}
 \mcal{M}_{\bmx_1\bmx_2\bmz}
 \lan \hat{S}_{\bmx_1\bmz} \hat{S}_{\bmz\bmx_2}
 -\hat{S}_{\bmx_1\bmx_2} \ran_Y\,.
 \eeq
This equation has the following physical interpretation: the first term
in the right hand side, which is quadratic in $\hat{S}$ and has been
generated by the `real' piece of the Hamiltonian, cf. \eqn{real},
describes the splitting of the original dipole $(\bmx_1,\,\bmx_2)$ into
two new dipoles $(\bmx_1,\,\bmz)$ and $(\bmz,\,\bmx_2)$, which then
scatter off the target. More precisely, the evolution step consists in
the emission of a soft gluon, so the original dipole gets replaced by a
quark--antiquark--gluon system which is manifest in the first line of
\eqn{real}, but in large--$N_c$ limit (to which refers the first term in
the second line of \eqn{real}), this emission is equivalent to the dipole
splitting alluded to above. As for the second term in \eqn{BK}, {\em
i.e.} the negative term linear in $\hat{S}$ which has been produced by
$H_{\rm virt}$, it describes the reduction in the probability that the
dipole survive in its original state --- that is, the probability for the
dipole {\em not} to emit. In what follows, we shall often refer to the
terms produced by $H_{\rm virt}$ ($H_{\rm real}$) as the `virtual'
(`real') terms, but one should keep in mind that not all such terms are
actually visible in the evolution equation in their standard form in the
literature (to be also used in this paper): some of these terms may have
canceled between `real' and `virtual' contributions.

A similar discussion applies to the evolution equation for the
quadrupole, which reads
 \begin{align}\label{Qevol}
 \hspace*{-.6cm}\frac{\del \lan\hat{Q}_{\bmx_1\bmx_2\bmx_3\bmx_4} \ran_Y}{\del Y} =
 \frac{\abar}{4\pi}  \int_{\bmz} &
 \Big[(\mcal{M}_{\bmx_1\bmx_2\bmz} +
 \mcal{M}_{\bmx_1\bmx_4\bmz} -
 \mcal{M}_{\bmx_2\bmx_4\bmz})
 \lan
 \hat{S}_{\bmx_1\bmz}\hat{Q}_{\bmz\bmx_2\bmx_3\bmx_4}
 \ran_Y
 \nn
 &+(\mcal{M}_{\bmx_1\bmx_2\bmz} +
 \mcal{M}_{\bmx_2\bmx_3\bmz} -
 \mcal{M}_{\bmx_1\bmx_3\bmz})
 \lan
 \hat{S}_{\bmz\bmx_2}\hat{Q}_{\bmx_1\bmz\bmx_3\bmx_4}
 \ran_Y
 \nn
 &+(\mcal{M}_{\bmx_2\bmx_3\bmz} +
 \mcal{M}_{\bmx_3\bmx_4\bmz} -
 \mcal{M}_{\bmx_2\bmx_4\bmz})
 \lan
 \hat{S}_{\bmx_3\bmz}\hat{Q}_{\bmx_1\bmx_2\bmz\bmx_4}
 \ran_Y
 \nn
 &+(\mcal{M}_{\bmx_1\bmx_4\bmz} +
 \mcal{M}_{\bmx_3\bmx_4\bmz} -
 \mcal{M}_{\bmx_1\bmx_3\bmz})
 \lan
 \hat{S}_{\bmz\bmx_4}\hat{Q}_{\bmx_1\bmx_2\bmx_3\bmz}
 \ran_Y
 \nn
 &-(\mcal{M}_{\bmx_1\bmx_2\bmz} + \mcal{M}_{\bmx_3\bmx_4\bmz}
 +\mcal{M}_{\bmx_1\bmx_4\bmz} + \mcal{M}_{\bmx_2\bmx_3\bmz})
 \lan
 \hat{Q}_{\bmx_1\bmx_2\bmx_3\bmx_4}
 \ran_Y
 \nn
 &-(\mcal{M}_{\bmx_1\bmx_2\bmz} + \mcal{M}_{\bmx_3\bmx_4\bmz}
 -\mcal{M}_{\bmx_1\bmx_3\bmz} - \mcal{M}_{\bmx_2\bmx_4\bmz})
 \lan
 \hat{S}_{\bmx_1\bmx_2}\hat{S}_{\bmx_3\bmx_4}
 \ran_Y
 \nn
 &-(\mcal{M}_{\bmx_1\bmx_4\bmz} + \mcal{M}_{\bmx_2\bmx_3\bmz}
 -\mcal{M}_{\bmx_1\bmx_3\bmz} - \mcal{M}_{\bmx_2\bmx_4\bmz})
 \lan
 \hat{S}_{\bmx_3\bmx_2}\hat{S}_{\bmx_1\bmx_4}
 \ran_Y\Big].
 \end{align}
Namely, the terms involving $\lan \hat{S} \hat{Q}\ran_Y$ in the right
hand side are `real' terms describing the splitting of the original
quadrupole into a new quadrupole plus a dipole, and have been all
generated by the action of the last two terms in the Hamiltonian
\eqref{H}. The `virtual' terms involving $\lan \hat{Q}\ran_Y$ and $\lan
\hat{S} \hat{S}\ran_Y$ are necessary for probability conservation, and
have been generated by the first two terms in the Hamiltonian. Once
again, all the terms subleading at large $N_c$ (as separately generated
by $H_{\rm virt}$ and $H_{\rm real}$) have canceled in the final
equation.

The above features are generic: they apply to the evolution equations
obeyed by all the single--trace observables like \eqn{S2n}. As visible on
Eqs.~\eqref{BK} and \eqref{Qevol}, these equations are generally not
closed : they couple single--trace observables with the multi--trace
ones. E.g., the equation for the quadrupole also involves the 4--point
function $\lan \hat{S} \hat{S}\ran_Y$ and the 6--point function $\lan
\hat{S} \hat{Q}\ran_Y$, which in turn are coupled (via the respective
evolution equations) to even higher--point correlators. The equations
obeyed by the multi--trace observables exhibit an interesting new
feature: they involve genuine $1/N_c^2$ corrections, as generated when
the two functional derivatives in \eqn{H} act on Wilson lines which
belong to different traces (see e.g. Appendix F in
\cite{Triantafyllopoulos:2005cn} for an example). At large $N_c$, these
corrections can be neglected and then it is easy to check that the
hierarchy admits the factorized solution
 \beq\label{fact}
 \lan\hat{\mcal{O}}\ran_Y \simeq
 \Big\lan\frac{1}{N_c}\,\rmtr({V}^{\dagger}_{\bmx_1}
 {V}_{\bmx_2}...)\Big\ran_Y
 \Big\lan\frac{1}{N_c}\,\rmtr({V}^{\dagger}_{\bmy_1}
  {V}_{\bmy_2}...)\Big\ran_Y
 \Big\lan\frac{1}{N_c}\,\rmtr({V}^{\dagger}_{\bmz_1}
  {V}_{\bmz_2}...)\Big\ran_Y...\,,
 \eeq
provided this factorization is already satisfied by the initial
conditions. Then the hierarchy drastically simplifies: it breaks into a
set of equations which can be solved one after the other (at least in
principle). Namely, \eqn{BK} becomes a closed equation for
$\lan\hat{S}\ran_Y$ (the BK equation
\cite{Balitsky:1995ub,Kovchegov:1999yj}), \eqn{Qevol} becomes an
inhomogeneous equation for $\lan \hat{Q} \ran_Y$ with coefficients which
depend upon $\lan\hat{S}\ran_Y$ \cite{JalilianMarian:2004da}, and so on.
In practice, however, the resolution of these equations is hindered by
their strong non--locality in the transverse coordinates. So far, only
the (numerical) solution to the BK equation has been explicitly
constructed.

\subsection{The mirror symmetry}
\label{sect:mirror}

In this subsection, we shall discuss a symmetry property of the JIMWLK
equation, which has not been noticed in the previous literature and which
has far--reaching consequences: the symmetry of the target field
distribution (the CGC) under reflection in $x^-$.

To start with, we shall identify a {\em mirror symmetry} in the evolution
equation \eqref{Qevol} for the quadrupole, that can be easily
demonstrated in the large--$N_c$ limit, but is likely to hold for any
$N_c$. (It does so, at least, in the Gaussian approximation that we shall
later construct.) Specifically, if the quadrupole $S$--matrix $\lan
\hat{Q}_{\bmx_1\bmx_2\bmx_3\bmx_4}\ran_Y $ is symmetric under the
exchange of the two antiquark Wilson lines (that is, the Wilson lines at
$\bmx_2$ and $\bmx_4$) at the initial rapidity $Y_0$ --- a condition
which is indeed satisfied within the MV model \cite{Dominguez:2011wm}
---, then this symmetry is preserved by the evolution. That is, for any
$Y\ge Y_0$, one has
 \beq\label{Qsym}
 \lan \hat{Q}_{\bmx_1\bmx_2\bmx_3\bmx_4}\ran_Y =
 \lan \hat{Q}_{\bmx_1\bmx_4\bmx_3\bmx_2}\ran_Y.
 \eeq
A similar property holds for the exchange of the quark Wilson lines at
$\bmx_1$ and $\bmx_3$, but this is not independent of \eqn{Qsym}, since
$\hat{Q}_{\bmx_3\bmx_2\bmx_1\bmx_4}=\hat{Q}_{\bmx_1\bmx_4\bmx_3\bmx_2}$
by the cyclic symmetry of the trace. To demonstrate \eqn{Qsym}, let us
consider the respective {\em anti}--symmetric piece:
 \beq\label{Qasymm}
 \hat{Q}_{\bmx_1\bmx_2\bmx_3\bmx_4}^{\rm \,asym} \equiv
 \frac{1}{2 N_c}
 \big[\rmtr({V}^{\dagger}_{\bmx_1}
 {V}_{\bmx_2}{V}^{\dagger}_{\bmx_3} {V}_{\bmx_4})-
 \rmtr({V}^{\dagger}_{\bmx_1} {V}_{\bmx_4}
 {V}^{\dagger}_{\bmx_3} {V}_{\bmx_2})\big]\,.
 \eeq
By using \eqn{Qevol}, it is easy to see that the associated expectation
value obeys the following evolution equation:
 \begin{align}\label{Qasymevol}
 \hspace*{-.4cm}\frac{\del \lan\hat{Q}_{\bmx_1\bmx_2\bmx_3\bmx_4}
 ^{\rm \,asym} \ran_Y}{\del Y} =
 \frac{\abar}{4\pi}  \int_{\bmz} &
 \Big[(\mcal{M}_{\bmx_1\bmx_2\bmz} +
 \mcal{M}_{\bmx_1\bmx_4\bmz} -
 \mcal{M}_{\bmx_2\bmx_4\bmz})
 \lan
 \hat{S}_{\bmx_1\bmz}\hat{Q}_{\bmz\bmx_2\bmx_3\bmx_4}^{\rm \,asym}
 \ran_Y
 \nn
 &+(\mcal{M}_{\bmx_1\bmx_2\bmz} +
 \mcal{M}_{\bmx_2\bmx_3\bmz} -
 \mcal{M}_{\bmx_1\bmx_3\bmz})
 \lan
 \hat{S}_{\bmz\bmx_2}\hat{Q}_{\bmx_1\bmz\bmx_3\bmx_4}^{\rm \,asym}
 \ran_Y
 \nn
 &+(\mcal{M}_{\bmx_2\bmx_3\bmz} +
 \mcal{M}_{\bmx_3\bmx_4\bmz} -
 \mcal{M}_{\bmx_2\bmx_4\bmz})
 \lan
 \hat{S}_{\bmx_3\bmz}\hat{Q}_{\bmx_1\bmx_2\bmz\bmx_4}^{\rm \,asym}
 \ran_Y
 \nn
 &+(\mcal{M}_{\bmx_1\bmx_4\bmz} +
 \mcal{M}_{\bmx_3\bmx_4\bmz} -
 \mcal{M}_{\bmx_1\bmx_3\bmz})
 \lan
 \hat{S}_{\bmz\bmx_4}\hat{Q}_{\bmx_1\bmx_2\bmx_3\bmz}^{\rm \,asym}
 \ran_Y
 \nn
 &-(\mcal{M}_{\bmx_1\bmx_2\bmz} + \mcal{M}_{\bmx_3\bmx_4\bmz}
 +\mcal{M}_{\bmx_1\bmx_4\bmz} + \mcal{M}_{\bmx_2\bmx_3\bmz})
 \lan
 \hat{Q}_{\bmx_1\bmx_2\bmx_3\bmx_4}^{\rm \,asym}
 \ran_Y\Big].
 \end{align}
At large $N_c$, where one can factorize $\lan \hat{S} \hat{Q}^{\rm
\,asym}\ran_Y \simeq \lan \hat{S}\ran_Y \lan\hat{Q}^{\rm \,asym}\ran_Y$,
\eqn{Qasymevol} becomes a homogeneous equation which implies that $
\lan\hat{Q}^{\rm \,asym}\ran_Y=0$ at any $Y$ provided this condition was
satisfied at $Y_0$. In turn, this implies the symmetry property
\eqref{Qsym}.

\begin{figure}
\centerline{\includegraphics[scale=0.7]{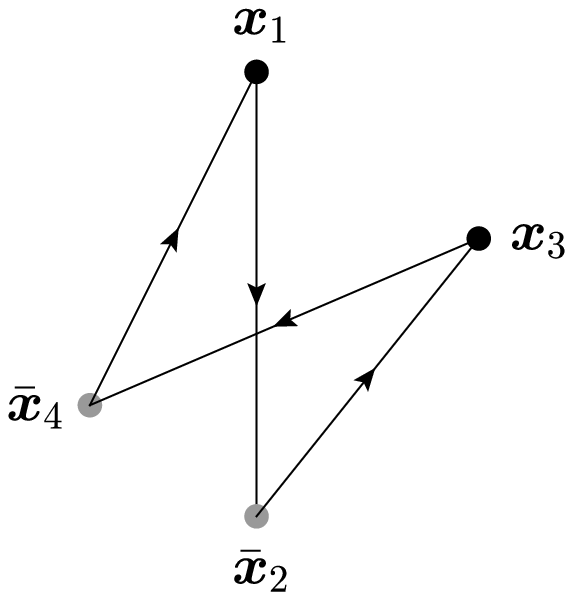}\hspace*{2cm}
\includegraphics[scale=0.7]{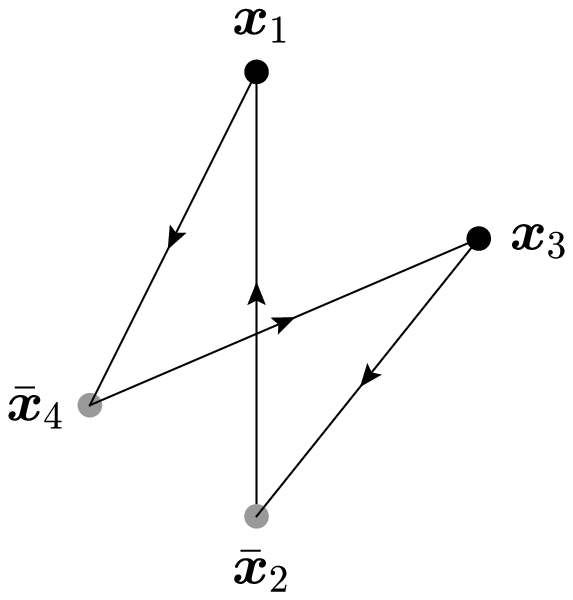}}
\caption{\sl A pictorial representation of the color flow with the
operator $\hat{Q}_{\bmx_1\bmx_2\bmx_3\bmx_4}$ (left) and respectively
$\hat{Q}_{\bmx_1\bmx_4\bmx_3\bmx_2}$ (right).}
\label{Fig:mirror}
\end{figure}

By inspection of the higher equations in the Balitsky--JIMWLK equations,
one can check that a similar symmetry holds for all the $n$--point
functions of the Wilson lines. For instance, the equation obeyed by the
sextupole $S$--matrix $\lan\hat{S}^{(6)}\ran_Y$ is explicitly shown in
Appendix B of Ref.~\cite{Iancu:2011ns}. From this equation, one can read
the following symmetry property:
 \beq\label{6perm}
 \lan\hat{S}_{\bmx_1\bmx_2\bmx_3\bmx_4\bmx_5\bmx_6}^{(6)}\ran_Y\,=\,
 \lan\hat{S}_{\bmx_1\bmx_6\bmx_5\bmx_4\bmx_3\bmx_2}^{(6)}\ran_Y\,.\eeq
The generalization of this property to the $2n$--point function shown in
\eqn{S2n} reads
 \beq\label{perm}
 \lan\hat{S}_{\bmx_1\bmx_2 ...\bmx_{2n\!-\!2}
 \bmx_{2n\!-\!1}\bmx_{2n}}^{(2n)}\ran_Y\,=\,\lan
 \hat{S}_{\bmx_1\bmx_{2n}\bmx_{2n\!-\!1}\bmx_{2n\!-\!2} ...\bmx_2}^{(2n)}
 \ran_Y\,.
 \eeq

To better understand the content of this symmetry, it is useful to give a
pictorial representation for it. To that aim, consider a generic
configuration of the quadrupole operator
$\hat{Q}_{\bmx_1\bmx_2\bmx_3\bmx_4}$ in the transverse plane, as
illustrated in Fig.~\ref{Fig:mirror}, and join the four points by
oriented lines, which follow the direction of color multiplication. In
this way, one constructs a closed, oriented, contour, whose orientation
indicates the flow of color within the operator. By repeating this
procedure for the `permuted' operator
$\hat{Q}_{\bmx_1\bmx_4\bmx_3\bmx_2}$, one obtains a similar contour,
where however the orientation of the color flow is reversed. One can
similarly check that, for a general $n$--point function, the symmetry
property \eqref{perm} refers to changing the contour orientation, say
from clockwise to counterclockwise. Such a change would also result from
the reflection in a mirror, so we shall refer to the symmetry property
\eqref{perm} as the `mirror symmetry'. Additional arguments in the favor
of this name will be given below.

There are several reasons why this this symmetry is so important for us
here. First, as we shall shortly argue, this corresponds to an important
symmetry property of the scattering amplitudes: their invariance under
{\em time--reversal}. Second, the way how this symmetry is actually
preserved by the JIMWLK evolution is very interesting as it sheds light
on the physical picture of the target field distribution: with increasing
$Y$, the color glass condensate expands {\em symmetrically} around
$x^-=0$. Third, this symmetry will later guide us in the construction of
a {\em mean field approximation} to the Balitsky--JIMWLK hierarchy.

\begin{figure}
\centerline{\includegraphics[scale=0.7]{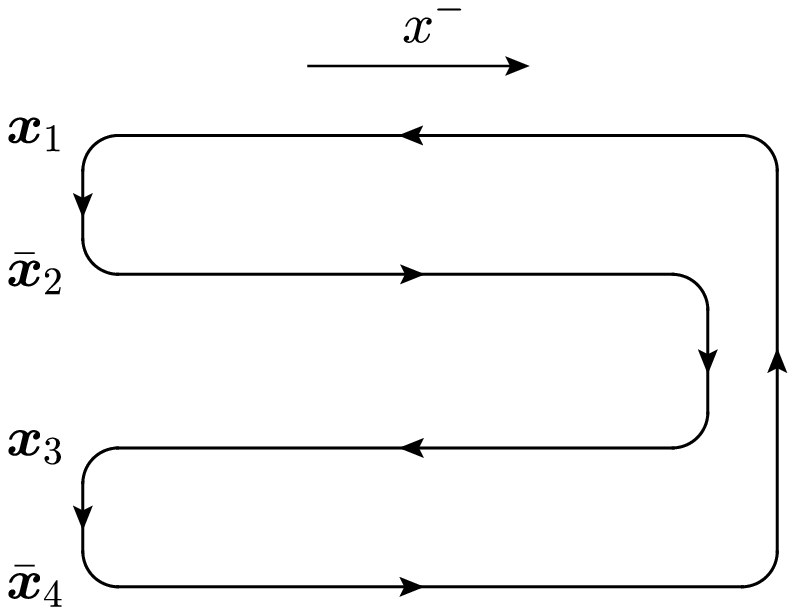}\hspace*{2cm}
\includegraphics[scale=0.7]{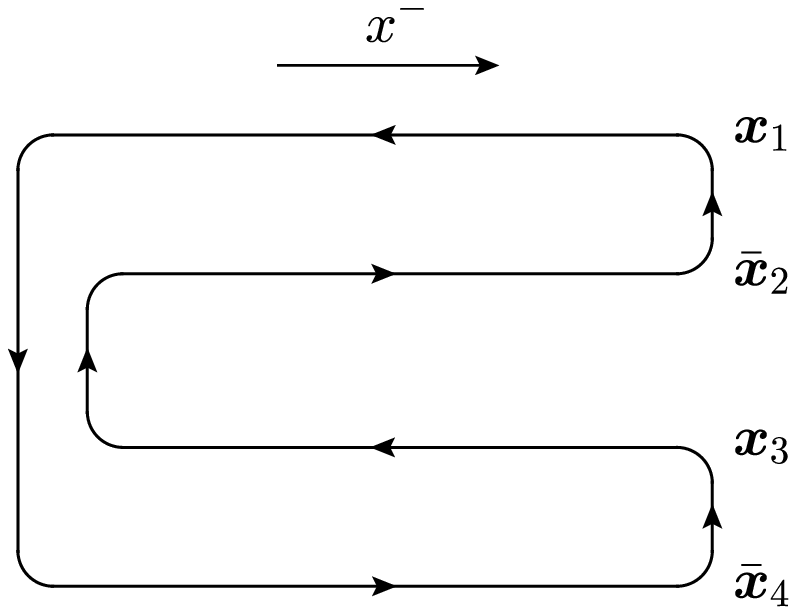}}
\caption{\sl A different pictorial representation of the
operators $\hat{Q}_{\bmx_1\bmx_2\bmx_3\bmx_4}$ (left) and
$\hat{Q}_{\bmx_1\bmx_4\bmx_3\bmx_2}$ (right), which emphasizes the fact that
they get exchanged with each other via time reversal, with `time'\,$=x^-$.}
\label{Fig:flow}
\end{figure}

To understand the relation to time--reversal, let us present another
pictorial representation for the two quadrupole operators which enter
\eqn{Qsym}: this is shown in Fig.~\ref{Fig:flow}, where the transverse
space is schematically represented by the vertical axis, whereas the
horizontal axis refers to $x^-$ --- the {\em light--cone time for the
projectile}. The Wilson lines are now explicitly shown, as the oriented
horizontal lines extending along the $x^-$ axis and connected with each
other, via matrix multiplication, at $x^-\to\pm \infty$. Once again, the
orientation of these lines corresponds to the direction of the color
flow. Clearly, these two figures get exchanged with each other when
inverting the arrow of time. In the left figure, corresponding to
$\hat{Q}_{\bmx_1\bmx_2\bmx_3\bmx_4}$, the 4--body system starts at
$x^-\to -\infty$ as a set of 2 dipoles, $(\bmx_1,\bmx_2)$ and
$(\bmx_3,\bmx_4)$, which then exchange color with each other at $x^-\to
\infty$ and thus reconnect into the new dipoles $(\bmx_1,\bmx_4)$ and
$(\bmx_3,\bmx_2)$. In the right figure, the opposite process happens: the
system starts with the dipoles $(\bmx_1,\bmx_4)$ and $(\bmx_3,\bmx_2)$,
which then reconnect at  $x^-\to \infty$ into the dipoles
$(\bmx_1,\bmx_2)$ and $(\bmx_3,\bmx_4)$, thus yielding the quadrupole
$\hat{Q}_{\bmx_1\bmx_4\bmx_3\bmx_2}$. Hence, the symmetry property
\eqref{Qsym} corresponds indeed to invariance under time--reversal, as
anticipated.

We now turn to the physical interpretation of the mirror symmetry in the
context of the JIMWLK evolution. One can check that the symmetric
structure of the virtual terms in \eqn{Qevol} stems from the combined
action of the first two terms in \eqn{H}. Half of the `virtual' terms are
generated by the first term, proportional to the color unity matrix, but
by themselves these terms do {\em not} show the mirror symmetry; this
symmetry is recovered only after adding the other half of the `virtual'
terms, as generated by the second term in \eqn{H}, proportional to
$\wt{V}^{\dagger}_{\bmu} \wt{V}_{\bmv}$. As an example, consider two of
the `virtual' terms in the r.h.s. of \eqn{Qevol} whose coefficients get
exchanged with each other under the exchange $\bmx_2\leftrightarrow
\bmx_4$ : $\mcal{M}_{\bmx_1\bmx_4\bmz}
\lan\hat{Q}_{\bmx_1\bmx_2\bmx_3\bmx_4}\ran$ and
$\mcal{M}_{\bmx_1\bmx_2\bmz} \lan\hat{Q}_{\bmx_1\bmx_2\bmx_3\bmx_4}\ran$.
The first of them is generated when acting with the first term in the
Hamiltonian on the pair ($\bmx_1,\bmx_4$) of the quadrupole, whereas the
second one emerges from the action of the second term in $H$ on the pair
($\bmx_1,\bmx_2$).

Hence, to elucidate this symmetry, one needs to better understand the
action of the JIMWLK Hamiltonian. As manifest from \eqn{donV}, the
functional derivatives within the Hamiltonian act as generators of
infinitesimal color rotations of the Wilson lines at their upper end
point in $x^-$; that is, they act as {\em Lie derivatives} for the color
group SU$(N_c)$. These color rotations express the evolution of the
target color field $\alpha_a(x^-,\bmx)$ with increasing rapidity:
performing one infinitesimal step in the evolution, from $Y$ to $Y+\rmd
Y$, amounts to `integrating out' one layer of quantum fluctuations within
the target wavefunction --- the gluons with longitudinal momentum
fractions between $x=\rme^{-Y}$ and $x'=\rme^{-(Y+\rmd Y)}$ --- and
results in adding one additional layer to the classical color field
$\alpha_a(x^-,\bmx)$. The fact that the JIMWLK Hamiltonian acts on the
Wilson lines via color rotations at the largest value in $x^-$ means that
the new layer of color field is located at larger $x^-$ as compared to
the previous layers.

\comment{This statement can be made more precise by reexpressing the
light--cone longitudinal coordinate $x^-$ in terms of the {\em
space--time rapidity} $\rmy$, defined as (for positive $x^-$)
 \beq\label{spacetimey}
\rmy\equiv Y_0+\,\ln\frac{x^-}{x^-_0}\,,\eeq with $x_0^-$ the
longitudinal extent of the target at the initial rapidity $Y_0$. The
space--time rapidity $\rmy$ is a more convenient variable since it is
directly identified with the momentum rapidity $Y$ of the quantum modes
which are integrated out \cite{Iancu:2002xk}: by integrating out one
layer of quantum gluons with momentum rapidity $Y'$ within $Y<Y'<Y+\rmd
Y$, one generates an additional contribution $\delta\alpha^a$ to the
target field with support in the space--time rapidity bin $Y<\rmy<Y+\rmd
Y$.}

%This additional contribution $\delta\alpha^a(x^-,\bmx)$ is a Gaussian
%random variable, whose 1-- and 2--point functions depend upon the color
%field created in the previous steps via Wilson lines
%\cite{Blaizot:2002xy}.

This argument makes it tempting to conclude that, with increasing $Y$,
the support of the target field $\alpha_a(x^-,\bmx)$ extends only towards
increasing $x^-$, thus yielding a field distribution which is asymmetric
in $x^-$. This was indeed the prevailing viewpoint in the original
literature on the JIMWLK evolution (see e.g.
\cite{Iancu:2002xk,Iancu:2003xm}), but now we shall argue that this is
actually not quite right: although the functional derivatives in \eqn{H}
have a one--sided action which amounts to color rotations at the largest
value of $x^-$ alone, the overall structure of the Hamiltonian is such
that the target field is nevertheless built {\em symmetrically} in $x^-$.
In fact, it is precisely this symmetry of the target field distribution
under reflection in $x^-$ which is responsible for the mirror symmetry in
the evolution equations.

To see that, it is useful to notice that \eqn{H} can be alternatively
rewritten as \cite{Kovner:2005jc,Kovner:2005en}
  \beq\label{HLR}
 H = \frac{g^2}{16 \pi^3} \int_{\bmu\bmv\bmz}
 \mcal{M}_{\bmu\bmv\bmz}\Big(J_{L\bmu}^aJ_{L\bmv}^a
+J_{R\bmu}^aJ_{R\bmv}^a + 2 \wt{V}_{\bmz}^{ab} J_{R\bmu}^a
J_{L\bmv}^b\Big),
 \eeq
where $J_{L\bmu}^a$ and $J_{R\bmv}^a$ are functional differential
operators acting as `left' and `right' Lie derivatives --- that is, the
generators of infinitesimal color rotations at the largest and,
respectively, smallest value of $x^-$. They are defined as
 \beq\label{JLJR}
 J_{L\bmu}^a\,\equiv\,-
 \frac{1}{\rmi g}\,\frac{\delta}{\delta \alpha^a_{\bmu}}
 \,,\qquad
 J_{R\bmu}^a\,\equiv\,\frac{1}{\rmi g}\,\wt{V}_{\bmu}^{ab}
 \frac{\delta}{\delta \alpha^b_{\bmu}}\,,\eeq
and satisfy
 \beq\label{funcder}
 J_{L\bmu}^a\,
 V_{\bmx}^{\dagger}\, =\,-
 \delta_{\bmx\bmu}\, t^a V_{\bmx}^{\dagger},
 \qquad J_{R\bmu}^a\,
 V_{\bmx}^{\dagger}\,=\,
  \delta_{\bmx\bmu}\, V_{\bmx}^{\dagger}t^a,\,
 \eeq
where the second equation follows from the first one after using
\eqn{adjtofun}. These equations imply the following commutation relations
 \beq\label{algebra}
 [J_{L\bmu}^a,\, J_{L\bmv}^b] = \rmi f^{abc}J_{L\bmu}^c
 \delta_{\bmu\bmv},\quad
 [J_{R\bmu}^a,\, J_{R\bmv}^b] = \rmi f^{abc}J_{R\bmu}^c
 \delta_{\bmu\bmv},\quad[J_{L\bmu}^a,\, J_{R\bmv}^b]=0\,,\eeq
showing that the two sets of generators satisfy two independent SU$(N_c)$
Lie algebras.

The physical interpretation of the various terms in \eqn{HLR} is quite
transparent: the action of $J_{L\bmu}^a$ on the quark Wilson line
$V_{\bmx}^{\dagger}$ (the first equation in \eqn{funcder}) amounts to the
addition of an infinitesimal layer of color field at the {\em largest}
values of $x^-$, whereas the action of $J_{R\bmu}^a$ is tantamount to a
corresponding addition at the {\em smallest} values of $x^-$. Hence, the
manifest symmetry of  \eqn{HLR} under the exchange $L\leftrightarrow R$
implies that, during the high--energy evolution, the {\em distribution}
of the target color field $\alpha^a(x^-,\bmx)$ --- by which we mean its
support and correlations --- is built symmetrically in $x^-$ around
$x^-=0$. Moreover, this is also the origin of the mirror symmetry since,
as previously noticed, the latter follows from the combined action of the
first two terms in the Hamiltonian \eqref{H}, or \eqref{HLR} --- those
which get interchanged with each other under the permutation
$L\leftrightarrow R$ of the Lie derivatives.

%By this, we more precisely mean that the longitudinal support of this
%field expands symmetrically around $x^-=0$ and that the additional
%(1--point and 2--point) correlations induced in one step of the evolution
%are even functions of $x^-$.

To better appreciate the differences between an evolution which is
symmetric in $x^-$ and one which is not, it is instructive to consider
the evolution of a Wilson line, say for a quark projectile. When
computing the target average \eqref{average} with the CGC weight function
at rapidity $Y$, the support of the color field $\alpha_a(x^-,\bmx)$ is
restricted to $-x^-_M \le x^- \le x^-_M$ with
$x^-_M(Y)=x^-_0\exp(Y-Y_0)$. (This follows from the uncertainty
principle: the softest gluon modes that have been integrated over have
longitudinal momentum $p^+ = xP$ with $x=\rme^{-Y}$, with $P=$ the total
hadron momentum; hence, they are delocalized in $x^-$ over a distance
$\sim 1/p^+\propto \rme^{Y}$.) Thus, the quark Wilson line can be
equivalently rewritten as
 \beq
  \label{Vfdt}
  {V}^{\dagger}_{\bmx}\,\equiv\,{\mbox P}\exp \left[
 \rmi g\int_{-x^-_M}^{x^-_M}
 \rmd x^-\alpha_a(x^-,\bmx)t^a \right].
 \eeq
This makes it manifest that, with increasing $Y$, the Wilson line `grows'
simultaneously at its both endpoints. To visualize the effect of one step
in the evolution ($Y\to Y+\rmd Y$), it is useful to discretize rapidity
by writing $Y_n=n\epsilon$, with $\epsilon$ an infinitesimal rapidity
interval. Then under one additional step $n\to n+1$, the upper bound of
the support extends as $x_n^-\equiv x^-_M(Y_n)\to x^-_{n+1}=
x_n^-(1+\epsilon)$ and the Wilson lines evolves as ${V}^{\dagger}_n\to
{V}^{\dagger}_{n+1}$, with
 \beq\label{Vevol}
 {V}^{\dagger}_{n+1}(\bmx)=\exp[\rmi g \epsilon \alpha_{n+1}(\bmx)]\,
 V_{n}^{\dagger}(\bmx)\,\exp[\rmi g \epsilon \alpha_{-(n+1)}(\bmx)]
 \eeq
where
 \beq
 \alpha_{n+1}(\bmx)\equiv x_n^-\alpha_a(x^-_{n+1},\bmx)t^a\quad
 \mbox{and}\quad
 \alpha_{-(n+1)}(\bmx)\equiv x_n^-\alpha_a(-x^-_{n+1},\bmx)t^a
 \eeq
represent the additional fields generated in this evolution step {per
unit of space--time rapidity}. The infinitesimal gauge rotations
associated with these new fields can be expanded in powers of $\epsilon$.
Strictly speaking, this expansion must be pushed to quadratic order in
$\epsilon$, to match with the fact that the evolution Hamiltonian
\eqref{HLR} involves second order functional derivatives. However, the
quadratic terms arising from the expansion of a given Wilson line do not
contribute to the evolution of gauge--invariant observables: they would
yield `tadpole' contributions $\propto \delta_{\bmu\bmv}$, but the dipole
kernel in the Hamiltonian vanishes when $\bmu=\bmv$. In other terms, the
two functional derivatives within $H$ must act on {\em different} Wilson
lines within the observable to give a non--zero result. So, we can
restrict the expansion of \eqref{Vevol} to linear order in $\epsilon$,
which yields
 \beq\label{Vevol2}
 {V}^{\dagger}_{n+1}(\bmx)-V_{n}^{\dagger}(\bmx)
 = \rmi g \epsilon \big[\alpha_{n+1}(\bmx)\,V_{n}^{\dagger}(\bmx)\,+
 \,V_{n}^{\dagger}(\bmx)\alpha_{-(n+1)}(\bmx)\big] +
 \mathcal{O}(\epsilon^2)\,.
 \eeq
Clearly, the two terms in the r.h.s. correspond to the infinitesimal,
`left' and `right', color rotations in \eqn{funcder}. If instead of the
symmetric evolution above, one would have considered an {\em asymmetric}
one, where the target fields expands towards positive values of $x^-$
alone, the analog of Eqs.~\eqref{Vevol}--\eqref{Vevol2} would have
involved the `left' infinitesimal color precession alone.

In Ref.~\cite{Blaizot:2002xy}, the JIMWLK evolution has been reformulated
as a random walk in the space of Wilson lines, which is formally such
that one additional step corresponds to an infinitesimal rotation of
${V}^{\dagger}(\bmx)$ on the `left' alone. However, by inspection of the
manipulations there, one can check that the additional contribution
$\alpha_{n+1}(\bmx)$ to the target field in the $(n+1)$th step is such
that, in reality, that step simultaneously generate a color precession on
the `left' {\em and} on the `right'. That is, the Langevin process
introduced in Ref.~\cite{Blaizot:2002xy} does in fact describe a
symmetric evolution for the Wilson lines (or for the target field
distribution), although this has not been recognized there.

\section{The Gaussian approximation}
\setcounter{equation}{0} \label{sect:gaussian}

In this section we shall demonstrate that the JIMWLK equation for the CGC
weight function admits an approximate Gaussian solution which properly
captures both the BFKL dynamics in the dilute regime at $k_\perp\gg
Q_s(Y)$ and the approach towards the black disk limit in the saturation
regime at $k_\perp\ll Q_s(Y)$. Our analysis improves over previous,
related, constructions in the literature
\cite{Iancu:2002xk,Iancu:2002aq,Kovchegov:2008mk} at two important
levels: \texttt{(i)} we actually {\em justify} the Gaussian approximation
--- including for the description of the higher--point correlation functions
and for finite $N_c$ ---, on the basis of the Balitsky--JIMWLK equations;
\texttt{(ii)} we implement the `mirror' symmetry discussed in
Sect.~\ref{sect:mirror}, that is, we construct a Gaussian distribution
which is symmetric in $x^-$ at any $Y$. As we shall see, this last
condition is in fact compulsory to achieve a faithful description of the
JIMWLK dynamics deeply at saturation.

The material of this section is organized as follows: the Gaussian weight
function is introduced in Sect.~\ref{sect:CGC}, compared to the MV model
in Sect.~\ref{sect:MV}, and justified in Sects.~\ref{sect:BFKL} and
\ref{sect:RPA} by comparison with piecewise approximations to the JIMWLK
equations in the limiting regimes alluded to above.

%Finally, in Sect.~\ref{sect:self} we present a self--consistent procedure
%to construct a smooth, global, approximation for the kernel of the
%Gaussian which has the right limiting behaviours and which holds for any
%$N_c$.

\subsection{The Gaussian weight function}
\label{sect:CGC}

The most general Gaussian weight function which is consistent with gauge
symmetry\footnote{By `gauge symmetry' we more precisely have in mind here
the class of gauges within which the target color field has the structure
$A^\mu_a =\delta^{\mu+}\alpha_a$. Some gauge artifacts, which are
inherent in \eqn{Gauss} but turn out to be harmless in practice, will be
later discussed.} and describes a target field distribution which is
symmetric in $x^-$ reads
 \beq\label{Gauss}
 W_Y[\alpha]=\mathcal{N}_Y\exp\left[
 -\frac{1}{2}\int_{-x^-_M}^{x^-_M}\rmd x^-\!\int_{\bmx_1\bmx_2}
 \,{\alpha_a(x^-,\bmx_1)\barg^{-1}(x^-,\bmx_1,\bmx_2)
 \alpha_a(x^-,\bmx_2)}
 \right]\delta_Y[\alpha]\,,
 \eeq
where $x^-_M(Y)=x^-_0\exp(Y-Y_0)$ and the kernel
$\barg^{-1}(x^-,\bmx_1,\bmx_2)$ is an even function of $x^-$, assumed to
be invertible. The functional $\delta$--function $\delta_Y[\alpha]$
ensures that the target field vanishes at larger longitudinal coordinates
$|x^-|>x^-_M(Y)$ :
 \beq
 \delta_Y[\alpha]\,\equiv\,\prod_{|x^-|>x^-_M}\,\prod_{\bmx}\,\prod_a
 \,\delta(\alpha_a(x^-,\bmx))\,.\eeq
Here, $\delta(\alpha_a(x^-,\bmx))$ is the usual $\delta$--function and a
discretization of the space--time is understood. Finally, the overall
normalization factor $\mathcal{N}_Y$ in \eqn{Gauss} is such that $\int
{\mathcal D}\alpha \, W_Y[\alpha]=1$.

\eqn{Gauss} implies that the only non--trivial correlation of the target
fields is their 2--point function, which is moreover local in $x^-$ :
\beq\label{gamma}
 \lan\alpha_a(x^-_1,\bmx_1)
 \alpha_b(x^-_2,\bmx_2)\ran_Y\,=\,\delta^{ab}\,
 \Theta\big(x^-_M(Y)-|x^-_1|\big)\,
 \delta(x^-_1-x^-_2)\,
 \barg(x^-_1,\bmx_1,\bmx_2)\,.\eeq
Within this Gaussian approximation, the locality in $x^-$ is required by
gauge symmetry: to preserve the latter, any non--locality in $x^-$ should
be accompanied by gauge links (Wilson lines) built with the field
$\alpha^a$, which would spoil Gaussianity.

The Gaussian distribution \eqref{Gauss} is manifestly symmetric in $x^-$
around $x^-=0$ and this symmetry is preserved by the high energy
evolution. In fact, \eqn{Gauss} depends upon $Y$ only via the two
endpoints, $x^-_M(Y)$ and $-x^-_M(Y)$, of the support in $x^-$, meaning
that the high--energy evolution proceeds via the symmetric expansion of
the color field distribution towards both larger and smaller values of
$x^-$. Specifically, by using the methods in
Refs.~\cite{Iancu:2002xk,Blaizot:2002xy}, one can check that the Gaussian
weight function (\ref{Gauss}) obeys the following evolution equation:
\beq\label{RGMFA}
{\del  W_Y[\alpha] \over {\del Y}}\,=\,{1 \over 2}
\int_{\bmu\bmv}\!\barg_Y(\bmu,\bmv)\, \left({\delta \over
\delta\alpha^a_{L \bmu}} {\delta \over\delta \alpha^a_{L\bmv}} +{\delta
\over \delta\alpha^a_{R \bmu}} {\delta \over\delta
\alpha^a_{R\bmv}}\right)W_Y[\alpha]\,,
 \eeq
where the `left' (`right') functional derivatives act on the target field
at the largest (smallest) value of $x^-$, that is, at $x^- =x^-_M(Y)$ and
respectively $x^- = -x^-_M(Y)$. Also $\barg_Y(\bmx_1,\bmx_2)$ denotes the
field correlator per unit space--time rapidity as produced in the last
step of the evolution, %({\em i.e.}, in the upper most bin in rapidity):
  \beq\label{gammayM}
\barg_Y(\bmx_1,\bmx_2)\equiv x^-_M\ \barg(\pm x^-_M,\bmx_1,\bmx_2)\,.
 \eeq
\eqn{RGMFA} makes it manifest that the momentum rapidity and the
space--time rapidity are identified with each other by the high--energy
evolution. In order to solve \eqn{RGMFA}, one also needs the
generalization of \eqn{gammayM} to intermediate values $\rmy < Y$ for the
space--time rapidity, that is
  \beq\label{gammay}
\barg_\rmy(\bmx_1,\bmx_2)\equiv |x^-|\ \barg(x^-,\bmx_1,\bmx_2)\,,\qquad
 \mbox{with}\qquad \rmy\equiv Y_0+\,\ln\frac{|x^-|}{x^-_0}\,.
 \eeq

\eqn{RGMFA} should be viewed as a mean field approximation to the JIMWLK
equation \eqref{jimwlk}. It shows the same `left--right' symmetry as the
original equation, cf. \eqn{HLR}, and hence it is consistent with the
mirror symmetry discussed in Sect.~\ref{sect:mirror}. Clearly, this would
not be the case if, instead of the symmetric Gaussian \eqref{Gauss}, one
would consider an asymmetric one, say with support at $0\le x^-\le
x^-_M(Y)$, as in the previous literature
\cite{Iancu:2002xk,Iancu:2002aq,Kovchegov:2008mk}: the corresponding
evolution equation would contain only the `left' functional derivatives
--- i.e., only the first term inside the brackets in \eqn{RGMFA}.

To justify the Gaussian {\em Ansatz} \eqref{Gauss} for the CGC weight
function, we shall shortly compare the associated evolution equation
\eqref{RGMFA} to the actual JIMWLK equation, in different kinematical
regimes. In this process, we shall deduce piecewise approximations for
the kernel $\barg_Y(\bmx_1,\bmx_2)$, valid at high ($k_\perp\gg Q_s(Y)$)
and low ($k_\perp\ll Q_s(Y)$) momenta, respectively.

\subsection{The McLerran--Venugopalan model}
\label{sect:MV}

Before we turn to the JIMWLK evolution, let us briefly discuss the
McLerran--Venugopalan (MV) model \cite{McLerran:1993ni,McLerran:1993ka}
that we shall take as our initial condition at rapidity $Y_0$. Besides
providing the initial conditions, this model (and its {\em ad--hoc}
extrapolation towards high--energy) will serve as a baseline of
comparison for the mean--field results that we shall later obtain. Its
discussion will also give us an opportunity to clarify some subtle
aspects of the Gaussian approximation, like the gauge artifacts in the
`$\alpha$--representation', cf. \eqn{Gauss}.

In the MV model one assumes that the color charges in the nucleus are
uncorrelated valence quarks. Accordingly the distribution of the color
charge density $\rho^a(x^-,\bmx)$ is a Gaussian with a kernel which is
local in the transverse plane:
 \beq\label{MVW}
 W_{Y_0}[\rho]\,=\,\mathcal{N}_{Y_0}\,\exp\left[
 -\frac{1}{2}\int_{-x^-_0}^{x^-_0}\rmd x^-\!\int_{\bmx}
 \,\frac{\rho^a(x^-,\bmx)
 \rho^a(x^-,\bmx)}{\lambda(x^-,\bmx)}
 \right]\,\delta_{Y_0}[\rho]\,,
 \eeq
where of course $\lambda(-x^-,\bmx)=\lambda(x^-,\bmx)$. \eqn{MVW}
implies:
 \beq\label{MV}
 \lan\rho_a(x^-_1,\bmx_1)
 \rho_b(x^-_2,\bmx_2)\ran_Y\,=\,\delta_{ab}\,
 \Theta\big(x^-_0-|x^-_1|\big)\,
 \delta(x^-_1-x^-_2)\,\delta_{\bmx_1\bmx_2}\,
 \lambda(x^-_1,\bmx_1)\,.\eeq
The quantity $\lambda(x^-,\bmx)$ has the meaning of color charge squared
per unit transverse area per unit longitudinal distance. In general the
nucleus is assumed to be homogeneous in the transverse plane, {\em i.e.}
the kernel in \eqn{MVW} is taken to be independent of $\bmx$. Under that
assumption, the calculation of expectation values in the MV model is not
sensitive to the detailed dependence of the kernel upon $x^-$, but only
to its integral
 \beq\label{intlam}
 \mu^2\equiv \int_{-x^-_0}^{x^-_0}\rmd x^-
 \lambda(x^-)\,=\,2
  \int_{-\infty}^{Y_0}\rmd \rmy\, \lambda_\rmy=
  \,\frac{g^2A}{2\pi R_A^2}\,,\eeq
which physically represents the color charge squared per unit area. In
the above equation, the quantity $\lambda_\rmy\equiv |x^-| \lambda(x^-)$
(the strength of the charge correlator per unit space--time rapidity) has
been defined by analogy with \eqn{gammay}. The last equality follows
after counting the color charges of the valence quarks within a nucleus
with atomic number $A$ and transverse area $\pi R_A^2$ (see e.g.
\cite{Iancu:2003xm}).  The fact that it is only the integrated quantity
\eqref{intlam} which matters arises from the fact that, under the present
assumptions, the charge correlator \eqref{MV} is {\em separable} as a
function of $x^-$ and the transverse coordinates. We shall return to this
issue in Sects.~\ref{sect:eqsMF} and \ref{sect:largeNc}.

\eqn{MVW} is gauge invariant, but in order to make contact with the
$\alpha$--representation that we use throughout this paper, we shall
henceforward consider it within the class of gauges where the target
field is of the form $A^\mu_a =\delta^{\mu+}\alpha_a$. Then
$\alpha_a(x^-,\bmx)$ is related to the color charge density
$\rho_a(x^-,\bmx)$ via the 2--dimensional Coulomb equation:
$-\nabla^2_\perp\alpha_a=\rho_a$. So, for a homogeneous target, \eqn{MV}
implies the following expression for the 2--point function for the color
field, in transverse momentum space (we denote ${\bm r}=\bmx_1-\bmx_2$
and $k_\perp=|\bmk|$):
  \beq\label{gammaMV}
  \barg_\rmy(\bmk)\,\equiv
  \int {\rmd^2{\bm r}}\,\rme^{\rmi\bmk\,\cdot\,{\bm r}}\,\barg_\rmy({\bm r})
   \,=\,\frac{\lambda_\rmy}{k_\perp^4}\quad\mbox{for}
   \quad\rmy\,\le\,Y_0\,.
 \eeq
Here and from now on, we prefer to work with the expressions of the
various correlators {\em per unit space--time rapidity}, cf.
\eqn{gammay}, since these are the expressions which most directly enter
the mean--field evolution equations like \eqref{RGMFA}.

\eqn{gammaMV} raises a potential problem: the Fourier transform of this
expression back to the transverse coordinate space is not well defined,
as it involves a (quadratic) infrared divergence. This problem reflects
the fact that, by itself, the field $\alpha_a(x^-,\bmx)$ is not invariant
under the residual gauge transformations which preserve the structure
$A^\mu_a =\delta^{\mu+}\alpha_a$ for the target field. The infinitesimal
version of such a transformation reads $\alpha_a(x^-,\bmx)\to
\alpha_a(x^-,\bmx)+ \del^+\omega_a(x^-)$, with $\omega_a(x^-)$ an
arbitrary function \cite{Hatta:2005as}; so, clearly, the color charge
density $\rho_a(x^-,\bmx)$ is invariant under this transformation.
Strictly speaking, the general weight function $W_Y$ (and, in particular,
its Gaussian approximation, \eqn{Gauss}) should be written as a
functional of $\rho$, to make gauge symmetry manifest. On the other hand,
observables like scattering amplitudes are built with Wilson lines, which
are path--ordered exponentials of $\alpha$. Taken separately, one Wilson
line is not gauge invariant (rather, it transforms via color rotations
\cite{Hatta:2005as}), but the physically relevant operators, which
involve a product of such lines, cf. \eqn{S2n}, {\em are} invariant.
Whenever computing the expectation value of such a gauge--invariant
operator, there is no problem with using the weight function in the
$\alpha$--representation, as given in \eqn{Gauss}: all the gauge
artifacts cancel out in the final result.

As an example, consider the calculation of the dipole $S$--matrix within
the MV model. The corresponding result is well known and reads (see also
Sect.~\ref{sect:eqsMF})
 \beq\label{SMV}
 \lan \hat{S}_{\bmx_1\bmx_2}\ran_{Y_0} =
 \rme^{-\Gamma_{Y_0}(\bmx_1,\bmx_2)}
 \eeq
where we have assumed the MV model to apply at all the rapidities $Y\le
Y_0$ and we defined
 \beq\label{GammaMV}
 \Gamma_{Y_0}(\bmx_1,\bmx_2) =
 {g^2 C_F}\int^{Y_0}_{-\infty}
 \dif \rmy \,
 \left[\barg_\rmy(\bmx_1,\bmx_1)
 +\barg_\rmy(\bmx_2,\bmx_2)
 - 2 \barg_\rmy(\bmx_1,\bmx_2)\right],
 \eeq
with $C_F={(N_c^2-1)}/{2N_c}$. \eqn{GammaMV} involves only the following
linear combination of the target field correlators
 \beq \label{gammacor}
 \hspace*{-0.6cm}\gamma_y(\bmx_1,\bmx_2)\equiv
  -\barg_y(\bmx_1,\bmx_2)+\frac{1}{2}\big[
 \barg_y(\bmx_1,\bmx_1)+\barg_y(\bmx_2,\bmx_2)\big]=
 \int \frac{\rmd^2\bmk}{(2\pi)^2}\frac{\lambda_y}
 {k_\perp^4}\Big[1-\rme^{\rmi\bmk\,\cdot\,(\bmx_1-\bmx_2)}\Big].
 \eeq
which {\em is} gauge--invariant, since under a residual gauge
transformation the target field $\alpha^a(x^-,\bmx)$ changes by a
$\bmx$--independent quantity. The sign in the r.h.s. of \eqn{gammacor} is
such that $\gamma_y(\bmx_1,\bmx_2)$ be positive--semidefinite. The last
equality in \eqn{gammacor}, which involves the color charge correlator
$\lambda_y$, illustrates the fact that the infrared divergences due to
gauge artifacts cancel out in the linear combination \eqref{gammacor}.
Strictly speaking, the above integral over $\bmk$ still has a logarithmic
infrared divergence, but this is milder than the quadratic divergence
appearing in the Fourier transform of $\barg_\rmy(\bmk)$ in
\eqn{gammaMV}. The remaining divergence is not a gauge artifact anymore,
but a `physical' singularity of this model: it reflects the lack of
correlations among the color sources. After taking into account the
high--energy evolution, transverse correlations get built which screen
out this divergence, as we shall shortly see. For completeness, let us
estimate the final integral in \eqn{GammaMV}: introducing an infrared
cutoff $\Lambda$ to regularize the remaining infrared divergence and
writing $r=|\bmx_1-\bmx_2|$, one finds
 \beq\label{Gamma0}
 \Gamma_{Y_0}(r) =
 {g^2 C_F}\,\frac{g^2A}{2\pi R_A^2}\,
 \int \frac{\rmd^2\bmk}{(2\pi)^2}\frac{1-\rme^{\rmi\bmk\,\cdot\,{\bm r}}}
 {k_\perp^4}\,\simeq\,\frac{r^2 Q_0^2}{4}\,\ln\frac{1}{r^2\Lambda^2}\,,
 \eeq
where $Q_0^2\equiv 2\alpha_s^2 C_F A/R_A^2$ is essentially the nuclear
saturation scale\footnote{More precisely, $Q_s(Y_0)$ is defined by the
condition $\Gamma_{Y_0}(r=2/Q_s(Y_0))\sim\order{1}$.} (as probed by a
quark--antiquark dipole) at the initial rapidity $Y_0$. Although obtained
within the MV model, the above results are generic in the following
sense: all the gauge--invariant observables computed in the Gaussian
approximation involve the kernel $\barg_y(\bmx_1,\bmx_1)$ of the Gaussian
(the correlator of the target color field) only via the linear
combination shown in \eqn{gammacor}. So, in practice, there is no problem
with using the $\alpha$--representation, as shown in \eqn{Gauss}.

Let us conclude this subsection with a remark on the calculation of
expectation values within the MV model. The similarity between the
respective weight function, \eqn{MVW}, and the Gaussian approximation to
the JIMWLK evolution, \eqn{Gauss}, makes it clear that one can consider
the MV model as the result of a fictitious `evolution' in which the
target charge distribution is built in layers of $x^-$, from $x^-=0$ up
to $|x^-|=x^-_0$. Specifically, let $W_{X^-}[\rho]$ denote the
generalization of \eqn{MVW} in which $x^-_0$ is replaced by $X^-$ and
assume the nucleus to be homogeneous in the transverse plane. Then
\eqn{MVW} is the solution to the following, functional, evolution
equation (compare to \eqn{RGMFA})
 \beq\label{RGMV}
{\del  W_{X^-}[\rho] \over {\del {X^-}}}\,=\,{1 \over 2}\,\lambda_{X^-}
\int_{\bmu\bmv}\! \left({\delta \over \delta\rho^a_{L \bmu}} {\delta
\over\delta \rho^a_{L\bmv}} +{\delta \over \delta\rho^a_{R \bmu}} {\delta
\over\delta \rho^a_{R\bmv}}\right)W_{X^-}[\rho]\,,
 \eeq
integrated from $X^-=0$ up to $X^-=x^-_0$. In this equation, $\rho^a_{L}$
and $\rho^a_{R}$ refer to the color charge densities at $x^-=X^-$ and
$x^-=-X^-$, respectively. When applied to the evolution of the
Wilson--line correlations, \eqn{RGMV} amounts to constructing the Wilson
lines via {\em symmetric} iterations, {\em i.e.} via infinitesimal color
precessions which proceed simultaneously `on the left' and `on the
right', as shown in \eqn{Vevol}. However, within the context of the MV
model, this symmetric iteration is merely a choice of a discretization
prescription and any other choice is equally good. As a matter of fact,
the common choice in the literature in this context (see e.g.
\cite{Kovner:2001vi,Blaizot:2004wv,JalilianMarian:2004da,Dominguez:2011wm})
is to perform asymmetric iterations `on the left'\,:
 \beq\label{Vevolleft}
 {V}^{\dagger}_{n}(\bmx)
 \,\to\,{V}^{\dagger}_{n+1}(\bmx)=\exp[\rmi g \epsilon \alpha_{n+1}(\bmx)]\,
 V_{n}^{\dagger}(\bmx),\eeq
where this time $n$ refers to a discretization of the $x^-$ axis. This
procedure is tantamount to solving the following evolution equation
 \beq\label{RGMV2}
{\del  W_{X^-}[\rho] \over {\del {X^-}}}\,=\,{1 \over 2}\,\lambda_{X^-}
\int_{\bmu\bmv} {\delta \over \delta\rho^a_{L \bmu}} {\delta \over\delta
\rho^a_{L\bmv}} W_{X^-}[\rho]\,,
 \eeq
from $X^-=-x_0^-$ up to $X^-=x^-_0$. In practice, one often takes
$x^-_0\to\infty$, since the results are anyway insensitive to the actual
value of $x^-_0$, but only depend upon the integral $\int \rmd X^-
\lambda(X^-)$.

The above discussion sheds more light on the role of the `left--right'
symmetry in the evolution equations. So long as the CGC weight function
is {\em given} (like in the MV model) and the associated evolution
equations are merely used as a convenient device to compute expectation
values, the symmetric discretization in \eqn{Vevol} is not compulsory and
it might not even be the most convenient one in practice. However, for
the JIMWLK equation and any (mean field) approximation to it, the
symmetric iteration is the only one to be correct, since this is how the
target field distribution gets actually built via quantum evolution: the
`outer' layers (those located at larger values of $|x^-|$) are
constructed {\em after} the `inner' ones (those at smaller $|x^-|$), and
the new correlations built in one step depend upon the color field
produced in all the previous steps. Hence, it would make no sense to
consider an asymmetric evolution, like \eqn{RGMV2}, since this would
violate causality in the domain of negative $x^-$.

\subsection{Weak--scattering regime: the BFKL dynamics}
\label{sect:BFKL}

In what follows, we shall study the JIMWLK evolution in two limiting
regimes --- large transverse momenta $k_\perp\gg Q_s(Y)$ in this section
and relatively small momenta $k_\perp\ll Q_s(Y)$ in the next subsection
--- with the purpose of showing that, in both regimes, the evolution is
consistent with a mean field approximation of the type shown in
\eqn{RGMFA}. We recall that $Q_s(Y)$ is the saturation momentum in the
target (in a frame in which the target carries most of the total rapidity
separation $Y$) and it increases with $Y$ very fast. For a multi--point
correlation function like the quadrupole \eqref{Squadrupole}, the
statement that the `transverse momenta are much larger than $Q_s$' means
that all the transverse separations $r_{ij}\equiv |\bmx_i-\bmx_j|$
between the external points are much smaller than $1/Q_s(Y)$. Similarly,
by `momenta much smaller than $Q_s$', we mean that $r_{ij}\gg 1/Q_s(Y)$
for any pair $(\bmx_i,\bmx_j)$ of external points. Very asymmetric
configurations, where some of the distances $r_{ij}$ are much larger than
$1/Q_s(Y)$ while the others are much smaller, are strictly speaking not
covered by the present analysis and must be separately studied. We shall
discuss some examples of that kind in Sect.~\ref{sect:finiteNc} below.

For high transverse momenta $k_\perp\gg Q_s(Y)$, the gluon density in the
target is low, meaning that the corresponding color field is weak: $g
\int \rmd x^- \alpha\ll 1$. It is then possible to expand the Wilson
lines to lowest non--trivial order in the field in their exponent, within
both the JIMWLK Hamiltonian and the operators defining the observables.
For an operator like the dipole $S$--matrix \eqn{Sdipole}, we need to
push the expansion in $g\alpha$  up to the second order, since the linear
terms vanish after averaging. Introducing the dipole $T$--matrix operator
$\hat{T}_{\bmx_1\bmx_2} \equiv 1 - \hat{S}_{\bmx_1\bmx_2}$, whose
expectation value represents the corresponding scattering amplitude, this
expansion yields
 \beq\label{Tdipdil}
 \lan\hat{T}_{\bmx_1\bmx_2}\ran_Y
 \,\simeq\,\frac{g^2}{4N_c}\lan(\alpha_{\bmx_1}^a-\alpha_{\bmx_2}^a)^2
 \ran_Y\qquad\mbox{with}\qquad
 \alpha^a_{\bmx}\,\equiv\,\int
 \rmd x^-\alpha_a(x^-,\bmx)\,.\eeq
%and we observe that to this order of accuracy $\hat{T}_{\bmx_1\bmx_2}$ is
%invariant under $\bmx_1 \leftrightarrow\bmx_2$.
The weak scattering regime corresponds to $\lan \hat{T} \ran_Y\ll 1$.
Note that \eqn{Tdipdil} involves only the linear combination
\eqref{gammacor} of the target field correlators, in agreement with the
discussion in Sect.~\ref{sect:MV}. The similar expansion for the
quadrupole $S$--matrix, \eqn{Squadrupole}, yields
 \beq\label{QBFKL} 1 -\lan \hat{Q}_{\bmx_1\bmx_2\bmx_3\bmx_4}
\ran_Y\simeq
  \lan\hat{T}_{\bmx_1\bmx_2}
 -\hat{T}_{\bmx_1\bmx_3}
 +\hat{T}_{\bmx_1\bmx_4}
 +\hat{T}_{\bmx_2\bmx_3}
 -\hat{T}_{\bmx_2\bmx_4}
 +\hat{T}_{\bmx_3\bmx_4} \ran_Y,
 \eeq
where it is understood that $\lan \hat{T} \ran_Y$ is evaluated according
to \eqn{Tdipdil}. More generally, in this dilute regime, all the
$n$--point functions of the type shown in Eqs.~\eqref{S2n} or
\eqref{multitrace} reduce to linear combinations of dipole amplitudes.
This already shows that a Gaussian approximation for the CGC weight
function should be indeed possible, to the accuracy of interest. To
identify this approximation, let us also consider the weak--field limit
of the JIMWLK Hamiltonian. Its obtention is facilitated by observing that
\eqn{H} can be rewritten as
 \beq\label{HWilson}
 1 + \wt{V}^{\dagger}_{\bmu} \wt{V}_{\bmv}
 -\wt{V}^{\dagger}_{\bmu} \wt{V}_{\bmz}
 -\wt{V}^{\dagger}_{\bmz} \wt{V}_{\bmv}\,=\,
 \left(1-\wt{V}^{\dagger}_{\bmu} \wt{V}_{\bmz}\right)\,
 \left(1-\wt{V}^{\dagger}_{\bmz} \wt{V}_{\bmv}\right)\,.\eeq
The leading order terms in the dilute regime are then obtained by
expanding the Wilson lines within each of the two parentheses above to
{\em linear} order in $g\alpha$. (This amounts to an expansion of the
original structure in the l.h.s. of \eqn{HWilson} up to {\em quadratic}
order.) For instance,
 \beq
\label{expV} 1-\wt{V}^{\dagger}_{\bmu} \wt{V}_{\bmz} \,\simeq\,-\rmi
g\big(\alpha^a_{\bmu} -
 \alpha^a_{\bmz}\big) T^a\,,\eeq
with $\alpha^a_{\bmu}$ as defined in \eqn{Tdipdil}. After also using
$(T^a)_{bc} = \rmi f^{abc}$, one finds $H\simeq H_{\rm BFKL}$ with
 \beq\label{Hweak}
 H_{\rm BFKL} = -\, \frac{g^2}{16\pi^3}
 \int_{\bmu\bmv\bmz}
 {\mathcal  M}_{\bmu\bmv\bmz}
 \big(\alpha^a_{\bm{u}}-\alpha^a_{\bm{z}}\big)
 \big(\alpha^b_{\bm{z}}-\alpha^b_{\bm{v}}\big)\,
 f^{acf} f^{bfd}
 \frac{\delta}{\delta\alpha^c_{\bm{u}}}\frac{\delta}
 {\delta\alpha^d_{\bm{v}}}\,.
 \eeq
This Hamiltonian is supposed to act on operators which are themselves
evaluated in the weak--scattering regime and hence are quadratic
functions of the field $\alpha_{\bmx}^a$, as illustrated in
\eqref{Tdipdil} and \eqref{QBFKL}. Clearly, the only evolution equation
of interest for us here is that obeyed by the dipole scattering amplitude
\eqref{Tdipdil}. This is readily obtained as
 \beq\label{BFKL}
 \frac{\del \lan \hat{T}_{\bmx_1\bmx_2} \ran_Y}{\del Y}=
 \atpi\, \int_{\bmz}
 \mcal{M}_{\bmx_1\bmx_2\bmz}\,
 \big\lan \hat{T}_{\bmx_1\bmz} + \hat{T}_{\bmz\bmx_2}
 -\hat{T}_{\bmx_1\bmx_2} \big\ran_Y\,,
 \eeq
and is recognized as the BFKL equation
\cite{Lipatov:1976zz,Kuraev:1977fs,Balitsky:1978ic}, that is, the
equation obtained after linearizing \eqn{BK} with respect to $\lan
\hat{T} \ran_Y$. By using its solution, one can compute any other
$n$--point function of the Wilson lines, like \eqn{QBFKL}, in this dilute
regime.

We now construct the Gaussian approximation which reproduces the BFKL
equation. To that aim, we shall compare the mean--field equation for
$\lan \hat{T}_{\bmx_1\bmx_2} \ran_Y$ generated by \eqn{RGMFA} with
\eqn{BFKL} and thus deduce an approximate expression for
$\barg_Y(\bmu,\bmv)$ valid in this linear regime. Notice that the left
and right functional derivatives yield identical results when acting on
the field $\alpha_{\bmx}^a$ which is integrated over $x^-$. Hence,
Eqs.~\eqref{RGMFA} and \eqref{Tdipdil} imply
 \begin{align}\label{BFKLMFA}
 \frac{\del \lan \hat{T}_{\bmx_1\bmx_2} \ran_Y}{\del Y}\bigg |_{\rm MFA}
 &=
 \frac{g^2}{4N_c}
\int_{\bmu\bmv}\!\barg_Y(\bmu,\bmv)\, \Big\lan{\delta \over
\delta\alpha^b_{\bmu}} {\delta \over\delta \alpha^b_{\bmv}}
 (\alpha_{\bmx_1}^a-\alpha_{\bmx_2}^a)^2
 \Big\ran_Y\nn
 &=
 \frac{g^2}{4N_c}\,2\delta^{aa}\,\left[\barg_Y(\bmx_1,\bmx_1)
 +\barg_Y(\bmx_2,\bmx_2)
 - 2 \barg_Y(\bmx_1,\bmx_2)\right]\nn
 &= g^2
 \frac{N_c^2-1}{N_c}\,\gamma_Y(\bmx_1,\bmx_2)\,,
 \end{align}
with $\gamma_Y(\bmx_1,\bmx_2)$ defined as in \eqn{gammacor}. The last
equation can be integrated to yield
 \beq\label{fdef}
 \lan \hat{T}_{\bmx_1\bmx_2} \ran_Y\Big |_{\rm MFA}= 2g^2 C_F\,f_Y(\bmx_1,\bmx_2)
\quad\mbox{with}\quad f_Y(\bmx_1,\bmx_2)\equiv \int_{-\infty}^Y
 \rmd \rmy\,\gamma_\rmy(\bmx_1,\bmx_2)\,.
 \eeq
It is easy to check that the same expression for $\lan \hat{T} \ran_Y$
would be obtained by directly evaluating the expectation value in
\eqn{Tdipdil} with the help of \eqn{gamma}. But its above derivation via
the mean--field equation of motion has the merit to emphasize that the
evolution equations for gauge--invariant observables generated by the
Gaussian approximation involve the well--behaved kernel
$\gamma_Y(\bmx_1,\bmx_2)$ in spite of the fact that the corresponding
functional equation \eqref{RGMFA} features the (generally ill defined)
kernel $\barg_Y(\bmx_1,\bmx_2)$. This property is generic: it holds
beyond the present, BFKL, approximation. Thus, for all practical purposes
one can replace $\barg_Y(\bmu,\bmv)\to - \gamma_Y(\bmu,\bmv)$ within
\eqn{RGMFA}. This replacement works in the same way as that of the
original kernel in the JIMWLK equation
\cite{Weigert:2000gi,Iancu:2001ad,Ferreiro:2001qy} by the dipole kernel
in \eqn{H}: the new kernel is to be used only when acting on
gauge--invariant observables and it has the property to vanish at
$\bmu=\bmv$.

Returning to the mean--field expression \eqref{fdef} for $\lan \hat{T}
\ran_Y$, this must be consistent with the BFKL equation \eqref{BFKL}.
This is clearly the case provided the function  $f_Y(\bmx_1,\bmx_2)$
itself satisfies the BFKL equation:
 \beq\label{BFKLf}
 \frac{\del f_Y(\bmx_1,\bmx_2)}{\del Y}=
 \atpi\, \int_{\bmz}
 \mcal{M}_{\bmx_1\bmx_2\bmz}\left[f_Y(\bmx_1,\bmz)+f_Y(\bmz,\bmx_2)
 - f_Y(\bmx_1,\bmx_2)\right]\,.\eeq
The initial conditions for the above equations can be taken from the MV
model, which yields (for $r\ll 1/Q_0$) : $\lan \hat{T}(r) \ran_{Y_0}=
2g^2 C_F\,f_{Y_0}(r)=\Gamma_{Y_0}(r)$, with $\Gamma_{Y_0}$ given in
\eqn{Gamma0}.

%The Fourier transform of $f_Y(\bmx_1,\bmx_2)$ is essentially the
%unintegrated gluon distribution.

The solution to \eqn{BFKLf} is by now well understood. Here we will just
remind that the BFKL evolution introduces transverse correlations between
the `color sources' (radiated gluons) which ensure that the solution
$f_Y(\bmx_1,\bmx_2)$ becomes infrared finite after a rapidity evolution
$Y-Y_0\sim 1/\abar$. In particular, in the window for `extended geometric
scaling' \cite{Iancu:2002tr,Mueller:2002zm,Munier:2003vc}, which holds
for transverse momenta relatively close to (but still larger than) the
saturation momentum $Q_s(Y)$, one has\footnote{Notice that $k_\perp^4
\gamma_Y (\bmk)=k_\perp^4 \barg_Y (\bmk)$ since in momentum space the
difference between $\gamma_Y (\bmk)$ and $\barg_Y (\bmk)$ is proportional
to $\delta^{(2)}(\bmk)$.} $\lambda_Y (\bmk)= k_\perp^4 \gamma_Y (\bmk)
\propto k_\perp^{2(1-\gamma_s)}$ with $\gamma_s\approx 0.63$ (the `BFKL
anomalous dimension at saturation'). Then, clearly, the integral over
$\bmk$ in \eqn{gammacor} is well defined when computed within the BFKL
approximation.

To summarize, the mean--field equation \eqref{RGMFA}, where it is
understood that the kernel can be replaced as $\barg_Y\to - \gamma_Y$
with the function $\gamma_Y(\bmu,\bmv)$ determined by \eqn{BFKLf},
properly encodes the BFKL evolution of the dipole amplitude in the weak
scattering regime. This conclusion holds for any value of the number of
colors $N_c$ and it extends to all the $n$--point functions like
\eqref{S2n} and \eqref{multitrace} which, in this regime, reduce to
linear combinations of dipole amplitudes.

Before concluding this section, let us recall that there are also other
aspects of the BFKL dynamics, which cannot be encoded into a Gaussian
weight function. They refer to operators more complicated than those in
\eqn{S2n}, which already at weak scattering involve more than two gluon
exchanges; that is, to lowest order in the weak field expansion, they
involve polynomials in $\alpha$ of a degree higher than two. (Such
operators can be obtained e.g. by subtracting the dipolar contributions
to the Wilson--line operators in Eqs.~\eqref{S2n}--\eqref{multitrace}.)
An example of that type is the `odderon' operator, which describes
$C$--odd exchanges and which in perturbation theory starts with three
gluon exchanges. The corresponding evolution equation is correctly
encoded (to leading logarithmic accuracy) in the JIMWLK equation
\cite{Hatta:2005as}
--- in particular, its low--density limit, known as the `BKP equation'
\cite{Bartels:1980pe,Kwiecinski:1980wb,Jaroszewicz:1980mq}, is generated
by the weak--field limit \eqref{Hweak} of the JIMWLK Hamiltonian
\cite{Hatta:2005as} --- but this description goes beyond the purpose of a
Gaussian approximation, which by construction can encode only the
2--point function of the $\alpha$ field. For instance, to describe
odderon effects in the initial conditions, one needs an extension of the
MV model allowing for a non--trivial 3--point function
\cite{Jeon:2005cf}.

What is however remarkable about the Gaussian approximation that we
pursue here is its capacity to encode non--trivial correlations among $n$
Wilson lines with arbitrary $n$ in the {\em strong} scattering regime,
where the linear relation between the $n$--point functions and the
2--point function does not hold anymore. This will be discussed in the
next subsection.

\subsection{Strong--scattering regime: the dominance of the `virtual' terms}
\label{sect:RPA}

For relatively low transverse momenta $k_\perp\lesssim Q_s(Y)$, the gluon
occupation numbers in the target wavefunction saturate at a  large value
of order $1/\alpha_s$, meaning that $g \int \rmd x^-
\alpha\sim\order{1}$. This in turn implies that the scattering is strong
for projectiles with transverse sizes $r\gtrsim 1/Q_s$. For instance, the
dipole scattering amplitude $\lan\hat{T}_{\bmx_1\bmx_2}\ran_Y$ becomes of
order one when $|\bmx_1-\bmx_2|\gtrsim 1/Q_s$. Then \eqn{QBFKL} implies
that, for generic configurations at least, the quadrupole scattering
becomes strong when at least one (which necessarily means at least three)
of the six transverse distances $r_{ij}=|\bmx_i-\bmx_j|$ is of order
$1/Q_s$, or larger. Similar considerations apply to the higher--point
correlations. In this regime, the Wilson lines cannot be expanded out
anymore. Rather, they resum multiple scattering to all orders in the
eikonal approximation.

To correctly describe the high--energy evolution in the presence of gluon
saturation and multiple scattering, it is of course essential to keep the
non--linear terms in the Balitsky--JIMWLK equations, so like $\lan
\hat{S} \hat{S}\ran_Y$ in the equation \eqref{BK} for the dipole
$S$--matrix and $\lan \hat{S} \hat{Q}\ran_Y$ in the r.h.s. of \eqn{Qevol}
for the quadrupole. In fact, these are precisely the terms responsible
for the approach towards saturation in the gluon distribution and towards
unitarity in the scattering of the projectile. Accordingly, in the {\em
transition} regime towards saturation/unitarity ({\em i.e.} for
$k_\perp\sim Q_s(Y)$), one has to deal with the whole, infinite,
hierarchy of coupled evolution equations: no simple mean--field
approximation (like a Gaussian) is possible in that regime. However,  the
situation drastically simplifies {\em deeply at saturation} ($k_\perp\ll
Q_s(Y)$), where the only role of the non--linear terms in the equation is
to {\em forbid} further evolution --- or, more correctly, to limit the
transverse phase--space for the high--energy evolution: gluons with soft
momenta $k_\perp\ll Q_s(Y)$ can (almost) not be emitted anymore, meaning
that domains separated by transverse distances $r\gg 1/Q_s(Y)$ evolve
independently from each other. This leads to considerable simplifications
in the Balitsky--JIMWLK equations, which can be most directly recognized
by inspection of the projectile evolution.

For multi--partonic projectiles which are such that all the interparticle
separations $r_{ij}$ are much larger than $1/Q_s(Y)$, the associated
$S$--matrices are very small (close to zero) --- the more so the larger
the number of partons. Roughly speaking, and up to subtleties related to
the $1/N_c^2$ corrections to which we shall later return, a 2--dipole
projectile scatters more strongly than a single--dipole one, $\lan
\hat{S} \hat{S}\ran_Y\ll \lan \hat{S}\ran_Y$, a projectile made with a
dipole plus a quadrupole scatters more strongly than the quadrupole
alone, $\lan \hat{S} \hat{Q} \ran_Y\ll \lan \hat{Q}\ran_Y$, etc. When
this happens, the `virtual' terms dominate the evolution, whereas the
`real' terms can be simply replaced with a lower cutoff $\sim 1/Q_s(Y)$
on the transverse separation $|\bmz-\bmx_i|$ between the newly emitted
gluon at $\bmz$ and any of the preexisting partons at $\bmx_i$. Once this
is done, the resulting evolution equations are {\em linear} and hence
admit a Gaussian solution. This is of course related to our previous
observation in Sect.~\ref{sect:mirror} that the only effect of the
`non--linear terms' (Wilson lines) within $H_{\rm virt}$ is to transform
`left' color precessions into `right' ones and thus ensure the symmetric
expansion of the target field distribution in $x^-$. This also shows
that, in this high density regime, where the Wilson lines cannot be
expanded anymore and `left' and `right' functional derivatives have
different mathematical consequences, it is essential to keep trace of the
`mirror' symmetry of the evolution, by using a {\em symmetric} Gaussian,
as shown in \eqref{Gauss}.

To render these considerations more precise and construct the
corresponding Gaussian approximation, we shall develop our mathematical
arguments in two steps: \texttt{(i)} at large $N_c$, and \texttt{(ii)} at
finite $N_c$.

\bigskip
\noindent \texttt{(i) Large $N_c$ :} Within the context of the
large--$N_c$ approximation, the prominence of the `virtual' terms in the
approach towards the black disk limit is quite obvious and has been
pointed out at several places in the literature
\cite{Levin:1999mw,Iancu:2001md,Iancu:2003zr,Iancu:2011ns}. Specifically,
the `real' terms which survive at large $N_c$ involve double--trace
operators, which can be factorized to the accuracy of interest: $\lan
\hat{S} \hat{S}\ran_Y\simeq \lan \hat{S}\ran_Y \lan \hat{S}\ran_Y$, $\lan
\hat{S} \hat{Q}\ran_Y\simeq \lan \hat{S}\ran_Y \lan \hat{Q}\ran_Y$, etc.
Then we can write e.g.
 \beq\label{separ}
  \lan \hat{S}_{\bmx_1\bmz}\ran_Y \lan\hat{S}_{\bmz\bmx_2}\ran_Y
 \,\ll\,\lan \hat{S}_{\bmx_1\bmx_2}\ran_Y\qquad\mbox{when}\qquad
 |\bmz-\bmx_i|\,\gg\,1/Q_s(Y)\,.\eeq
Now, in equations like \eqref{BK} or \eqref{Qevol}, the transverse
position $\bmz$ of the emitted gluon is integrated over, so it can become
close to one of the external points $\bmx_i$, in which case \eqn{separ}
does not hold anymore. However, in the high density regime under
consideration, such special configurations are disfavoured by the
phase--space for the transverse integration. Namely, assuming
$|\bmx_i-\bmx_j|\gg 1/Q_s(Y)$ for all the pairs $(i,j)$, one can check
that the integrals over $\bmz$ receive their dominant contributions from
points relatively far apart from all the external points, which satisfy
  \beq\label{conds}
 1/Q_s\,\ll\, |\bmz-\bmx_i|\,\ll\, |\bmx_i-\bmx_j|.\eeq
Indeed, the contribution of such a range is enhanced by the large
transverse logarithm
 \beq\label{LLsat}
 \frac{1}{2\pi}\int_{\bmz}
 \mcal{M}_{\bmx_i\bmx_j\bmz}\,\simeq\,\int_{1/Q_s^2}^{|\bmx_i-\bmx_j|^2}
 \frac{\dif z^2}{z^2}\,=\,\ln \left[(\bmx_i-\bmx_j)^2 Q_s^2 \right].
 \eeq
Hence, to leading logarithmic accuracy in the sense of \eqn{LLsat}, one
can indeed neglect the `real' terms in the Balitsky--JIMWLK equations at
large $N_c$, as anticipated.

\bigskip
\noindent \texttt{(ii) Finite $N_c$ :} The physical argument at finite
$N_c$ is the same as at large $N_c$ except that, now, one has to take
into account the fact that the evolution described by the `real' terms
truly corresponds to the emission of a {\em gluon}, and not just to the
splitting of, say, one dipole into two dipoles. When this new gluon is
sufficiently soft, in the sense that $|\bmz-\bmx_i|\gtrsim 1/Q_s(Y)$ for
any $i$, its emission leads to a partonic system with a wider
distribution of color charge in the transverse plane, which therefore
interacts stronger with the target than the original projectile. But in
order to rigorously justify this, one needs to actually estimate the
$S$--matrix for, say, a quark--antiquark--gluon ($q\bar q g$) system
deeply at saturation and show that this is indeed much smaller than the
$S$--matrix of the dipole ($q\bar q$). To appreciate how subtle this is,
let us recall that, when rewriting the `real' terms in terms of Wilson
lines in the {\em fundamental} representation (as customary in the
Balitsky--JIMWLK equations), one generates single--trace pieces
proportional to $1/N_c^2$, which by themselves count on the same footing
as the `virtual' terms near the unitarity limit. For instance, the
contribution of the `real' terms to the dipole equation involves the
following expectation value (cf. the second line in \eqn{real})
 \beq\label{SSreal}
  \Big\lan\hat{S}_{\bmx_1\bmz} \hat{S}_{\bmz\bmx_2}
 -\frac{1}{N_c^2}\, \hat{S}_{\bmx_1\bmx_2}\Big\ran_Y\,,
 \eeq
where one may naively think that the second, single--trace, term
dominates over the first one when all the transverse separations are much
larger than $1/Q_s(Y)$. As another example, we show here some `real'
terms from the evolution equation \eqref{Qevol} for the quadrupole,
namely those arising when acting with $H_{\rm real}$ on the two quarks at
$\bmx_1$ and $\bmx_3$ :
 \begin{align}\label{Qthird}
 \hspace{-1cm}\big\lan H_{\rm real} \hat{Q}_{\underline{\bmx_1}\bmx_2
 \underline{\bmx_3}\bmx_4}
 \big\ran_Y&=
 -\frac{g^2}{8 \pi^3 N_c}
 \int_{\bmz} \mcal{M}_{\bmx_1\bmx_3\bmz}\big\lan\wt{V}_{\bmz}^{ab}
 \big[
 \rmtr({V}^{\dagger}_{\bmx_1} t^a {V}_{\bmx_2}
 t^b {V}^{\dagger}_{\bmx_3} {V}_{\bmx_4})
 +
 \rmtr(t^b {V}^{\dagger}_{\bmx_1} {V}_{\bmx_2}{V}_{\bmx_3}^{\dagger}
 t^a {V}_{\bmx_4})\big]\big\ran_Y,\nn
 &= -\frac{\abar}{4\pi} \int_{\bmz}
 \mcal{M}_{\bmx_1\bmx_3\bmz}
 \Big\lan\hat{S}_{\bmz\bmx_2}\hat{Q}_{\bmx_1\bmz\bmx_3\bmx_4}
 + \hat{S}_{\bmz\bmx_4}\hat{Q}_{\bmx_1\bmx_2\bmx_3\bmz}
 -\frac{2}{N_c^2}\, \hat{Q}_{\bmx_1\bmx_2\bmx_3\bmx_4}\Big\ran_Y\,,
 \end{align}
where the second line follows from the first one after using the Fierz
identity \eqref{Fierz1}. Once again, one may think that the last term in
\eqn{Qthird},  proportional to $(1/N_c^2)\lan \hat{Q}\ran_Y$, is the
dominant term for large transverse separations $\gg 1/Q_s(Y)$ (and for
finite $N_c$). If that was indeed the case, there would be a mixing
between `real' and `virtual' terms deeply at saturation, which would
prevent a Gaussian approximation (since the latter could not accommodate
the `real' terms beyond the BFKL approximation).

The situation becomes even more confusing if one recalls that, in the
equations obeyed by the single--trace observables, the terms subleading
at large $N_c$ precisely cancel between `real' and `virtual'
contributions. In view of this, one may be tempted to argue that the
finite--$N_c$ corrections are totally irrelevant. But that would be
wrong, since there is no similar cancelation in the equations obeyed by
the multi--trace operators, like $\lan \hat{S} \hat{S}\ran_Y$ or $\lan
\hat{S} \hat{Q}\ran_Y$.

What `saves' the Gaussian approximation, is the fact that, in spite of
appearance, the single--trace components in equations like \eqref{SSreal}
or \eqref{Qthird} do {\em not} dominate over the respective double--trace
ones, but merely subtract fake `single--trace contributions' from the
latter, that have been artificially introduced via the Fierz identity.
That is, the expression in the first line of \eqn{real}, which involves
an {\em adjoint} Wilson line and describes a $q\bar q g$ system, vanishes
very fast in the approach towards the black disk limit, where it is
suppressed with respect to the corresponding `virtual' term
$\lan\hat{S}_{\bmx_1\bmx_2}\ran_Y$. But this is {\em not} the case for
the 2--dipole $S$--matrix in the second line of \eqn{real}, which in that
regime approaches to $(1/N_c^2)\lan\hat{S}_{\bmx_1\bmx_2}\ran_Y$. A
similar discussion refers to \eqn{Qthird}: deeply at saturation, the
observable in the first line, which describes a $q\bar q q\bar q g$
partonic system, is suppressed compared to the respective `virtual'
terms, that is, the quadrupole and the pair of dipoles.

In order to demonstrate this while dealing with an infinite hierarchy, we
shall provide a self--consistent argument. That is, we start by assuming
that the JIMWLK evolution deeply at saturation is controlled by $H_{\rm
virt}$ alone and we prove that, under this assumption, the `real' terms
in \eqn{SSreal} and \eqn{Qthird} vanish exponentially faster than the
respective `virtual' terms in the vicinity of the unitarity limit. We
shall give the details of the proof for the dipole evolution, {\em i.e.}
for the operator in \eqn{SSreal}, and then briefly discuss its
generalization to the quadrupole and higher $n$--point functions. In this
context, by `$H_{\rm virt}$' we mean, of course, the first two terms in
the JIMWLK Hamiltonian \eqref{H} {\em together} with the phase--space
restriction $|\bmz-\bmx_i|\gg 1/Q_s(Y)$ as introduced by the `real'
terms. That is, we work in the leading--logarithmic approximation in
Eqs.~\eqref{conds}--\eqref{LLsat}, which enables us to write
  \beq\label{Hsat}
 H_{\rm virt} \simeq
 -\frac{1}{8 \pi^2} \int_{\bmu\bmv}
 \ln \left[(\bmu-\bmv)^2 Q_s^2(Y) \right]
 \left(1 + \wt{V}^{\dagger}_{\bmu} \wt{V}_{\bmv}
 \right)^{ab}
 \frac{\delta}{\delta \alpha_{\bmu}^a}
 \frac{\delta}{\delta \alpha_{\bmv}^b}
 \,,
 \eeq
to the accuracy of interest.

So, let us calculate the action of $H_{\rm virt}$ on the combination of
the operators appearing in \eqn{SSreal}. This action on the second term
has been already computed in \eqn{virt}, that we here rewrite for
convenience as
\beq
- \frac{1}{N_c^2}\,H_{\rm virt}\, \hat{S}_{\bmx_1\bmx_2} =
\frac{\abar}{2\pi N_c^2}\,\left(1 - \frac{1}{N_c^2} \right) \int_{\bmw}
\mcal{M}_{\bmx_1\bmx_2\bmw} \hat{S}_{\bmx_1\bmx_2},
 \eeq
with the integral over $\bmw$ understood in the sense of \eqn{LLsat}.
Now, when both derivatives act on the {\em same} (either the first or the
second) dipole of the first term in \eqn{SSreal}, we get the following,
`diagonal', contribution
\beq
H_{\rm virt}\, \hat{S}_{\bmx_1\bmz}\hat{S}_{\bmz\bmx_2} \big|_{\rm diag}=
-\frac{\abar}{2 \pi}\, \left(1 - \frac{1}{N_c^2}\right) \int_{\bmw}
(\mcal{M}_{\bmx_1\bmz\bmw} + \mcal{M}_{\bmx_2\bmz\bmw})
\hat{S}_{\bmx_1\bmz}\hat{S}_{\bmz\bmx_2},
\eeq
and when they act on different dipoles we find the cross term
\beq
H_{\rm virt}\, \hat{S}_{\bmx_1\bmz}\hat{S}_{\bmz\bmx_2} \big|_{\rm
cross}= -\frac{\abar}{2 \pi N_c^2} \int_{\bmw} (\mcal{M}_{\bmx_1\bmz\bmw}
+ \mcal{M}_{\bmx_2\bmz\bmw} - \mcal{M}_{\bmx_1\bmx_2\bmw})
(\hat{S}_{\bmx_1\bmz}\hat{S}_{\bmz\bmx_2} - \hat{S}_{\bmx_1\bmx_2}).\
\eeq
Putting everything together we arrive at
\begin{align}\label{HSSreal}
H_{\rm virt}&
\left(
\hat{S}_{\bmx_1\bmz}\hat{S}_{\bmz\bmx_2}
- \frac{1}{N_c^2}\,\hat{S}_{\bmx_1\bmx_2}
\right)
 =
 \nn
&-\frac{\abar}{2 \pi}
\int_{\bmw}
\left(\mcal{M}_{\bmx_1\bmz\bmw}
+ \mcal{M}_{\bmx_2\bmz\bmw}
- \frac{1}{N_c^2}\,\mcal{M}_{\bmx_1\bmx_2\bmw}\right)
\left(
\hat{S}_{\bmx_1\bmz}\hat{S}_{\bmz\bmx_2}
- \frac{1}{N_c^2}\,\hat{S}_{\bmx_1\bmx_2}
\right),
\end{align}
where it is crucial to notice that the operator of interest has been
reconstructed in the r.h.s.~of the equation. It should be clear from the
above derivation that this would have not happened without the
subtraction of the $1/N_c^2$--suppressed dipole. By assumption, the above
equation describes the approach towards unitarity of the `real' piece in
the evolution equation \eqref{BK} for the dipole $S$--matrix. This should
be compared to \eqn{virt}, which describes the corresponding approach for
the `virtual' piece (the dipole itself). Clearly, the kernel in
\eqn{HSSreal} is `twice as large' than that in \eqn{virt}, showing that,
deeply at saturation, the expectation value of the `real' operator in
\eqn{SSreal} vanish exponentially faster than the `virtual' term
$\propto\lan \hat{S}_{\bmx_1\bmx_2}\ran_Y$. Hence, the latter dominates
in the evolution equation and in this regime, as anticipated.

This self--consistent argument can be generalized to higher--point
correlators, as we now show  for the operator
 \beq\label{QSreal}
 \hat{S}_{\bmx_1\bmz}\hat{Q}_{\bmz\bmx_2\bmx_3\bmx_4}
 -\frac{1}{N_c^2}\,\hat{Q}_{\bmx_1\bmx_2\bmx_3\bmx_4},
 \eeq
which appears in \eqn{Qthird} and counts for the evolution of the
quadrupole. Acting with $H_{\rm virt}$, we see that the only new element
appearing, when comparing to \eqn{HSSreal}, is operator mixing. Indeed,
one finds that we also need to consider the operators
 \beq\label{SSSreal}
 \hat{S}_{\bmx_1\bmz} \hat{S}_{\bmz\bmx_2} \hat{S}_{\bmx_3\bmx_4} -
 \frac{1}{N_c^2}\, \hat{S}_{\bmx_1\bmx_2} \hat{S}_{\bmx_3\bmx_4},
 \eeq
and
 \beq\label{S6real}
 \hat{S}^{(6)}_{\bmz \bmx_2 \bmx_1 \bmz \bmx_3\bmx_4} -
 \hat{S}_{\bmx_1\bmx_2} \hat{S}_{\bmx_3\bmx_4},
 \eeq
plus permutations of all the operators appearing in Eqs.~\eqref{QSreal},
\eqref{SSSreal} and \eqref{S6real}. Without going into too much detail,
one understands that the action of $H_{\rm virt}$ on the above operators
leads to
 \beq
 \hspace*{-1.0cm}
 H_{\rm virt}\!
 \begin{bmatrix}
 \hat{S}_{\bmx_1\bmz}\hat{Q}_{\bmz\bmx_2\bmx_3\bmx_4}
 -{\displaystyle\frac{1}{N_c^2}}\,\hat{Q}_{\bmx_1\bmx_2\bmx_3\bmx_4}\\*[0.3cm]
 \hat{S}_{\bmx_1\bmz} \hat{S}_{\bmz\bmx_2} \hat{S}_{\bmx_3\bmx_4} -
 {\displaystyle\frac{1}{N_c^2}}\, \hat{S}_{\bmx_1\bmx_2} \hat{S}_{\bmx_3\bmx_4}\\*[0.3cm]
 \hat{S}^{(6)}_{\bmz \bmx_2 \bmx_1 \bmz \bmx_3\bmx_4} -
 \hat{S}_{\bmx_1\bmx_2} \hat{S}_{\bmx_3\bmx_4}\\
 \vdots
 \end{bmatrix}
  =
 \begin{bmatrix}
 \mathbb{M} & \cdots \\
 \vdots & \ddots
 \end{bmatrix}\!
 \begin{bmatrix}
 \hat{S}_{\bmx_1\bmz}\hat{Q}_{\bmz\bmx_2\bmx_3\bmx_4}
 -{\displaystyle\frac{1}{N_c^2}}\,\hat{Q}_{\bmx_1\bmx_2\bmx_3\bmx_4}\\*[0.3cm]
 \hat{S}_{\bmx_1\bmz} \hat{S}_{\bmz\bmx_2} \hat{S}_{\bmx_3\bmx_4} -
 {\displaystyle\frac{1}{N_c^2}}\, \hat{S}_{\bmx_1\bmx_2} \hat{S}_{\bmx_3\bmx_4}\\*[0.3cm]
 \hat{S}^{(6)}_{\bmz \bmx_2 \bmx_1 \bmz \bmx_3\bmx_4} -
 \hat{S}_{\bmx_1\bmx_2} \hat{S}_{\bmx_3\bmx_4}\\
 \vdots
 \end{bmatrix}\!,
 \eeq
where the elements of the $3 \times 3$ matrix $\mathbb{M}$ are
proportional to $-\abar/2 \pi$ times an integral over $\bmw$ of linear
combinations of the dipole kernel. The counting is such that the
integrand in the diagonal elements is the sum of three dipole kernels
which enter all with a plus sign, plus terms proportional to $1/N_c^2$
(in analogy with \eqn{HSSreal}). Furthermore, the integrand in the
non--diagonal elements is the sum of dipole kernels with equal number of
plus and minus signs, plus again terms proportional to $1/N_c^2$.
Clearly, the diagonal components are those which control the approach
towards the black disk limit and they are larger than those which control
the corresponding evolution for the `virtual' terms in \eqn{Qevol}, that
is $\lan \hat{Q}\ran_Y$ and $\lan \hat{S} \hat{S}\ran_Y$.

Incidentally, the above argument also shows that the two operators in
Eqs.~\eqref{SSSreal} and \eqref{S6real} vanish faster than the quadrupole
and the 2--dipole system in the approach towards unitarity. This is
interesting since these are precisely the `real' terms in the evolution
equation for $\lan\hat{S}_{\bmx_1\bmx_2} \hat{S}_{\bmx_3\bmx_4}\ran_Y$,
whereas $\lan \hat{Q}\ran_Y$ and $\lan \hat{S} \hat{S}\ran_Y$ are the
corresponding `virtual' terms. So, we have also demonstrated the property
of interest (the dominance of the `virtual' terms deeply at saturation)
for the evolution of a system of two dipoles with arbitrary coordinates.
We are confident that a similar proof applies to the higher--point
(single--trace or multi--trace) correlation functions.

It is furthermore instructive to check these arguments via explicit
calculations within the Gaussian approximation \eqref{Gauss}. Via methods
to be described later, this yields e.g. \cite{Kovchegov:2008mk}
 \beq\label{S13S32G}
 \hspace*{-0.5cm}\Big\lan \hat{S}_{\bmx_1 \bmx_3} \hat{S}_{\bmx_3 \bmx_2}-
 \frac{1}{N_c^2}\,
 \hat{S}_{\bmx_1 \bmx_2}\Big\ran_Y =
 \frac{N_c^2-1}{N_c^2}
 \left[
 \frac{\lan \hat{S}_{\bmx_1 \bmx_3}\ran_Y
 \lan \hat{S}_{\bmx_3\bmx_2}\ran_Y}
 {\lan \hat{S}_{\bmx_1 \bmx_2}\ran_Y}
 \right]^{\textstyle\frac{1}{(N_c^2-1)}}
 \lan \hat{S}_{\bmx_1 \bmx_3}\ran_Y
 \lan \hat{S}_{\bmx_3\bmx_2}\ran_Y,
 \eeq
where we have assumed that $\lan \hat{S}_{\bmx_i \bmx_j} \ran$ and $\lan
\hat{S}_{\bmx_1 \bmx_3} \hat{S}_{\bmx_3 \bmx_2}\ran$ are equal to 1 as an
initial condition, to simplify writing. This formula makes it clear that
the operator in the l.h.s. vanishes, roughly, as a `dipole squared' in
the approach towards the unitarity limit. A corresponding argument for
the operator \eqref{QSreal} which enters the evolution of the quadrupole
will be given in Sect.~\ref{sect:finiteNc}.

We thus conclude that the JIMWLK evolution deeply at saturation is indeed
correctly described by the `virtual' Hamiltonian in \eqn{Hsat}. When
acting on operators built with Wilson lines, the two terms in $H_{\rm
virt}$ amount to `left' and `right' Lie derivatives, in the sense of
\eqn{JLJR}. So, clearly, the Hamiltonian \eqref{Hsat} is of the
`symmetric Gaussian' form in \eqn{RGMFA}, with the following kernel
 \beq\label{Coulomb}
 \gamma_Y(\bmu,\bmv)\,=\,\frac{1}{4\pi^2}
 \ln \left[(\bmu-\bmv)^2 Q_s^2 \right]\quad\Longrightarrow\quad
 \gamma_Y
 (\bmk)\,=\,\frac{1}{\pi k_\perp^2}
 \,.\eeq
This applies for $k_\perp\ll Q_s(Y)$ and is recognized as the
2--dimensional Coulomb propagator. In turn this implies that the
charge--charge correlator $\lambda_Y (\bmk)= k_\perp^4 \gamma_Y (\bmk)$
vanishes like $k_\perp^{2}$ when $k_\perp\to 0$, which is the expression
of color shielding due to gluon saturation
\cite{Mueller:2002pi,Iancu:2002aq}: the average color charge squared
vanishes when integrated over a transverse area $\gg 1/Q_s^2(Y)$.

Notice that in some previous versions of the mean field approximation
\cite{Iancu:2001md,Iancu:2002xk,Iancu:2002aq}, one has assumed that the
JIMWLK Hamiltonian takes an even simpler form in the vicinity of the
black disk limit, namely it reduces to the first term in \eqn{Hsat},
which involves the `left' derivatives alone. That simplification was
motivated \cite{Iancu:2001md} by a `random phase approximation', which
assumed that, in the strong field regime deeply at saturation, all the
Wilson lines within the Hamiltonian are rapidly oscillating and thus
average out to zero. As shown by our present manipulations, this argument
is qualitatively correct, but only for the `real' terms (the last 2
terms) in the JIMWLK Hamiltonian.

To summarize the arguments in this section, the JIMWLK evolution in the
two limiting regimes --- the weak--scattering regime at low gluon density
and the approach towards the black--disk limit deeply at saturation ---
can be properly encoded, for any value of $N_c$, into a symmetric
Gaussian weight function of the type \eqref{Gauss}. In turn, this
Gaussian is tantamount to the functional evolution equation shown in
\eqn{RGMFA}, or to the following, mean field, Hamiltonian:
 \beq\label{HMFA}
 \hspace*{-0.6cm}
 H_{\rm MFA} = -\frac{1}{2} \int_{\bmu\bmv}
 \gamma_Y(\bmu,\bmv)
 \big(1 + \wt{V}^{\dagger}_{\bmu} \wt{V}_{\bmv}\big)^{ab}
 \frac{\delta}{\delta\alpha_{\bmu}^a}\,
 \frac{\delta}{\delta\alpha_{\bmv}^b}
 = \frac{g^2}{2} \int_{\bmu\bmv}
 \gamma_Y(\bmu,\bmv)
 \big(
 J^a_{L\bmu} J^a_{L\bmv}
 +
 J^a_{R\bmu} J^a_{R\bmv}
 \big)\,.
 \eeq
This has the same operator structure as the `virtual' piece of the JIMWLK
Hamiltonian, but with a different, $Y$--dependent, kernel, which is
essentially the 2--point function of the target color field. This kernel
interpolates between the solution to the BFKL equation \eqref{BFKLf} at
small transverse separations $|\bmu-\bmv|\ll 1/Q_s(Y)$ and the Coulomb
propagator \eqref{Coulomb} at relatively large distances $|\bmu-\bmv|\gg
1/Q_s(Y)$. Remarkably, the kernel is independent of $N_c$, in agreement
with the corresponding property of the `dipole' kernel in the JIMWLK
Hamiltonian. (This can be checked e.g. on \eqn{Coulomb} and on the
relation \eqref{BFKLMFA} between this kernel and the dipole amplitude at
weak coupling.) Any smooth function $\gamma_Y(\bmu,\bmv)$ with the
correct limiting behaviors can in principle be used to define the
Gaussian; indeed, different such functions can differ from each other
only in the transition region around $Q_s$, which is not under control
within the present approximation. In practice, however, a proper choice
for the kernel is probably essential in order to achieve a good global
accuracy. In the following section, we shall propose two such choices.

\section{Evolution equations in the Gaussian approximation}
\setcounter{equation}{0} \label{sect:MFA}

In this section we shall describe two methods for constructing smooth,
global, expressions for the kernel $\gamma_Y(\bmu,\bmv)$, which are
equivalent with each other to the accuracy of interest. Then we shall
derive the evolution equations associated with the mean field Hamiltonian
\eqref{HMFA}, first for generic $N_c$ (in Sect.~\ref{sect:eqsMF}), then
at large $N_c$ (in Sect.~\ref{sect:largeNc}). As before, we shall mostly
focus on the evolution of the dipole and of the quadrupole. In the
large--$N_c$ limit we shall recover the equations previously proposed in
Ref.~\cite{Iancu:2011ns}. At finite $N_c$, the general equations are more
complicated, but explicit solutions will be presented for special
configurations in Sect.~\ref{sect:finiteNc}.

\subsection{Self--consistent constructions of the kernel}
\label{sect:self}

Within the Gaussian approximation, it is always possible to trade the
kernel $\gamma_Y(\bmx_1,\bmx_2)$ for the dipole $S$--matrix  $\lan
\hat{S}_{\bmx_1\bmx_2}\ran_Y$, for which it is easier to construct
global, smooth, approximations in practice. The expression of $\lan
\hat{S}_{\bmx_1\bmx_2}\ran_Y$ in the Gaussian approximation is well known
in the literature and will be rederived, for completeness, in the next
subsection. Here it is preferable to work with the corresponding
evolution equation, which reads
 \beq\label{dipevolR}
 \frac{\del \big\lan \hat{S}^R_{\bmx_1\bmx_2} \big\ran_Y}{\del Y} =
 -2g^2 C_R\, \gamma_Y(\bmx_1,\bmx_2)
 \big\lan \hat{S}^R_{\bmx_1\bmx_2} \big\ran_Y.
 \eeq
for a dipole in an arbitrary representation $R$ of the color group.
Hence, if one disposes of a numerical solution to the JIMWLK equation,
like in Refs.~\cite{Rummukainen:2003ns,Lappi:2011ju,Dumitru:2011vk}, then
one can use the respective estimate for the dipole $S$--matrix, say, in
the fundamental representation, together with \eqn{dipevolR} to deduce a
corresponding estimate for the kernel.

In practice, solving the full JIMWLK evolution is quite tedious, so it is
customary to rely on its large--$N_c$ approximation (for the dipole
evolution), namely the BK equation. This is equally good for the present
purposes (including at finite $N_c$) since, as noticed at the end of the
previous section, the kernel $\gamma_Y(\bmx_1,\bmx_2)$ is independent of
$N_c$. Hence, its limiting behaviors are correctly reproduced by the
large--$N_c$ version of \eqn{dipevolR}. The latter implies
  \beq\label{gammaS}
 {g^2 N_c}\,\gamma_Y(\bmx_1,\bmx_2) =-\,
 \frac{\del \ln\big \lan \hat{S}_{\bmx_1\bmx_2}^{\rm BK}\big\ran_Y}{\del
 Y}\,,
  \eeq
with $\lan \hat{S}_{\bmx_1\bmx_2}^{\rm BK}\ran_Y$ denoting the solution
to the BK equation with an initial condition itself evaluated at large
$N_c$. (This is important in order to preserve finite--$N_c$ accuracy in
the limiting kinematical domains where \eqn{gammaS} is correct as it
stands for any value of $N_c$.) For instance, within the MV model, this
is provided by Eqs.~\eqref{SMV}--\eqref{GammaMV} with $C_F\simeq N_c/2$.
The function $\lan \hat{S}_{\bmx_1\bmx_2}^{\rm BK}\ran_Y$ can be obtained
either as an exact, numerical, solution to the BK equation, or as an
analytic approximation to it with the correct limiting behaviors.

Although correct to the accuracy of interest, the construction in
\eqn{gammaS} may look a bit unesthetic at a conceptual level, as it
requires an input --- the solution of the BK equation --- which seems
external to the Gaussian approximation. However, as we now explain,
\eqn{gammaS} can be also understood as a {\em self--consistency
condition} internal to the mean field approximation
\cite{Kovchegov:2008mk}. Specifically, let us start with the first
equation in the Balitsky--JIMWLK hierarchy, that is \eqn{BK} for the
dipole, and evaluate all the expectation values there with the Gaussian
weight function \eqref{Gauss}, that is, by using \eqn{dipevolR} with
$C_R=C_F$ together with \eqn{S13S32G}. This leads to an equation for the
kernel $\gamma_Y(\bmx_1,\bmx_2)$ which is precisely equivalent to solving
the BK equation and then computing the kernel according to \eqn{gammaS}
(see Ref.~\cite{Kovchegov:2008mk} for details).

This procedure, which in Ref.~\cite{Kovchegov:2008mk} has been dubbed
`the Gaussian truncation', is of course not unique: one can similarly
start with {\em any} equation in the Balitsky--JIMWLK hierarchy, compute
all the expectation values there with  the Gaussian weight function
\eqref{Gauss}, and thereby transform the original equation into an
equation for $\gamma_Y(\bmx_1,\bmx_2)$. A different self--consistency
condition, which in practice is not more difficult to use than
\eqn{gammaS}, has been originally proposed in Ref.~ \cite{Iancu:2002xk}.
It amounts to requiring the mean--field Hamiltonian \eqref{HMFA} to
coincide with the JIMWLK Hamiltonian \eqref{H} {\em on the average} :
 \beq\label{gammab}
 \frac{1}{16 \pi^3} \int_{\bmz}
 \mcal{M}_{\bmu\bmv\bmz}
 \big\lan 1 + \wt{V}^{\dagger}_{\bmu} \wt{V}_{\bmv}
 -\wt{V}^{\dagger}_{\bmu} \wt{V}_{\bmz}
 -\wt{V}^{\dagger}_{\bmz} \wt{V}_{\bmv}\big\ran_Y^{ab} =
 \frac{1}{2}\,\gamma_Y(\bmu,\bmv)\,
 \big\lan 1 + \wt{V}^{\dagger}_{\bmu} \wt{V}_{\bmv}\big\ran_Y^{ab}.
 \eeq
The average here refers, of course, to the Gaussian weight function,
which implies that both the l.h.s. and the r.h.s. in the above equation
are proportional to $\delta^{ab}$. Introducing the $S$--matrix for the
gluonic dipole operator
 \beq\label{Sadj}
 \hat{S}^A_{\bmx_1\bmx_2} \,\equiv\,\frac{1}{N_c^2-1}\,
 \rmTr(\wt{V}^{\dagger}_{\bmx_1} \wt{V}_{\bmx_2})\,=\,
 \frac{1}{N_c^2-1}\,\Big(
 N_c^2\, \hat{S}_{\bmx_1\bmx_2}\hat{S}_{\bmx_2\bmx_1} - 1\Big)\,
 \eeq
which it related to the respective fermionic operator as shown in the
second equality above, and multiplying \eqn{gammab} by $\delta^{ab}$, we
can rewrite the latter as
 \beq\label{gammab2}
 \gamma_Y(\bmu,\bmv) =
 \frac{1}{8 \pi^3 \big\lan 1+ \hat{S}^A_{\bmu\bmv}\big\ran_Y}\,
 \int_{\bmz} \mcal{M}_{\bmu\bmv\bmz}
 \big\lan 1 + \hat{S}^A_{\bmu\bmv}
 -\hat{S}^A_{\bmu\bmz} -\hat{S}^A_{\bmz\bmv}
 \big\ran_Y.
 \eeq
This relation immediately implies $\gamma_Y(\bmu,\bmu) = 0$ because of
the corresponding property of the dipole kernel
$\mcal{M}_{\bmu\bmv\bmz}$. It furthermore implies that
$\gamma_Y(\bmu,\bmv)$ is symmetric under $\bmu \leftrightarrow \bmv$,
because so is the gluonic dipole $S$--matrix, as obvious from its
definition \eqref{Sadj}.

The self--consistency condition \eqref{gammab2} is most conveniently
written as an equation for $\lan\hat{S}^A\big\ran_Y$: by using
\eqn{dipevolR} in the adjoint representation ($C_R=C_A\equiv N_c$), one
finds
 \beq\label{self-consist}
 \frac{\del \big\lan \hat{S}^A_{\bmx_1\bmx_2} \big\ran_Y}{\del Y} =
 \frac{\abar}{\pi}
 \int_{\bmz} \mcal{M}_{\bmx_1\bmx_2 \bmz}\,
 \frac{\big\lan \hat{S}^A_{\bmx_1\bmx_2}\big\ran_Y}
 {\big\lan 1+ \hat{S}^A_{\bmx_1\bmx_2}\big\ran_Y}
 \big\lan \hat{S}^A_{\bmx_1\bmz}
 +\hat{S}^A_{\bmz\bmx_2}
 -\hat{S}^A_{\bmx_1\bmx_2} -1
 \big\ran_Y.
 \eeq
The initial condition should be taken from the MV model applied to a {\em
gluonic} dipole, that is, Eqs.~\eqref{SMV}--\eqref{GammaMV} with $C_F\to
C_A= N_c$.

To summarize, by using either \eqn{gammaS} together with the solution to
the BK equation, or \eqn{dipevolR} with $R=A$ together with the solution
to \eqn{self-consist}, one obtains two global approximations for the
kernel $\gamma_Y(\bmx_1,\bmx_2)$, which can differ from each other only
in the transition region around saturation. In principle, these two
approximations are equivalent with each other to the accuracy of
interest. In practice, any (numerical) difference between them will have
an impact on the calculation of the higher $n$--point functions, to be
described in the remaining part of this section. It is likely that one
can optimize the reliability of this whole scheme by choosing a `good'
approximation for the dipole $S$--matrix used to compute the kernel, like
the exact, numerical, solution to the JIMWLK equation, or to the BK
equation at least.

\subsection{Mean--field equations for the dipole and the quadrupole}
\label{sect:eqsMF}

The evolution equations associated with the mean--field Hamiltonian
\eqref{HMFA} are straightforward to obtain, by using the same techniques
as for the JIMWLK Hamiltonian \eqref{H}. Alternatively, given that
$H_{\rm MFA}$ has the same operator structure as $H_{\rm virt}$, the
mean--field equations can be directly inferred from the corresponding
Balitsky--JIMWLK equations, by keeping only the `virtual' terms in the
latter and replacing everywhere the dipole kernel according to
 \beq\label{replace}
  \frac{1}{8 \pi^3} \int_{\bmz}
 \mcal{M}_{\bmu\bmv\bmz}\,\to\,\gamma_Y(\bmu,\bmv)\,.\eeq
In doing that, one should be careful to restore {\em all} the `virtual'
terms in the Balitsky--JIMWLK equations, including those which may have
canceled against similar `real' contributions and hence are not manifest
in the final equations (cf. the discussion in Sect.~\ref{sect:eqs}).
Clearly, the resulting equations will inherit the relatively simple
structure characteristic of the `virtual' terms. They form closed systems
of equations, which connect only correlation functions with the same
number of Wilson lines (or external points) and are local in the
transverse coordinates, meaning that they do not mix different transverse
configurations. Note also that, although linear, these new equations
respect unitarity by construction: the tame of the BFKL growth by the
non--linear physics of gluon saturation is already encoded in the kernel
of the Gaussian. The fact that the $S$--matrices for the various
projectiles approach the right limit at strong scattering is ensured by
the unitarity of the Wilson lines which appear within the respective
operators.

Consider first the dipole equation. Using the substitution rule
\eqref{replace} and the respective `virtual' term in \eqn{virt}, one
finds
 \beq\label{effevolS}
 \frac{\del \lan \hat{S}_{\bmx_1\bmx_2}\ran_Y}{\del Y} =-
 {2g^2 C_F}\,\gamma_Y(\bmx_1,\bmx_2)
 \lan \hat{S}_{\bmx_1\bmx_2}\ran_Y\,.
 \eeq
This is easily solved to give
 \beq\label{SMFA}
 \lan \hat{S}_{\bmx_1\bmx_2}\ran_Y =
 \rme^{-\Gamma_Y(\bmx_1,\bmx_2)}
 ,\qquad \Gamma_Y(\bmx_1,\bmx_2) =
 {2g^2 C_F}\int_{-\infty}^{Y}
 \dif \rmy \,\gamma_\rmy(\bmx_1,\bmx_2)\,,
 \eeq
where it is understood that $\Gamma_{Y_0}$ is given by the MV model, cf.
Eqs.~\eqref{SMV}--\eqref{GammaMV}. The corresponding expressions for a
color dipole $\big\lan \hat{S}^R_{\bmx_1\bmx_2} \big\ran_Y$ in an
arbitrary representation $R$ are obtained by replacing $C_F\to C_R$ in
the equations above (cf. \eqn{dipevolR}). In particular, in the
`BK--representation' in which the Gaussian kernel is computed according
to \eqn{gammaS}, the dipole $S$--matrix in the Gaussian approximation and
in a generic representation $R$ of the color group is related to the
solution $\lan \hat{S}_{\bmx_1\bmx_2}^{\rm BK}\ran_Y$ to the BK equation
via
 \beq\label{SBK}
\ln\big\lan \hat{S}^R_{\bmx_1\bmx_2} \big\ran_Y\,=\,\frac{2C_R}{N_c}
\,\ln\big\lan \hat{S}^{\rm BK}_{\bmx_1\bmx_2} \big\ran_Y\,.\eeq

From the discussion in Sect.~\ref{sect:gaussian}, one expects \eqn{SMFA}
to have the correct limits at both weak and strong scattering, and it is
instructive to explicitly check that. At weak scattering, one has $\lan
\hat{S}\ran_Y = 1- \lan \hat{T}\ran_Y$ with $\lan \hat{T}\ran_Y\ll 1$, so
in particular one can replace $\lan \hat{S}\ran_Y\approx 1$ in the r.h.s.
of \eqn{effevolS}. Then the latter reduces to \eqn{BFKLMFA}, with
$\gamma_Y(\bmx_1,\bmx_2)$ determined by the solution to the BFKL equation
\eqref{BFKLf}. This is indeed the expected result. At strong scattering,
the kernel takes the form of the Coulomb propagator \eqref{Coulomb}, in
agreement with the limit $|\bmu-\bmv|\gg 1/Q_s(Y)$ of \eqn{replace}
(recall \eqn{LLsat}). So, in this regime, \eqn{effevolS} is identical to
the `virtual' part of the respective Balitsky--JIMWLK equation
\eqref{BK}, which is the part that controls the approach towards the
black disk limit. Let us study this approach in more detail. By using
\eqn{Coulomb} within \eqn{SMFA}, one obtains
 \begin{align}\label{SLT}
 \Gamma_{Y}(\bmx_1,\bmx_2)
 % &\equiv
 %\Gamma_{Y}(\bmx_1,\bmx_2) - \Gamma_{Y_s(r)}(\bmx_1,\bmx_2)
 %\nn
 &\simeq \frac{g^2 C_F}{2 \pi^2} \int_{Y_s(r)}^{Y}\dif \rmy
 \int^{Q_s^2(\rmy)}_{1/r^2}\frac{\rmd k_\perp^2}{k_\perp^2}
  ={\abar}\,\frac{N_c^2-1}{N_c^2}
  \int_{Y_s(r)}^{Y}\dif \rmy\,\ln\big(r^2Q_s^2(\rmy)\big)
  \nn
 &=\frac{\omega\abar^2}{2}\frac{N_c^2-1}{N_c^2}
 \big(Y-Y_s(r)\big)^2
 = \frac{1}{2\omega} \,\frac{N_c^2-1}{N_c^2}
 \ln^2\big(r^2Q_s^2(Y)\big)
 \end{align}
where $Y_s(r)$ is the rapidity at which saturation is reached over a
transverse size $r$ (that is, $Q_s(Y_s(r)) = 1/r$), $\omega \abar$ is the
logarithmic derivative of the saturation momentum, and we used
$\ln\big(r^2Q_s^2(\rmy)\big)=\omega \abar(\rmy-Y_s(r))$ for $\rmy\ge
Y_s(r)$. \eqn{SLT} holds in the leading logarithmic approximation in the
sense of \eqn{LLsat}. The large--$N_c$ version of this result has already
appeared in the literature \cite{Levin:1999mw,Iancu:2003zr}.

We now turn to the quadrupole. The corresponding evolution equation in
the MFA is obtained as
 \begin{align}\label{QevolMFA}
 \hspace*{-1.2cm}\frac{\del \lan
 \hat{Q}_{\bmx_1\bmx_2\bmx_3\bmx_4}\ran_Y}{\del Y}  =
 &-{g^2 C_F}
 [\gamma_Y(\bmx_1,\bmx_2)
 +\gamma_Y(\bmx_3,\bmx_2)
 +\gamma_Y(\bmx_3,\bmx_4)
 +\gamma_Y(\bmx_1,\bmx_4)]
 \lan \hat{Q}_{\bmx_1\bmx_2\bmx_3\bmx_4}\ran_Y
 \nn
 &\hspace*{-2.8cm}-\frac{g^2}{2 N_c}
 [2 \gamma_Y(\bmx_1,\bmx_3)
 +2 \gamma_Y(\bmx_2,\bmx_4)
 -\gamma_Y(\bmx_1,\bmx_2)
 -\gamma_Y(\bmx_3,\bmx_2)
 -\gamma_Y(\bmx_3,\bmx_4)
 -\gamma_Y(\bmx_1,\bmx_4)]
 \lan \hat{Q}_{\bmx_1\bmx_2\bmx_3\bmx_4}\ran_Y
 \nn
 &\hspace*{-2.8cm}-\frac{g^2 N_c}{2}
 [\gamma_Y(\bmx_1,\bmx_2)
 +\gamma_Y(\bmx_3,\bmx_4)
 -\gamma_Y(\bmx_1,\bmx_3)
 -\gamma_Y(\bmx_2,\bmx_4)]
 \lan \hat{S}_{\bmx_1\bmx_2} \hat{S}_{\bmx_3\bmx_4}\ran_Y
 \nn
 &\hspace*{-2.8cm}-\frac{g^2 N_c}{2}
 [\gamma_Y(\bmx_1,\bmx_4)
 +\gamma_Y(\bmx_3,\bmx_2)
 -\gamma_Y(\bmx_1,\bmx_3)
 -\gamma_Y(\bmx_2,\bmx_4)]
 \lan \hat{S}_{\bmx_1\bmx_4} \hat{S}_{\bmx_3\bmx_2}\ran_Y.
 \end{align}
Once again the BFKL limit is easy to check: when all the separations
$|\bmx_i-\bmx_j|$ are much smaller than $1/Q_s(Y)$, one can replace $\lan
\hat{Q} \ran_Y\approx 1$ and $\lan \hat{S}\hat{S}\ran_Y\approx 1$ in the
r.h.s. of the above equation, which then reduces to
 \begin{align}
 \label{QuadBFKL}
 \hspace*{-0.8cm}\frac{\del}{\del Y}
 \lan \hat{Q}_{\bmx_1\bmx_2\bmx_3\bmx_4} \ran_Y &\simeq
 -{2 g^2 C_F}
 \big[
 \gamma_Y(\bmx_2,\bmx_3)
 +\gamma_Y(\bmx_1,\bmx_4) +\gamma_Y(\bmx_1,\bmx_2)
 +\gamma_Y(\bmx_3,\bmx_4)
 \nn
 &\hspace*{2cm}
 -\gamma_Y(\bmx_1,\bmx_3)
 -\gamma_Y(\bmx_2,\bmx_4)
 \big]
 \nn
 &= -\frac{\del }{\del Y}\,
 \lan\hat{T}_{\bmx_2\bmx_3}
 +\hat{T}_{\bmx_1\bmx_4}
 +\hat{T}_{\bmx_1\bmx_2}
 +\hat{T}_{\bmx_3\bmx_4}
 -\hat{T}_{\bmx_1\bmx_3}
 -\hat{T}_{\bmx_2\bmx_4} \ran_Y\,.
 \end{align}
The second line, which follows from the first one after using
\eqn{BFKLMFA}, is the expected result for  $\lan \hat{Q} \ran_Y$ at weak
scattering, cf. \eqref{QBFKL}.

Let us also notice that, within the Gaussian approximation, the dipole
and quadrupole $S$--matrices are invariant under charge conjugation, that
is, under the exchange of the quarks with the antiquarks. More precisely,
from Eqs.~\eqref{effevolS} and \eqref{QevolMFA}, and using the fact that
the kernel $\gamma_Y(\bmx_i,\bmx_j)$ is symmetric, we easily deduce that
 \beq
 \lan \hat{S}_{\bmx_1\bmx_2} \ran_Y =
 \lan \hat{S}_{\bmx_2\bmx_1} \ran_Y
 \qquad \mathrm{and} \qquad
 \lan \hat{Q}_{\bmx_1\bmx_2\bmx_3\bmx_4} \ran_Y =
 \lan \hat{Q}_{\bmx_2\bmx_3\bmx_4\bmx_1} \ran_Y,
 \eeq
so long as the above conditions already hold at $Y_0$ (as is the case
within the MV model). Conversely, $C$--odd scattering amplitudes, like
the odderon, cannot be accounted by the Gaussian approximation, as they
would require non--Gaussian effects already at $Y_0$ (cf. the discussion
at the end of Sect.~\ref{sect:BFKL}).

Returning to the full equation \eqref{QevolMFA}, we notice that this is
consistent with mirror symmetry, as it should. (E.g. the r.h.s. is
symmetric under the exchange $\bmx_1 \leftrightarrow \bmx_3$.) That would
not have been the case\footnote{To see this, let us assume for the sake
of the argument the large--$N_c$ limit where the second term is absent.
Then we find that in the first term only $\gamma_Y(\bmx_3,\bmx_2)$ and
$\gamma_Y(\bmx_1,\bmx_4)$ are present and that the fourth term is absent.
Clearly the aforementioned symmetry of the evolution equation is lost.},
had we considered a Gaussian Hamiltonian with only left derivatives, as
proposed in Refs.~\cite{Iancu:2002xk,Iancu:2002aq,Kovchegov:2008mk}. In
fact, without any further assumption, this equation implies that the
evolution for the antisymmetric part of the quadrupole defined in
\eqn{Qasymm} is closed and homogeneous. In turn, this means that $\lan
\hat{Q}^{\rm asym} \ran_Y=0$ in the MFA provided this condition was
satisfied at $Y_0$. It is furthermore clear that the quadrupole couples
to the product of two dipoles, whose evolution in the MFA is in turn
determined by
 \begin{align}\label{SSevolMFA}
 \hspace*{-1cm}\frac{\del \lan \hat{S}_{\bmx_1\bmx_2}
 \hat{S}_{\bmx_3\bmx_4}\ran_Y}{\del Y} =
 &-2g^2 C_F
 [\gamma_Y(\bmx_1,\bmx_2)
 +\gamma_Y(\bmx_3,\bmx_4)]
 \lan \hat{S}_{\bmx_1\bmx_2}\hat{S}_{\bmx_3\bmx_4}\ran_Y
 \nn
 &\hspace*{-2.8cm}-\frac{g^2}{ N_c}
 [\gamma_Y(\bmx_1,\bmx_3)
 +\gamma_Y(\bmx_2,\bmx_4)
 -\gamma_Y(\bmx_1,\bmx_4)
 -\gamma_Y(\bmx_3,\bmx_2)]
 \lan \hat{S}_{\bmx_1\bmx_2}\hat{S}_{\bmx_3\bmx_4}\ran_Y
 \nn
 &\hspace*{-2.8cm}-\frac{g^2}{2 N_c}
 [\gamma_Y(\bmx_1,\bmx_4)
 +\gamma_Y(\bmx_3,\bmx_2)
 -\gamma_Y(\bmx_1,\bmx_3)
 -\gamma_Y(\bmx_2,\bmx_4)]
 \lan \hat{Q}_{\bmx_1\bmx_2\bmx_3\bmx_4}
 + \hat{Q}_{\bmx_1\bmx_4\bmx_3\bmx_2}\ran_Y.
 \end{align}
Since the equations above involve $\lan \hat{S}_{\bmx_1\bmx_4}
\hat{S}_{\bmx_3\bmx_2} \ran_Y$ and $\lan
\hat{Q}_{\bmx_1\bmx_4\bmx_3\bmx_2} \ran_Y$ we need also to consider them
with the their indices $\bmx_2$ and $\bmx_4$ interchanged. (Actually,
$\lan \hat{Q}_{\bmx_1\bmx_4\bmx_3\bmx_2} \ran_Y$ coincides with $\lan
\hat{Q}_{\bmx_1\bmx_2\bmx_3\bmx_4}$ if one assumes mirror symmetry at
$Y_0$, but here we prefer to keep the discussion general.) Thus we arrive
at a homogeneous system of first order differential equations
 \beq\label{mateq}
 \frac{\del}{\del Y}
 \begin{bmatrix}
 \lan \hat{Q}_{\bmx_1\bmx_2\bmx_3\bmx_4} \ran_Y \\
 \lan \hat{Q}_{\bmx_1\bmx_4\bmx_3\bmx_2} \ran_Y \\
 \lan \hat{S}_{\bmx_1\bmx_2} \hat{S}_{\bmx_3\bmx_4} \ran_Y \\
 \lan \hat{S}_{\bmx_1\bmx_4} \hat{S}_{\bmx_3\bmx_2} \ran_Y
 \end{bmatrix}
 = \begin{bmatrix}
 \mathbb{M}_Y(\bmx_i)
 \end{bmatrix}
 \begin{bmatrix}
 \lan \hat{Q}_{\bmx_1\bmx_2\bmx_3\bmx_4} \ran_Y \\
 \lan \hat{Q}_{\bmx_1\bmx_4\bmx_3\bmx_2} \ran_Y \\
 \lan \hat{S}_{\bmx_1\bmx_2} \hat{S}_{\bmx_3\bmx_4} \ran_Y \\
 \lan \hat{S}_{\bmx_1\bmx_4} \hat{S}_{\bmx_3\bmx_2} \ran_Y
 \end{bmatrix}
\eeq
with $\mathbb{M}_Y$ a $4 \times 4$ matrix. Its elements are proportional
to $g^2\gamma_Y(\bmx_i,\bmx_j)$ accompanied by color factors and can be
read from Eqs.~(\ref{QevolMFA}) and (\ref{SSevolMFA}). The difficulty
that appears now is that one cannot solve \eqn{mateq} for a generic
dependence of $\gamma_Y(\bmx_i,\bmx_j)$ on $Y$, since in general the
matrices $\mathbb{M}_Y$ at different rapidities $Y$ do not commute with
each other, that is $[\mathbb{M}_{Y_1},\mathbb{M}_{Y_2}] \neq 0$. (More
precisely, one could write down a formal solution which involves the
rapidity--ordered exponential of the mixing matrix, but we do not find
that very useful in practice.)

There are special cases where the rapidity integration in the equation
above can be explicitly performed, leading to a simpler expression. The
large--$N_c$ limit to be discussed in the next subsection is one such a
special case. Another one is when the kernel $\gamma_Y(\bmx_i,\bmx_j)$,
is a {\em separable} function of $Y$ and the transverse coordinates, plus
an arbitrary function of $Y$~:
 \beq\label{separable}
 \gamma_Y(\bmx_i,\bmx_j)\,=\,h_1(Y)\,g(\bmx_i,\bmx_j)+h_2(Y)\,.
 \eeq
This property is manifestly satisfied within the (homogeneous) MV model,
as noticed after \eqn{intlam}, and it is also approximately satisfied by
the solution to the BK equation, at least in particular kinematical
regimes: in the window for extended geometric scaling, where
$\gamma_Y(r)\propto (r^2 Q_s^2(Y))^{\gamma_s}$ with $\gamma_s\approx
0.63$ \cite{Iancu:2002tr,Mueller:2002zm,Munier:2003vc}, and also deeply
at saturation where $\gamma_Y(r)\propto  \ln[r^2 Q_s^2(Y)]$, cf.
\eqn{SLT}. Presumably, this is a reasonable approximation for all the
dipole sizes. Furthermore, this is also fulfilled in some widely used
dipole models, like the GBW model
\cite{GolecBiernat:1998js,GolecBiernat:1999qd}, where $\Gamma_Y(r)=r^2
Q_s^2(Y)/4$. The role of separability in simplifying the results of the
high--energy evolution in a similar context has been previously
recognized in Ref.~\cite{Marquet:2010cf}.

Within such a simplified scenario, the $Y$--dependence factorizes out
from the mixing matrix and the resulting, $Y$--independent, matrix
$\mathbb{M}$ can be diagonalized. Then one can explicitly solve the
system of equations. Within the context of the MV model this was done in
Refs.~\cite{Blaizot:2004wv,Dominguez:2011wm,Marquet:2010cf}. In that
case, one had to deal with a $2 \times 2$ system only, because the Wilson
lines were constructed via `left' iterations alone, cf.
Eqs.~\eqref{Vevolleft}--\eqref{RGMV2}. That is, the associated,
effective, evolution equations were not explicitly `mirror--symmetric',
unlike the above equations (\ref{QevolMFA}) and (\ref{SSevolMFA}).
However, this symmetry is recovered in the final results, because the
color charge distribution in the MV model in symmetric in $x^-$ (cf. the
discussion after \eqn{RGMV2}). Moreover, for a separable kernel, the
final results for all the $n$--point functions depend only upon the
integral $\int^Y \rmd\rmy\, \gamma_\rmy$ (a property which looks natural
in the case of the dipole, cf. \eqn{SMFA}, but which in general requires
separability). Hence, the results obtained in
\cite{Blaizot:2004wv,Dominguez:2011wm,Marquet:2010cf} within the MV model
can be transposed to a more general Gaussian which is still `separable',
by simply replacing the kernel in the final formul\ae{} according to
\eqn{separable}. We shall not write here the respective general results,
but refer to \cite{Dominguez:2011wm} for $\lan
\hat{Q}_{\bmx_1\bmx_2\bmx_3\bmx_4} \ran_Y$ and to \cite{Marquet:2010cf}
for $\lan \hat{S}_{\bmx_1\bmx_2} \hat{S}_{\bmx_3\bmx_4} \ran_Y$.

But separability is not always needed in order to obtain explicit
solutions at finite $N_c$: for special configurations of the 4 external
points $\bmx_i$ in the transverse plane, the matrix $\mathbb{M}_Y$ in
\eqn{mateq} may happen to simplify independently of the structure of the
kernel $\gamma_Y(\bmx_i,\bmx_j)$. As an example consider the evolution of
the 2--dipole $S$--matrix $\lan
\hat{S}_{\bmx_1\bmx_3}\hat{S}_{\bmx_3\bmx_2}\ran_Y$, in which the quark
in one dipole and the antiquark in the other dipole are located at the
same point $\bmx_3$. The expression of this correlator in the Gaussian
approximation has been shown in \eqn{S13S32G} and we would like to check
that here. By identifying $\bmx_2$ with $\bmx_3$ in \eqn{SSevolMFA} and
then relabeling $\bmx_4$ as $\bmx_2$ for convenience, we see that the two
quadrupole operators in the r.h.s. there reduce to single dipoles and
then this equation decouples from the evolution of the quadrupole :
\begin{align}
 \frac{\del \lan \hat{S}_{\bmx_1 \bmx_3} \hat{S}_{\bmx_3 \bmx_2}\ran_Y}{\del Y}
  =
 - \Big\{&2g^2 C_F \left[\gamma_Y(\bmx_1,\bmx_3)
 + \gamma_Y(\bmx_3,\bmx_2) \right]
 \nn
 + &\frac{g^2}{ N_c}
 \left[\gamma_Y(\bmx_1,\bmx_3)
 + \gamma_Y(\bmx_3,\bmx_2) - \gamma_Y(\bmx_1,\bmx_2) \right]
 \Big\}
 \lan \hat{S}_{\bmx_1 \bmx_3} \hat{S}_{\bmx_3 \bmx_2}\ran_Y
 \nn
  +
 &\frac{g^2}{N_c}
 \left[\gamma_Y(\bmx_1,\bmx_3)
 + \gamma_Y(\bmx_3,\bmx_2) - \gamma_Y(\bmx_1,\bmx_2) \right]
  \lan \hat{S}_{\bmx_1 \bmx_2}\ran_Y.
 \end{align}
The last, inhomogeneous, term in the r.h.s. is particularly interesting,
as it describes a process in which the two dipoles $\hat{S}_{\bmx_1
\bmx_3}$  and $\hat{S}_{\bmx_3 \bmx_2}$ having one common leg merge with
each other into a single dipole $\hat{S}_{\bmx_1 \bmx_2}$. This process
is suppressed at large $N_c$ (as expected, since dipoles cannot merge
with each other in the large $N_c$ limit \cite{Iancu:2004es}) but for
finite $N_c$ it controls the approach towards the unitarity limit, since
a single dipole scatter less than a system of two dipoles.

Using \eqn{effevolS} to express $\gamma_Y(\bmx_i,\bmx_j)$ in terms of the
logarithmic derivative of $\lan \hat{S}_{\bmx_i\bmx_j}\ran_Y$ we can
rewrite the above equation as
 \begin{align}
 \frac{\del \lan \hat{S}_{\bmx_1 \bmx_3}
 \hat{S}_{\bmx_3 \bmx_2}\ran_Y}{\del Y} =
 &\left[
 \frac{\del}{\del Y}\,
 \ln \frac{\lan \hat{S}_{\bmx_1 \bmx_3}\ran_Y^{1+\varepsilon}
 \lan \hat{S}_{\bmx_3\bmx_2}\ran_Y^{ 1+\varepsilon}}
 {\lan \hat{S}_{\bmx_1 \bmx_2}\ran_Y^{\varepsilon}}
 \right]
 \lan \hat{S}_{\bmx_1 \bmx_3} \hat{S}_{\bmx_3 \bmx_2}\ran_Y
 \nn
 - \varepsilon
 &\left[
 \frac{\del}{\del Y}\,
 \ln \frac{\lan \hat{S}_{\bmx_1 \bmx_3}\ran_Y
 \lan \hat{S}_{\bmx_3\bmx_2}\ran_Y}
 {\lan \hat{S}_{\bmx_1 \bmx_2}\ran_Y}
 \right]
 \lan \hat{S}_{\bmx_1 \bmx_2}\ran_Y,
 \end{align}
where we temporarily defined $\varepsilon = 1/(N_c^2-1)$. This is a first
order inhomogeneous linear differential equation which can be readily
solved, with the result shown in \eqn{S13S32G}. Some other special
configurations, which in particular involve the quadrupole, will be
studied in Sect.~\ref{sect:finiteNc}.

The generalization of the above considerations to an arbitrary $n$--point
function like \eqn{S2n} is straightforward. For instance, the mean--field
version of the equation obeyed by the sextupole $S$--matrix
$\lan\hat{S}^{(6)}\ran_Y$ can be inferred from the results in Appendix B
of Ref.~\cite{Iancu:2011ns}.

\subsection{Quadrupole evolution at large $N_c$} %\setcounter{equation}{0}
\label{sect:largeNc}

In the large--$N_c$ limit, the hierarchy generated by the Gaussian
approximation drastically simplifies: it reduces to a triangular
hierarchy, in which the equations can be successively decoupled from each
other and explicitly solved. Let us illustrate that on the example of the
4--point functions --- the quadrupole $\lan
\hat{Q}_{\bmx_1\bmx_2\bmx_3\bmx_4} \ran_Y $ and the 2--dipole system
$\lan \hat{S}_{\bmx_1\bmx_2} \hat{S}_{\bmx_3\bmx_4} \ran_Y$ --- which in
general mix under evolution, as shown in \eqn{mateq}. At large $N_c$, one
can ignore the last two lines in the r.h.s. of \eqn{SSevolMFA}, meaning
that the 2--dipole system evolves independently of the quadrupole. The
corresponding entries in the mixing matrix in \eqn{mateq} are now equal
to zero, so that this matrix becomes triangular, as anticipated.
Specifically, \eqn{SSevolMFA} reduces to
 \beq
 \frac{\del \lan \hat{S}_{\bmx_1\bmx_2}
 \hat{S}_{\bmx_3\bmx_4}\ran_Y}{\del Y} =
 -g^2 N_c
 [\gamma_Y(\bmx_1,\bmx_2)
 +\gamma_Y(\bmx_3,\bmx_4)]
 \lan \hat{S}_{\bmx_1\bmx_2}\hat{S}_{\bmx_3\bmx_4}\ran_Y,
 \eeq
which is immediately solved as
  \beq\label{SSgaussian}
 \lan \hat{S}_{\bmx_1\bmx_2} \hat{S}_{\bmx_3\bmx_4} \ran_Y =
 \rme^{-\Gamma_{Y,Y_0}(\bmx_1,\bmx_2)-\Gamma_{Y,Y_0}(\bmx_3,\bmx_4)}
 \lan \hat{S}_{\bmx_1\bmx_2} \hat{S}_{\bmx_3\bmx_4} \ran_{Y_0},
 \eeq
and where we have defined
 \beq
 \Gamma_{Y,Y_0}(\bmx_i,\bmx_j)\equiv
  \Gamma_{Y}(\bmx_i,\bmx_j)-\Gamma_{Y_0}(\bmx_i,\bmx_j).
 \eeq
As expected, this implies that the 2--dipole $S$--matrix factorizes at
large $N_c$ provided it did so in the initial conditions at $Y_0$:
 \beq\label{SSfact}
 \lan \hat{S}_{\bmx_1\bmx_2} \hat{S}_{\bmx_3\bmx_4} \ran_Y =
 \lan \hat{S}_{\bmx_1\bmx_2} \ran_Y \lan \hat{S}_{\bmx_3\bmx_4} \ran_Y.
 \eeq

Consider now the evolution of the quadrupole: by keeping in
\eqref{QevolMFA} only the leading terms at large $N_c$ (which in
particular means using the factorization property \eqref{SSfact}), one
finds
\begin{align}\label{QuadMFAsymm}
 \hspace*{-1.1cm}\frac{\del}{\del Y}
 \lan \hat{Q}_{\bmx_1\bmx_2\bmx_3\bmx_4} \ran_Y=
 &-\frac{g^2 N_c}{2}
 \left[
 \gamma_Y(\bmx_1,\bmx_2)
 +\gamma_Y(\bmx_3,\bmx_2)
 +\gamma_Y(\bmx_3,\bmx_4)
 +\gamma_Y(\bmx_1,\bmx_4)
 \right]
 \lan \hat{Q}_{\bmx_1\bmx_2\bmx_3\bmx_4} \ran_Y
 \nn
 & -\frac{g^2 N_c}{2}
 \left[
 \gamma_Y(\bmx_1,\bmx_2)
 +\gamma_Y(\bmx_3,\bmx_4)
 -\gamma_Y(\bmx_1,\bmx_3)
 -\gamma_Y(\bmx_2,\bmx_4)
 \right]
 \lan \hat{S}_{\bmx_1\bmx_2}\ran_Y
 \lan\hat{S}_{\bmx_3\bmx_4} \ran_Y
 \nn
 & -\frac{g^2 N_c}{2}
 \left[
 \gamma_Y(\bmx_3,\bmx_2)
 +\gamma_Y(\bmx_1,\bmx_4)
 -\gamma_Y(\bmx_1,\bmx_3)
 -\gamma_Y(\bmx_2,\bmx_4)
 \right]
 \lan \hat{S}_{\bmx_3\bmx_2}\ran_Y
 \lan\hat{S}_{\bmx_1\bmx_4} \ran_Y\,,\nn
 \end{align}
which is an ordinary, first order, inhomogeneous differential equation.
The dipole $S$--matrix is given by \eqref{SMFA} with $C_F\simeq N_c/2$
(as appropriate at large $N_c$) and acts as a source for the evolution of
the quadrupole. As explained in Sect.~\ref{sect:self}, in practice it is
preferable to view \eqn{SMFA} as a {\em definition} of the Gaussian
kernel in terms of the dipole $S$--matrix, since the latter is a more
directly relevant physical quantity. By using \eqn{gammaS} to express
$\gamma_Y(\bmx_1,\bmx_2)$ in terms of the logarithmic derivative of $\lan
\hat{S}_{\bmx_1\bmx_2}\ran_Y$ we can rewrite \eqn{QuadMFAsymm} as
 \begin{align}\label{QevolS}
 \frac{\del \lan\hat{Q}_{\bmx_1\bmx_2\bmx_3\bmx_4} \ran_Y}{\del Y} & =
 \frac{1}{2}
 \left[\frac{\del}{\del Y}
 \ln \lan \hat{S}_{\bmx_1\bmx_2} \ran_Y
 \lan \hat{S}_{\bmx_3\bmx_2} \ran_Y
 \lan \hat{S}_{\bmx_3\bmx_4} \ran_Y
 \lan \hat{S}_{\bmx_1\bmx_4} \ran_Y
 \right]
 \lan\hat{Q}_{\bmx_1\bmx_2\bmx_3\bmx_4} \ran_Y
 \nn
 & +
 \frac{1}{2}
 \left[\frac{\del}{\del Y}
 \ln \frac{\lan \hat{S}_{\bmx_1\bmx_2} \ran_Y
 \lan \hat{S}_{\bmx_3\bmx_4} \ran_Y}
 {\lan \hat{S}_{\bmx_1\bmx_3} \ran_Y
 \lan \hat{S}_{\bmx_2\bmx_4} \ran_Y}
 \right]
 \lan \hat{S}_{\bmx_1\bmx_2} \ran_Y
 \lan \hat{S}_{\bmx_3\bmx_4} \ran_Y
 \nn
 & +
 \frac{1}{2}
 \left[\frac{\del}{\del Y}
 \ln \frac{\lan \hat{S}_{\bmx_1\bmx_4} \ran_Y
 \lan \hat{S}_{\bmx_3\bmx_2} \ran_Y}
 {\lan \hat{S}_{\bmx_1\bmx_3} \ran_Y
 \lan \hat{S}_{\bmx_2\bmx_4} \ran_Y}
 \right]
 \lan \hat{S}_{\bmx_1\bmx_4} \ran_Y
 \lan \hat{S}_{\bmx_3\bmx_2} \ran_Y.
 \end{align}
The general solution of this equation is easily found as
\cite{Iancu:2011ns}
 \begin{align}\label{QsolS}
 \hspace*{-1.5cm}\lan\hat{Q}_{\bmx_1\bmx_2\bmx_3\bmx_4} \ran_Y &=
 \sqrt{\lan \hat{S}_{\bmx_1\bmx_2} \ran_Y
 \lan \hat{S}_{\bmx_3\bmx_2} \ran_Y
 \lan \hat{S}_{\bmx_3\bmx_4} \ran_Y
 \lan \hat{S}_{\bmx_1\bmx_4} \ran_Y
 }
 \Bigg[
 \frac{\lan\hat{Q}_{\bmx_1\bmx_2\bmx_3\bmx_4} \ran_{Y_0}}
 {\sqrt{\lan \hat{S}_{\bmx_1\bmx_2} \ran_{Y_0}
 \lan \hat{S}_{\bmx_3\bmx_2} \ran_{Y_0}
 \lan \hat{S}_{\bmx_3\bmx_4} \ran_{Y_0}
 \lan \hat{S}_{\bmx_1\bmx_4} \ran_{Y_0}
 }}
 \nn
 &+\frac{1}{2}\,\int_{Y_0}^{Y} \dif y\,
 \frac{\lan \hat{S}_{\bmx_1\bmx_3} \ran_y
 \lan \hat{S}_{\bmx_2\bmx_4} \ran_y}
 {\sqrt{\lan \hat{S}_{\bmx_1\bmx_2} \ran_y
 \lan \hat{S}_{\bmx_3\bmx_2} \ran_y
 \lan \hat{S}_{\bmx_3\bmx_4} \ran_y
 \lan \hat{S}_{\bmx_1\bmx_4} \ran_y
 }}\,
 \frac{\del}{\del y}
 \frac{\lan \hat{S}_{\bmx_1\bmx_2} \ran_y
 \lan \hat{S}_{\bmx_3\bmx_4} \ran_y+
 \lan \hat{S}_{\bmx_1\bmx_4} \ran_y
 \lan \hat{S}_{\bmx_3\bmx_2} \ran_y}{\lan \hat{S}_{\bmx_1\bmx_3} \ran_y
 \lan \hat{S}_{\bmx_2\bmx_4} \ran_y}\Bigg].
 \end{align}
This solution is already explicit, but it takes an even simpler form if
one assumes separability, in the sense of the discussion after
\eqn{mateq}. In that case, the integral over $\rmy$ in \eqn{QsolS} can be
exactly performed, to yield
 \begin{align}\label{QL}
 \hspace*{-1.0cm}\lan\hat{Q}_{\bmx_1\bmx_2\bmx_3\bmx_4} \ran_Y & =
 \frac{L_{\bmx_1\bmx_2\bmx_3\bmx_4}}
 {L_{\bmx_1\bmx_2\bmx_4\bmx_3}}\,
 \lan \hat{S}_{\bmx_1\bmx_2} \ran_Y
 \lan \hat{S}_{\bmx_3\bmx_4} \ran_Y
 +\frac{L_{\bmx_1\bmx_4\bmx_3\bmx_2}}{L_{\bmx_1\bmx_4\bmx_2\bmx_3}}\,
 \lan \hat{S}_{\bmx_3\bmx_2} \ran_Y
 \lan \hat{S}_{\bmx_1\bmx_4} \ran_Y
 \nn
 & + \frac{\sqrt{\lan \hat{S}_{\bmx_1\bmx_2} \ran_Y
 \lan \hat{S}_{\bmx_3\bmx_2} \ran_Y
 \lan \hat{S}_{\bmx_3\bmx_4} \ran_Y
 \lan \hat{S}_{\bmx_1\bmx_4} \ran_Y}}
 {\sqrt{\lan \hat{S}_{\bmx_1\bmx_2} \ran_{Y_0}
 \lan \hat{S}_{\bmx_3\bmx_2} \ran_{Y_0}
 \lan \hat{S}_{\bmx_3\bmx_4} \ran_{Y_0}
 \lan \hat{S}_{\bmx_1\bmx_4} \ran_{Y_0}}}
 \nn
 & \times \left[\lan\hat{Q}_{\bmx_1\bmx_2\bmx_3\bmx_4} \ran_{Y_0}
 -\frac{L_{\bmx_1\bmx_2\bmx_3\bmx_4}}
 {L_{\bmx_1\bmx_2\bmx_4\bmx_3}}\,
 \lan \hat{S}_{\bmx_1\bmx_2} \ran_{Y_0}
 \lan \hat{S}_{\bmx_3\bmx_4} \ran_{Y_0}
  -\frac{L_{\bmx_1\bmx_4\bmx_3\bmx_2}}{L_{\bmx_1\bmx_4\bmx_2\bmx_3}}\,
 \lan \hat{S}_{\bmx_3\bmx_2} \ran_{Y_0}
 \lan \hat{S}_{\bmx_1\bmx_4} \ran_{Y_0}
 \right],
 \end{align}
where we have denoted
  \beq\label{Ldef}
  L_{\bmx_1\bmx_2\bmx_3\bmx_4}=
  \Gamma_{Y,Y_0}(\bmx_1,\bmx_2)
  +\Gamma_{Y,Y_0}(\bmx_3,\bmx_4)
  -\Gamma_{Y,Y_0}(\bmx_1,\bmx_3)
  -\Gamma_{Y,Y_0}(\bmx_2,\bmx_4).
  \eeq
Notice that the function $L$ also depends upon $Y$ and $Y_0$, but because
of separability the ratio between two $L$'s is a function of the
transverse coordinates alone. \eqn{QL} depends upon the kernel
$\gamma_\rmy(\bmx_i,\bmx_j)$ only via its integral over $\rmy$. This is a
consequence of separability, as already noticed in Sect.~\ref{sect:MV} in
the context of the MV model. As a matter of facts, \eqn{QL} is quite
similar to the corresponding expression in the MV model
\cite{JalilianMarian:2004da,Dominguez:2011wm} and it becomes formally
identical to it once we assume an initial condition of the MV type.
Specifically, \eqn{QL} reduces to
 \beq\label{QLMV}
 \lan\hat{Q}_{\bmx_1\bmx_2\bmx_3\bmx_4} \ran_Y =
 \frac{L_{\bmx_1\bmx_2\bmx_3\bmx_4}}
 {L_{\bmx_1\bmx_2\bmx_4\bmx_3}}\,
 \lan \hat{S}_{\bmx_1\bmx_2} \ran_Y
 \lan \hat{S}_{\bmx_3\bmx_4} \ran_Y
 +\frac{L_{\bmx_1\bmx_4\bmx_3\bmx_2}}{L_{\bmx_1\bmx_4\bmx_2\bmx_3}}\,
 \lan \hat{S}_{\bmx_3\bmx_2} \ran_Y
 \lan \hat{S}_{\bmx_1\bmx_4} \ran_Y,
 \eeq
provided this functional relation is already satisfied at $Y_0$, as is
indeed the case in the MV model and for large $N_c$
\cite{JalilianMarian:2004da,Dominguez:2011wm}. Note that there is an
alternative way to deduce \eqn{QLMV} from \eqn{QL}, which makes no
reference to the MV model. Namely, if one assumes \eqn{QL} to capture the
whole evolution from $Y\to -\infty$ (where the Wilson lines reduce to the
unit matrix) up to the rapidity $Y$ of interest, then one can use $\lan
\hat{Q} \ran_{Y_0}\to 1$ and $\lan \hat{S} \ran_{Y_0}\to 1$ for $Y_0\to
-\infty$ to check that the expression within the square brackets in
\eqn{QL} vanishes for that particular initial condition.

In general, {\em i.e.} without assuming separability, one expects the
ratio of two $L$'s to depend very weakly on $Y$. If so, it might be still
a good approximation to use the simpler formula \eqref{QLMV} for the
quadrupole rather than the general one in \eqn{QsolS}, which is more
involved. For that purpose, the function $L$ in \eqn{QLMV} should be
defined by \eqn{Ldef} with $Y_0\to -\infty$, and hence
 \beq
 \Gamma_{Y,Y_0}(\bmx_1,\bmx_2)\,\to\,\Gamma_{Y}(\bmx_1,\bmx_2)
 \,=\,-\ln \lan \hat{S}_{\bmx_1\bmx_2}\ran_Y\,.\eeq
Given a smooth approximation for  $\lan \hat{S}\ran_Y$, such as the
numerical solution to the BK equation, \eqn{QsolS} (or \eqref{QLMV})
provides a correspondingly smooth approximation for $\lan\hat{Q}\ran_Y$,
which is guaranteed to be correct whenever all the transverse separations
$r_{ij}\equiv |\bmx_i-\bmx_j|$ are either much smaller, or much larger,
than $1/Q_s(Y)$, and for large $N_c$. On the other hand, the present
approximations are strictly speaking not under control in the transition
region around saturation ($r_{ij}\sim 1/Q_s(Y)$), nor for very asymmetric
configurations, such that some distances $r_{ij}$ are much larger than
$1/Q_s$ while the other ones are much smaller. Some very asymmetric but
relatively simple configurations will be discussed in the next
subsection, directly for finite $N_c$.

\subsection{Special configurations at finite $N_c$} %\setcounter{equation}{0}
\label{sect:finiteNc}

In this subsection, we shall study some special configurations of the
4--point function and the 6--point function in the transverse plane,
which because of their degree of symmetry allow for explicit, and
relatively simple, solutions without additional assumptions like
separability or large--$N_c$.

%This includes some of the configurations considered in the numerical
%analysis in Ref.~\cite{Dumitru:2011vk}.

\begin{figure}
\begin{minipage}[b]{0.245\textwidth}
\begin{center}
\includegraphics[scale=0.75]{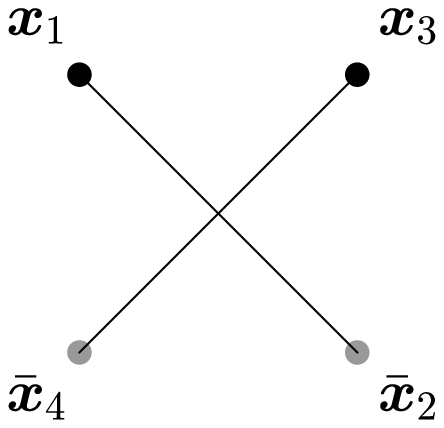}\\(a)
\end{center}
\end{minipage}
\begin{minipage}[b]{0.245\textwidth}
\begin{center}
\includegraphics[scale=0.75]{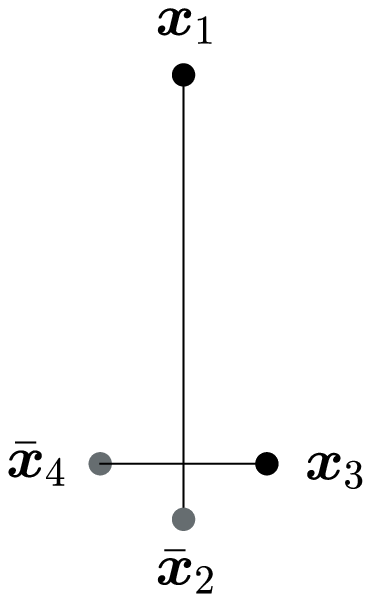}\\(b)
\end{center}
\end{minipage}
\begin{minipage}[b]{0.245\textwidth}
\begin{center}
\includegraphics[scale=0.75]{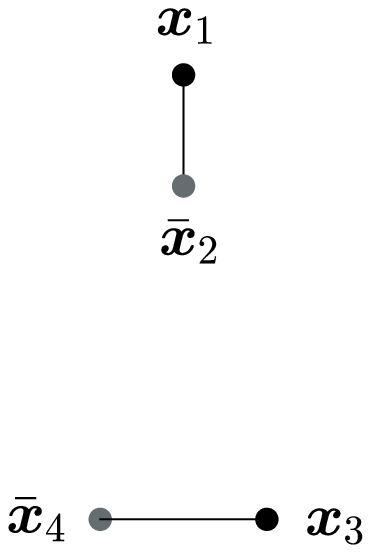}\\(c)
\end{center}
\end{minipage}
\begin{minipage}[b]{0.245\textwidth}
\begin{center}
\includegraphics[scale=0.75]{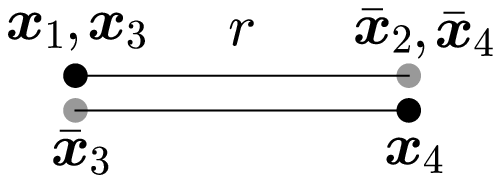}\\(d)
\end{center}
\end{minipage}
\caption{\sl (a), (b) and (c) correspond to special configurations of the quadrupole for which $r_{13}=r_{14}$ and $r_{23}=r_{24}$. In (a) $r_{12}=r_{34}$ and all $r_{ij}$ of the same order, in (b) $r_{34} \ll r_{12} \sim r_{14}$ and in (c) $r_{12} \sim r_{34} \ll r_{14}$. The average values of $\hat{Q}$ and $\hat{S}_6$ depend only on the distances depicted by straight lines. Figure (d) corresponds to the line configuration for the operator $\hat{S}_6$.}
\label{Fig:configs}
\end{figure}

First we shall consider the class of configurations introduced in
\cite{Iancu:2011ns} for which the only constraints are $r_{13}=r_{14}$
and $r_{23} = r_{24}$. For example, three such configurations are shown
in Figs.~\ref{Fig:configs}.a, \ref{Fig:configs}.b and
\ref{Fig:configs}.c. From these figures, it should be clear that there is
a high degree of variability (concerning both shapes and sizes) within
this particular class. By using the constraints aforementioned, it is
straightforward to see that Eqs.~(\ref{QevolMFA}) and (\ref{SSevolMFA})
reduce to
 \begin{align}\label{QevolMFAspecial}
 \hspace*{-0.5cm}\frac{\del \lan \hat{Q}_{\bmx_1\bmx_2\bmx_3\bmx_4}\ran_Y}{\del Y} =
 &-g^2 C_F
 [\gamma_Y(\bmx_1,\bmx_2)
 +\gamma_Y(\bmx_3,\bmx_2)
 +\gamma_Y(\bmx_3,\bmx_4)
 +\gamma_Y(\bmx_1,\bmx_4)]
 \lan \hat{Q}_{\bmx_1\bmx_2\bmx_3\bmx_4}\ran_Y
 \nn
 &\hspace*{-1cm}-\frac{g^2}{2 N_c}
 [\gamma_Y(\bmx_1,\bmx_3)
 +\gamma_Y(\bmx_2,\bmx_4)
 -\gamma_Y(\bmx_1,\bmx_2)
 -\gamma_Y(\bmx_3,\bmx_4)]
 \lan \hat{Q}_{\bmx_1\bmx_2\bmx_3\bmx_4}\ran_Y
 \nn
 &\hspace*{-1cm}-\frac{g^2}{2 N_c}
 [\gamma_Y(\bmx_1,\bmx_2)
 +\gamma_Y(\bmx_3,\bmx_4)
 -\gamma_Y(\bmx_1,\bmx_3)
 -\gamma_Y(\bmx_2,\bmx_4)]
 \lan \hat{S}_{\bmx_1\bmx_2} \hat{S}_{\bmx_3\bmx_4}\ran_Y
 \end{align}
and
 \beq\label{SSevolMFAspecial}
 \frac{\del \lan \hat{S}_{\bmx_1\bmx_2}\hat{S}_{\bmx_3\bmx_4}\ran_Y}{\del Y} =
 -2 g^2 C_F
 [\gamma_Y(\bmx_1,\bmx_2)
 +\gamma_Y(\bmx_3,\bmx_2)]
 \lan \hat{S}_{\bmx_1\bmx_2}\hat{S}_{\bmx_3\bmx_4}\ran_Y.
 \eeq
Thus, for such configurations, the evolution of a system of 2 dipoles
decouples from that of the quadrupole even without invoking separability.
By also using \eqn{effevolS} for the dipole $S$--matrix in the Gaussian
approximation at finite $N_c$, we can easily solve \eqn{SSevolMFAspecial}
to find
 \beq
 \lan \hat{S}_{\bmx_1\bmx_2}\hat{S}_{\bmx_3\bmx_4}\ran_Y =
 \frac{\lan \hat{S}_{\bmx_1\bmx_2}\hat{S}_{\bmx_3\bmx_4}\ran_{Y_0}}
 {\lan\hat{S}_{\bmx_1\bmx_2} \ran_{Y_0}
 \lan \hat{S}_{\bmx_3\bmx_4}\ran_{Y_0}}\,
 \lan \hat{S}_{\bmx_1\bmx_2} \ran_Y
 \lan \hat{S}_{\bmx_3\bmx_4}\ran_Y
 \eeq
which simplifies furthermore to
 \beq\label{SSMFAspecial}
 \lan \hat{S}_{\bmx_1\bmx_2}\hat{S}_{\bmx_3\bmx_4}\ran_Y =
 \lan \hat{S}_{\bmx_1\bmx_2} \ran_Y
 \lan \hat{S}_{\bmx_3\bmx_4}\ran_Y,
 \eeq
provided the latter holds in the initial condition at $Y_0$ (as is indeed
the case within the MV model, as one can similarly check). Then it is
clear that \eqn{QevolMFAspecial} can be solved as an inhomogeneous first
order differential equation and it gives
 \begin{align}
 &\hspace*{-1cm}\lan \hat{Q}_{\bmx_1\bmx_2\bmx_3\bmx_4}\ran_Y =
 \frac{\lan \hat{S}_{\bmx_1\bmx_2}\hat{S}_{\bmx_3\bmx_4}\ran_{Y_0}}
 {\lan\hat{S}_{\bmx_1\bmx_2} \ran_{Y_0}
 \lan \hat{S}_{\bmx_3\bmx_4}\ran_{Y_0}}\,
 \lan \hat{S}_{\bmx_1\bmx_2} \ran_Y
 \lan \hat{S}_{\bmx_3\bmx_4}\ran_Y
 +
 [\lan \hat{Q}_{\bmx_1\bmx_2\bmx_3\bmx_4}\ran_{Y_0} -
 \lan\hat{S}_{\bmx_1\bmx_2}
 \hat{S}_{\bmx_3\bmx_4}\ran_{Y_0}]
 \nn
 &\hspace*{-1cm}\times
 \sqrt{\frac{\lan \hat{S}_{\bmx_1\bmx_2} \ran_Y
 \lan \hat{S}_{\bmx_3\bmx_2} \ran_Y
 \lan \hat{S}_{\bmx_3\bmx_4} \ran_Y
 \lan \hat{S}_{\bmx_1\bmx_4} \ran_Y}
 {\lan \hat{S}_{\bmx_1\bmx_2} \ran_{Y_0}
 \lan \hat{S}_{\bmx_3\bmx_2} \ran_{Y_0}
 \lan \hat{S}_{\bmx_3\bmx_4} \ran_{Y_0}
 \lan \hat{S}_{\bmx_1\bmx_4} \ran_{Y_0}}}
 \left(
 \frac{\lan \hat{S}_{\bmx_1\bmx_4} \ran_Y
 \lan \hat{S}_{\bmx_3\bmx_2} \ran_Y}
 {\lan \hat{S}_{\bmx_1\bmx_2} \ran_Y
 \lan \hat{S}_{\bmx_3\bmx_4} \ran_Y}\,
 \frac{\lan \hat{S}_{\bmx_1\bmx_2} \ran_{Y_0}
 \lan \hat{S}_{\bmx_3\bmx_4} \ran_{Y_0}}
 {\lan \hat{S}_{\bmx_1\bmx_4} \ran_{Y_0}
 \lan \hat{S}_{\bmx_3\bmx_2} \ran_{Y_0}}
 \right)^{\textstyle{\frac{1}{2(N_c^2-1)}}}.
 \end{align}
Again, when the initial condition is given by the MV model (where
\eqn{QMFAspecial} below holds indeed, as one can explicitly check), the
above becomes very simple:
 \beq\label{QMFAspecial}
 \lan \hat{Q}_{\bmx_1\bmx_2 \bmx_3\bmx_4}\ran_Y =
 \lan \hat{S}_{\bmx_1\bmx_2} \ran_Y
 \lan \hat{S}_{\bmx_3\bmx_4}\ran_Y.
 \eeq
This result is truly remarkable: within the class of configurations at
hand, the quadrupole factorizes into two dipoles independently of how
small or large the various distances $r_{ij}$ are, and for any $N_c$, so
long as the two constraints $r_{13} = r_{14}$ and $r_{23} = r_{24}$ are
satisfied. It would be interesting to check this factorization via
numerical solutions to the JIMWLK equation, as a non--trivial test of the
present MFA.

Now let us proceed to our second example and consider the operator
 \beq\label{QSdef}
 \hat{Q}_{\bmx_1\bmx_2\bmx_3\bmx_4} \hat{S}_{\bmx_4\bmx_3}
 =\frac{1}{N_c}\,
 \rmtr({V}^{\dagger}_{\bmx_1} {V}_{\bmx_2}
 {V}^{\dagger}_{\bmx_3} {V}_{\bmx_4})\,
 \frac{1}{N_c}\,
 \rmtr({V}^{\dagger}_{\bmx_4} {V}_{\bmx_3}).
 \eeq
The choice is of direct phenomenological interest, since the operator
above is the most complicated quantity appearing in the calculation of
di--hadron production in proton--nucleus collisions\footnote{In fact, the
operator appearing in such a process is
$\hat{Q}_{\bmx_2\bmx_3\bmx_4\bmx_1} \hat{S}_{\bmx_3\bmx_4}$, but, due to
the invariance under charge conjugation, its expectation value is equal
to the one of the operator in \eqn{QSdef}.}
\cite{JalilianMarian:2004da,Nikolaev:2005dd,Baier:2005dv,Marquet:2007vb}.
Considering the same configuration as before, that is, taking
$r_{13}=r_{14}$ and $r_{23} = r_{24}$, we see that the evolution couples
the operator in \eqn{QSdef} to $\hat{S}_{\bmx_1\bmx_2}
\hat{S}_{\bmx_1\bmx_2}\hat{S}_{\bmx_2\bmx_1}$. After a straightforward
calculation, similar to the one leading at \eqn{QMFAspecial}, we arrive
at
 \begin{align}
 \hspace*{-0.5cm}\lan \hat{Q}_{\bmx_1\bmx_2\bmx_3\bmx_4} \hat{S}_{\bmx_4\bmx_3} \ran_Y
 =\lan \hat{S}_{\bmx_1\bmx_2} \hat{S}_{\bmx_3\bmx_4} \hat{S}_{\bmx_4\bmx_3} \ran_Y
 =
 \frac{N_c^2 -1}{N_c^2}\,
 \lan \hat{S}_{\bmx_1\bmx_2}\ran_Y
 \lan \hat{S}_{\bmx_3\bmx_4}\ran_Y^{\textstyle\frac{2 N_c^2}{N_c^2 - 1}}
 +\frac{1}{N_c^2}\,\lan \hat{S}_{\bmx_1\bmx_2}\ran_Y.
 \end{align}
(Once again, we have assumed this equation to hold already in the initial
condition at $Y_0$, which is in particular true within the MV model, as
one can check.) It is interesting that for this configuration, and
similar to \eqn{QMFAspecial}, the result depends only on $r_{12}$ and
$r_{34}$, but not on $r_{14}$ and $r_{23}$. Also, by keeping $N_c$
finite, we notice that the last term in the r.h.s. (the single dipole
$S$--matrix) dominates deeply at saturation. This is in agreement with
our earlier discussion in \ref{sect:RPA}, since the linear combination
 \beq
 \hat{Q}_{\bmx_1\bmx_2\bmx_3\bmx_4} \hat{S}_{\bmx_4\bmx_3} -
 \frac{1}{N_c^2}\, \hat{S}_{\bmx_1\bmx_2}
 \eeq
is a special case of the operator in \eqn{QSreal} for this particular
configuration. In fact, the operator that appears in the di--hadron cross
section is the combination
 \beq\label{S6}
 \hat{S}_{6\,\bmx_1\bmx_2\bmx_3\bmx_4} =
 \frac{N_c^2}{N_c^2 - 1}\,
 \hat{Q}_{\bmx_1\bmx_2\bmx_3\bmx_4} \hat{S}_{\bmx_4\bmx_3}
 - \frac{1}{N_c^2 - 1}\,\hat{S}_{\bmx_1\bmx_2},
 \eeq
which is also the quantity studied numerically in \cite{Dumitru:2011vk}.
Using the previous results one finds that, for our special configuration,
the expectation value of $\hat{S}_6$ is given by the relatively simple
expression
 \beq\label{S6MFAspecial}
 \lan \hat{S}_{6\,\bmx_1\bmx_2\bmx_3\bmx_4} \ran_Y
 = \lan \hat{S}_{\bmx_1\bmx_2}\ran_Y
 \Big[\lan \hat{S}_{\bmx_3\bmx_4}\ran_Y\Big]
 ^{\textstyle\frac{2 N_c^2}{N_c^2 - 1}}.
 \eeq
It is amusing to note that for this particular configuration and for
large $N_c$, the 4--point function relevant for di--hadron production
factorizes as $\lan \hat{S}_{6\,\bmx_1\bmx_2\bmx_3\bmx_4} \ran_Y
 = \lan \hat{S}_{\bmx_1\bmx_2}\ran_Y \hat{S}_{\bmx_3\bmx_4}\ran_Y^2$.
This is precisely the factorization formula used (for a {\em generic}
configuration) in the phenomenological study in \cite{Albacete:2010pg}
--- at that time, by lack of a better formula. Such a factorization
however has no deep justification and is merely a property of the
configuration at hand. As our next example will show, this
`factorization' can badly fail for other, equally simple, configurations.

Specifically, let us consider the expectation value of the operator
\eqref{QSdef} for the `line' configuration studied in
\cite{Dumitru:2011vk,Iancu:2011ns} and shown in Fig.~\ref{Fig:configs}.d.
Note that the two quarks of the quadrupole and the antiquark of the
dipole are put in a same point ($\bmx_1=\bmx_3$), and similarly for the
two antiquarks of the quadrupole and the quark in the dipole
($\bmx_2=\bmx_4$). Thus, only one non--trivial distance $r \equiv
r_{12}=r_{23}=r_{34}=r_{14}$ characterizes the configuration. Then one
can easily check that the evolution of
$\hat{Q}_{\bmx_1\bmx_2\bmx_1\bmx_2} \hat{S}_{\bmx_2\bmx_1}$ couples again
to $\hat{S}_{\bmx_1\bmx_2} \hat{S}_{\bmx_1\bmx_2}
\hat{S}_{\bmx_2\bmx_1}$, leading to a $2 \times 2$ inhomogeneous system
of equations. Expressing $\gamma_Y(r)$ in terms of the dipole $\lan
\hat{S}(r) \ran_Y$ and using an obvious shorthand notation we have
 \begin{align}
 &\frac{\del \lan \hat{Q}\hat{S} \ran_Y}{\del \ln \lan \hat{S} \ran_Y} =
 \frac{3 N_c^2 - 1}{N_c^2 - 1}\,
 \lan \hat{Q}\hat{S} \ran_Y
 +
 \frac{2 N_c^2}{N_c^2 - 1}\,
 \lan \hat{S}^3 \ran_Y
 -
 \frac{4}{N_c^2 - 1}\,
 \lan \hat{S} \ran_Y,
 \\
 &\frac{\del \lan \hat{S}^3 \ran_Y}{\del \ln \lan \hat{S} \ran_Y} =
 \frac{2}{N_c^2 - 1}\,
 \lan \hat{Q}\hat{S} \ran_Y
 +
 \frac{3 N_c^2 - 1}{N_c^2 - 1}\,
 \lan \hat{S}^3 \ran_Y
 -
 \frac{4}{N_c^2 - 1}\,
 \lan \hat{S} \ran_Y.
 \end{align}
The solution to this system is straightforward to obtain; so long as
$\lan \hat{Q}\hat{S} \ran_Y$ is concerned, one finds
 \beq\label{QSfull}
 \lan \hat{Q}\hat{S} \ran_Y =
 \frac{(N_c+2)(N_c-1)}{2 N_c}\,
 \lan \hat{S} \ran_Y^{\textstyle{\frac{3N_c-1}{N_c-1}}}
 -
 \frac{(N_c+1)(N_c-2)}{2 N_c}\,
 \lan \hat{S} \ran_Y^{\textstyle{\frac{3N_c+1}{N_c+1}}},
 \eeq
where we assumed that the above is already valid at $Y_0$, as is the case
in the MV model. Using this result together with \eqn{S6}, it is
straightforward to evaluate $\lan \hat{S}_6 \ran_Y$ for this particular
configuration and compare with the numerical results in
Ref.~\cite{Dumitru:2011vk}. We shall find it rewarding to plot $\lan
\hat{S}_6 \ran_Y$ in two different ways; first as a function of $r Q_s$
and then as a function of $1 - \lan \hat{S} \ran_Y$. To be in accordance
with \cite{Dumitru:2011vk}, the saturation momentum is defined by the
condition $\lan \hat{S} \ran_Y = 1/\sqrt{e}$ for $r Q_s = \sqrt{2}$.

\begin{figure}
\centerline{\includegraphics[width=0.52\textwidth]{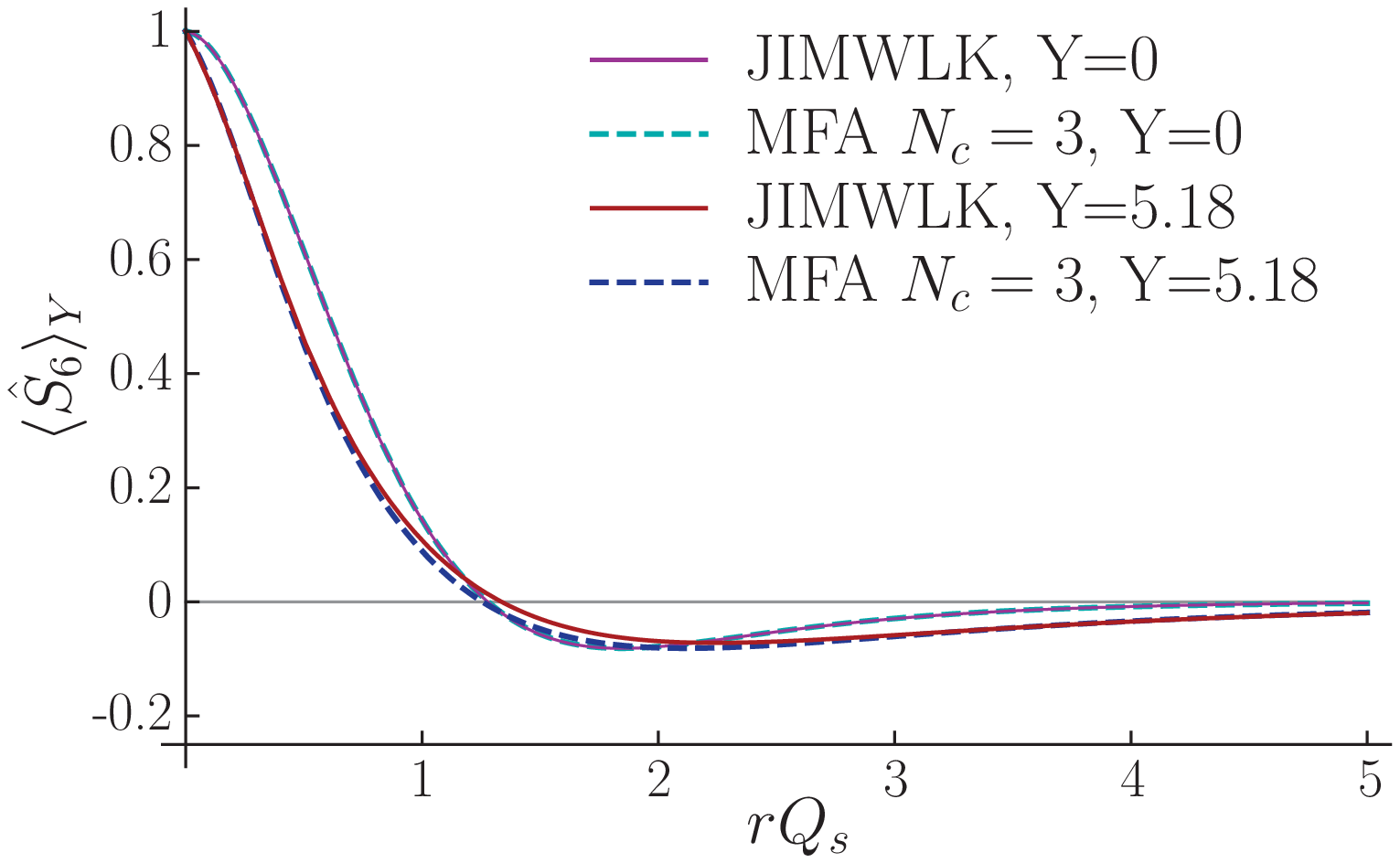}
\hspace*{-0.04\textwidth}
\includegraphics[width=0.52\textwidth]{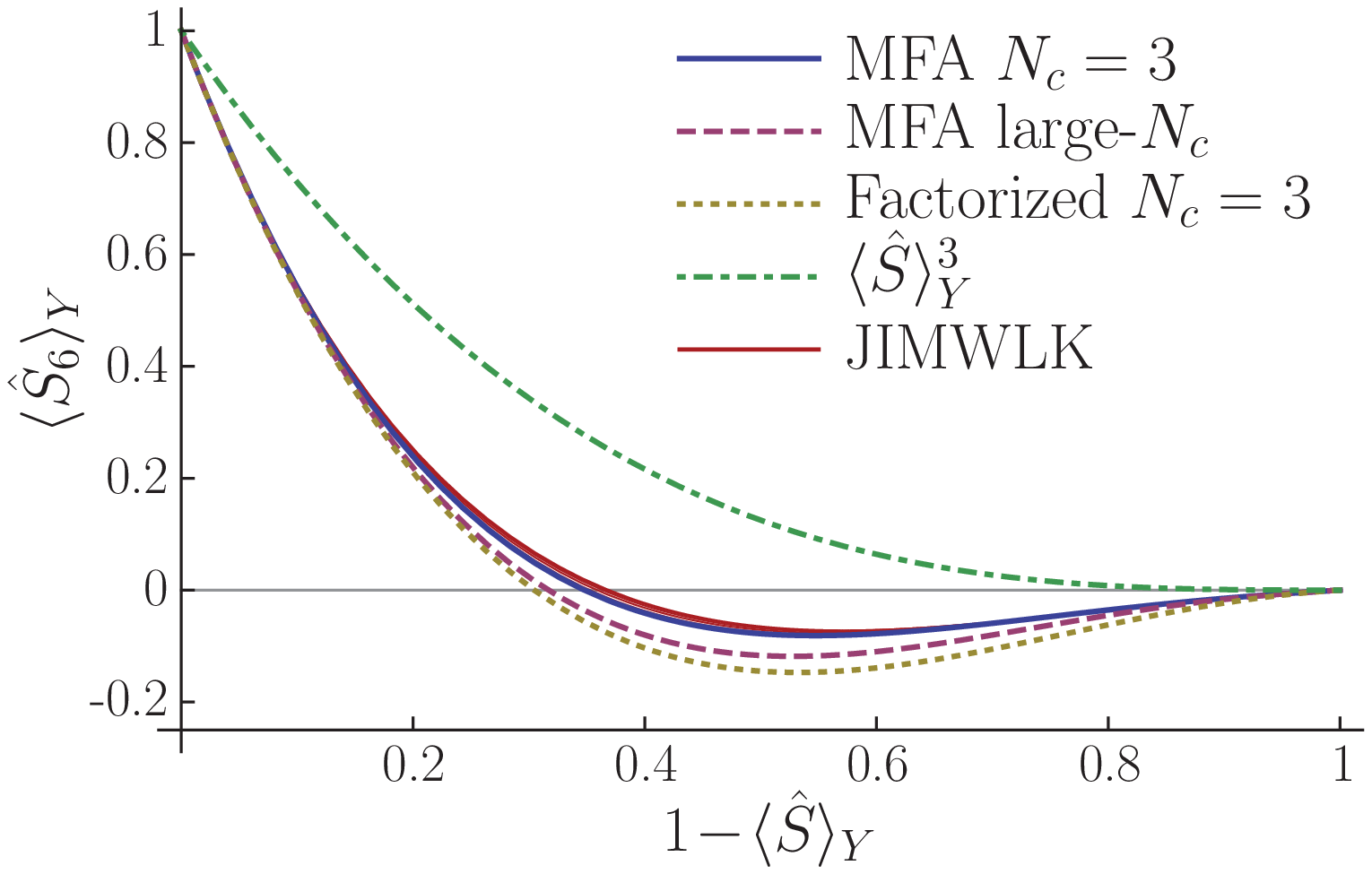}}
\caption{\sl The expectation value of $\hat{S}_6$, as defined in \eqn{S6},
for the `line' configuration. Left: as a function of the scaling variable
$r Q_s$. Continuous magenta: JIMWLK at Y=0. Dashed cyan: MFA for $N_c$=3 at Y=0. Continuous red: JIMWLK at Y=5.18. Dashed blue: MFA for $N_c$=3 at Y=5.18. Right: as a function of $1\!-\!\lan \hat{S} \ran_Y$.  Continuous red: JIMWLK at six different values of rapidity Y=0, 1.04, 2.07, 3.11, 4.14 and 5.18.  Continuous blue: complete result in the MFA for $N_c=3$. Dashed magenta: large--$N_c$ result in the MFA. Dotted gold: assuming factorization for the expectation value of $\hat{Q} \hat{S}$ and using the MFA for $N_c=3$ for the expectation value of the quadrupole $\hat{Q}$. Dotted dashed green: for illustrative purposes we also show $\lan \hat{S} \ran_Y^3$. JIMWLK curves are constructed from the numerical solution given in \cite{Dumitru:2011vk}. MFA curves are analytical expressions in terms of $\lan \hat{S} \ran$, which is again provided by the numerical solution in \cite{Dumitru:2011vk} for the purposes of the left figure.}
\label{Fig:S6Line}
\end{figure}

We show this comparison in Fig.~\ref{Fig:S6Line}, where for some of the
curves we have used the numerical data of \cite{Dumitru:2011vk}. On the
left we show the correlator of interest as a function of $r Q_s$ and for
two values of the rapidity: $Y=0$, which is where one starts the
evolution (with initial conditions of the MV type), and $Y=5.18$, which
is large enough for the effects of the evolution to be fully developed.
The curves denoted as `MFA' represent our present results, cf.
\eqn{QSfull} and \eqref{S6}, whereas the `JIMWLK' curves follow from the
numerical solution to the JIMWLK equation. For $Y=0$, the two types of
curves overlap with each by construction, as they both reduce to the
respective prediction of the MV model. What is remarkable though, is that
a very good agreement between the numerical solution and the MFA persists
for $Y=5.18$. In the limiting regimes of weak and respectively strong
scattering, the respectives curves are practically indistinguishable. In
the transition region around $r Q_s \sim 1$, the agreement is not that
perfect anymore, but the two curves are still very close to each other,
confirming that the MFA is also an excellent {\em global} approximation.

From Fig.~\ref{Fig:S6Line} (left) we also notice that the shape of the
curve changes as we evolve from $Y=0$ to higher values of rapidity
($Y=5.18$ in the figure). However, this change is mostly attributed to
the evolution of $\lan \hat{S} \ran_Y$ as a function of $r Q_s$ as the
rapidity grows. Indeed, in the right panel of Fig.~\ref{Fig:S6Line} we
show the correlator of interest as a function of $1 - \lan \hat{S}
\ran_Y$. One can see that the curves obtained from the numerical solution
to JIMWLK for various values of $Y$ form a very thin ``band'' whose
borderline on the ``lower'' side is the MFA. This ``band'' is practically
a line perfectly overlapping with the MFA when the scattering is either
weak or strong, and becomes just a bit wider in the transition region.
Thus, as we evolve in rapidity, the shape of the curve is barely
changing. In fact, if one expands out the plot in order to better
disentangle the various steps in the evolution, one can see that the
high--$Y$ curve stabilizes very close to the $Y=0$ curve after just a few
units in rapidity.

\comment{The stabilization in the shape of such a curve is a clear
manifestation of geometric scaling. Since both $\lan \hat{S}_6 \ran_Y$
and $1 - \lan \hat{S} \ran_Y$ are functions of only $rQ_s$ in the
asymptotic regime at high $Y$, we deduce that $\lan \hat{S}_6 \ran_Y$
(and of course any other higher--point correlator like the quadrupole
$\lan \hat{Q} \ran_Y$) should approach a universal shape as a function of
$1 - \lan \hat{S} \ran_Y$, independently of the initial conditions. Given
also the fact that the two limits of weak and strong scattering are very
well described in the Gaussian approximation, we understand that this
asymptotic curve should be very close to the initial curve, so long as
the latter is also given by a Gaussian wavefunction like the MV model.}

Still in the right plot we also show two different approximations for
$\lan \hat{S}_6 \ran_Y$, the one is the large-$N_c$ result, while the
other consists of factorizing $\lan \hat{Q} \hat{S} \ran_Y$ into $\lan
\hat{Q} \ran_Y \lan \hat{S} \ran_Y$, as it would be justified at large
$N_c$, but then using the finite--$N_c$ Gaussian approximation for $\lan
\hat{Q} \ran_Y$. This latter approximation takes into account some
$1/N_c^2$ corrections, but not in a systematic way, and was used in
\cite{Dumitru:2011vk} since the (MV--like) expression \eqref{QSfull} was
not available at the time. Comparing with the numerical findings in
\cite{Dumitru:2011vk}, we already saw that the complete result
\eqn{QSfull} at finite--$N_c$ is the one which shows the best agreement.
Even though it is not very significant, we note that the large--$N_c$
expression is the next one closer to the numerical data, perhaps because
it is at least a systematic approximation. For illustrative purposes we
also show $\lan \hat{S} \ran_Y^3$, which is based neither on a
large--$N_c$ approximation nor on a mean field one, but simply
corresponds to a `naive' counting of Wilson lines. It fails badly even in
the BFKL regime and clearly it has no chance to describe properly the
correlator of interest.

\section*{Acknowledgments}
We would like to thank the Galileo Galilei Institute for Theoretical
Physics for its hospitality and the INFN for partial support during the
completion of this work. We furthermore thank Fabio Dominguez and Al
Mueller for insightful remarks, in particular on the physical meaning of
the mirror symmetry. We acknowledge useful discussions with Bjoern
Schenke and Raju Venugopalan and we are indebted to them for providing us
with their numerical data. Figures \ref{Fig:mirror}, \ref{Fig:flow} and
\ref{Fig:configs} were created with Jaxodraw
\cite{Binosi:2003yf,Binosi:2008ig}.

%\appendix

%\bibliographystyle{utcaps}
%\bibliography{refs}
\providecommand{\href}[2]{#2}\begingroup\raggedright\endgroup

\end{document}